\documentclass[useAMS,usenatbib]{mn2e} 
\pdfoutput=1 
\usepackage{journals}
\usepackage[pdftex]{graphicx,color} 
\usepackage[latin1]{inputenc}
\usepackage{graphics} 
\usepackage{amsfonts} 
\usepackage{amsmath}
\usepackage{multicol} 
\usepackage{layout} 
\usepackage{amssymb}
\title[Cosmology in 2D]
{Cosmology in 2D: the concentration-mass relation for galaxy clusters}
\author[Giocoli et al. 2012]
{\parbox{\textwidth}{Carlo Giocoli$^{1,2,3}$\thanks{E-mail:
{cgiocoli@oabo.inaf.it}},
 Massimo Meneghetti$^{2,3}$, Stefano Ettori$^{2,3}$, 
Lauro Moscardini$^{1,2,3}$} \\ \\ 
 $^{1}$ Dipartimento di Astronomia, Universit\`a di Bologna,
 via Ranzani 1, 40127, Bologna, Italy \\
 $^{2}$ INAF - Osservatorio Astronomico di Bologna, via Ranzani 1, 40127, Bologna,
 Italy \\ 
 $^{3}$ INFN - Sezione di Bologna, viale Berti Pichat 6/2, 40127,
 Bologna, Italy}
\begin{document}
\date{}
\maketitle
\label{firstpage}
\pagerange{\pageref{firstpage}--\pageref{lastpage}} \pubyear{2012}
\begin{abstract}

  The  aim of  this  work is  to  perform a  systematic  study of  the
  measures  of the  mass and  concentration estimated  by  fitting the
  convergence profile  of a large  sample of mock galaxy  cluster size
  lenses, created with the  publicly available code \texttt{MOKA}.  We
  found  that  the  main contribution  to  the  bias  in mass  and  in
  concentration  is due  to the  halo  triaxiality and  second to  the
  presence of  substructures within the  host halo virial  radius.  We
  show that  knowing the  cluster elongation along  the line  of sight
  helps in correcting the mass  bias, but still keeps a small negative
  bias for the concentration.   If these mass and concentration biases
  will characterize the  galaxy cluster sample of a  wide field survey
  it  will  be  difficult  to   well  recover  within  one  sigma  the
  cosmological  parameters that mainly  influence the  $c-M$ relation,
  using  as reference  a 3D  $c-M$ relation  measured  in cosmological
  $N$-body simulation.   In this  work we propose  how to  correct the
  $c-M$ relation for projection  effects and for adiabatic contraction
  and  suggest to  use  these  as reference  for  real observed  data.
  Correcting mass and concentration  estimates, as we propose, gives a
  measurement   of  the   cosmological  parameter   within  $1-\sigma$
  confidence contours.

\end{abstract}

\section{Introduction}
Galaxy clusters  represent a  very important cosmic  laboratory. Their
abundance and  structural properties are correlated  with the content,
the  formation and  the  evolution history  of  the universe.   Recent
observational  analyses  have  favored  the dark  energy  as  dominant
component,  contributing  to  the  late accelerate  expansion  of  the
universe \citep{perlmutter98,perlmutter99}, and the dark matter as the
second one,  made up of  some kind of non-baryonic  weakly interacting
elementary particle, left over from  the Big Bang, this represents the
concordance $\mathrm{\Lambda  CDM}$ model.  Galaxy  clusters, the most
bound and  late forming structures  of the universe, possess  a matter
content    which     is    compatible    with     the    cosmic    one
\citep{ettori09}. About $85\%$ of the  total galaxy cluster mass is in
form of dark matter, while  $15\%$ is attributed to baryons: $75\%$ of
which is in form of hot  and $7\%$ of cold gas, other baryon fractions
count  for the  remaining  $18\%$.  For  this  reason, cluster  number
counts represent a very important  probe for the nature of dark energy
and matter, as like as the study of their mass and concentration, that
we will present and discuss in this work.  Observing them at different
wavelengths gives  the possibility to  study their whole  content: the
hot and cold gas, stars in galaxies and the presence of dark matter.

Numerical simulations  of structure formation  in the $\mathrm{\Lambda
  CDM}$ framework predict the halo concentration, defined as the ratio
between the virial and the scale radii of the density profile, to be a
monotonic function of  the host halo mass: smaller  haloes, forming at
higher redshifts than  larger ones, tend to possess  a larger value of
the  concentration.   Recently,  increasing  the force  and  the  mass
resolution in numerical simulation has given the possibility to span a
very  large  halo mass  range  down  to  the size  of  dwarf-galaxies.
\citet{dolag04,neto07,maccio07,maccio08,gao08},    interpreting    the
results  of  different  numerical  simulations,  agree  in  finding  a
power-law   relation  between   the   average  values   of  mass   and
concentration.  On the  other hand,  more  theoretical interpretations
have    been    given   to    the    average    $c-M$   relation    by
\citet{zhao09,giocoli12b}, that relate the concentration to the time at
which the main  halo progenitor assembles a certain  fraction of their
mass.  Since observationally the halo formation time, $t_f$ -- defined
as the  time at which the  main halo progenitor assembles  half of its
mass  \citep{lacey93,giocoli07a}   --  is  not   a  direct  measurable
quantity, we need  a good and unbiased estimate of  the host halo mass
and concentration to translate them in $t_f$ \citep{giocoli12b}.
The Cluster  Lensing And Supernova survey with  Hubble (CLASH) project
\citep{postman11},  together  with   X-ray  and  ground-based  optical
observations, is  providing the  2D mass distribution  of a  sample of
$25$  X-ray selected galaxy  clusters.  The  clusters are  selected as
dynamically relaxed and span a  range of redshifts going from $z=0.18$
up to $z=0.9$.  The cluster mass density distribution is reconstructed
combining both  strong and weak  lensing analyses. The  strong lensing
analysis is performed thanks  to the identification of multiple images
\citep{zitrin09} allowed by the excellent HST angular resolution.  The
reconstruction of the mass distribution is performed assuming that the
galaxy cluster light  traces the matter distribution made  up of three
components: the cluster  galaxies, a dark matter halo  and an external
shear to account for  additional ellipticity in the mass distribution.
The weak lensing analysis,  which probes the mass distribution outside
the  strong  lensing region,  is  conducted  using wider  ground-based
images   obtained  from   Subaru   \citep{umetsu09,okabe10},  combined
together  with   KPNO  Mayall   4-m  imaging  and   spectroscopy  from
MMT/Hctospec.   In   this  case,  selected   background  galaxies  are
identified  from the  wider cluster  images  and used  to measure  the
observed   shear  and  magnification.    Deriving  the   mass  profile
simultaneously combining the strong and weak lensing analyses allows a
very  good  determination of  both  the  halo  mass and  concentration
\citep{meneghetti10b,rasia12}. Unlike most other methods that are used
to measure the  mass and concentration of galaxy  clusters, like X-ray
analyses  that need  to  assume hydrostatic  equilibrium  for the  hot
intracluster medium or satellite galaxy velocity distribution function
that  needs virial equilibrium  \citep{becker07}, lensing  analyses do
not require any assumption of the dynamical state of the cluster.  The
interpretation  based  on the  assumption  of  hydrostatic and  virial
equilibrium is sometimes complicated by the presence of massive and/or
numerous  substructures and  by the  fact that  most clusters  are not
really relaxed.

Most of  the analyses  performed on combined  weak and  strong lensing
observations tend  to find clusters over-concentrated  with respect to
what is expected in a  $\mathrm{\Lambda CDM}$ universe.  This bias can
be due both  to intrinsic projection effects --  main halo triaxiality
\citep{meneghetti07b} and/or presence  of substructures along the line
of  sight -- and  to the  presence of  a massive  background structure
\citep{coe12}. To be  able to recover a good  and unbiased estimate of
the  dark mass  and  concentration is  important  to better  constrain
cosmological parameters.

\citet{meneghetti10b,rasia12}, studying galaxy clusters extracted from
numerical simulations,  have shown that  fitting projected quantities,
using a  \citet{navarro97} (hereafter; NFW) function,  recovers a mass
that could be  underestimated down to $10-15\%$.  The  halo mass tends
to be  more biased generally when  the 2D projected  mass is unrelaxed
due to presence of nearby structures.

Recently, \citet{bahe12}  have produced mock observations  of a sample
of massive clusters extracted  at redshift $z=0.2$ from the Millennium
Simulation  \citep{springel05b}, to  obtain a  projected  estimate for
both halo  mass and concentration through simulated  weak lensing only
analysis.  They  showed that  the slope and  the normalization  of the
derived 2D $c-M$ relation tend to be lower than the underlying 3D ones
by $40$ and $15$ percent respectively.  An underestimation of the mass
measured  from  weak gravitational  lensing  has  been  also found  by
\citet{becker11}, using cluster-size haloes extracted from a numerical
simulation.  Their  analysis includes correlated  and uncorrelated LSS
plus gaussian noise in the tangential shear measurement.

Light  travelling from  background  sources to  the  observer is  also
deflected by  the presence of  large scale structures (LSS)  along the
line of  sight: the  observed lensing signal  is a combination  of the
cluster  and   the  cosmic  shear   signal  \citep{hoekstra03}.   This
combination limits the accuracy with which both mass and concentration
can   be    estimated,   not   introducing    any   particular   bias.
\citet{hoekstra11},  also  using  the Millennium  Simulation  results,
derived  that  this  contribution   adds  an  error  budget  which  is
comparable to the  statistical errors.  In order to  bypass the effect
of LSS and also the errors  and the biases that could be introduced on
the  cluster  mass  reconstructions  by  the  data  analysis  software
pipelines, we build a large  sample of galaxy cluster convergence maps
using the publicly available software \texttt{MOKA} \citep{giocoli12},
and fit their derived spherical averaged profile to recover their mass
and concentration.

Throught this  work we use  a $\mathrm{\Lambda CDM}$  reference model,
according  to WMAP7-year  results  \citep{komatsu11}: $\Omega_m=0.272$
(DM+baryons),  $\Omega_{\Lambda}=0.728$, $w=-1$,  $\sigma_8=0.809$ and
$h=0.704$ at $z=0$.

The  paper  is organized  as  follow.   In  Section \ref{section1}  we
summarize the  ingredients used to create our  cluster size-haloes. In
Section  \ref{section2} we  quantify the  bias and  scatter  caused by
intrinsic halo properties.  Section \ref{section3} is devoted to study
how well  cosmological parameters can  be constrained using  the $c-M$
relation of  objects extracted from  a wide field survey.  Summary and
conclusions are discussed in Section \ref{lastsection}.

\section{The Method}
\label{section1}
\subsection{The MOKA Code}
\texttt{MOKA} \citep{giocoli12} is a new  and fast code able to create
realistic gravitational  lenses, whose size spans from  galaxies up to
clusters            of           galaxies\footnote{web           page:
  \texttt{cgiocoli.wordpress.com/research-interests/moka}}.         The
algorithm uses the recent  results obtained from numerical simulations
to model the structural properties of the lenses.  In creating cluster
2D-maps  \texttt{MOKA}  is  very  fast, its  utilization  is  friendly
allowing  the  possibility of  switching  on  and  off different  halo
properties.   The  user  can  also  easily  change  and  redefine  the
characteristics  of the  main halo,  of  subhaloes and  of the  bright
central galaxy.

\subsection{Presence of Substructures and of a Bright Central Galaxy}
In this  work we produced  a set of  simulations creating a  sample of
different galaxy  cluster size lenses  at various redshifts.   We have
divided the haloes in nine mass bins from $10^{13.5}$ up to $10^{15.5}
M_{\odot}/h$,      logarithmically       equally      spaced      with
$\mathrm{d}\log(M)=0.25$.  In  each mass bin, with  mass $M_{vir}$, we
have generated a sample of  $128$ lenses, that differ in concentration
$c_{vir}$,  randomly sampling  a log-normal  distribution,  around the
average value  for that  mass, with a  given $\sigma_{\ln c}$,  and in
subhalo  abundance.   In what  follows  we  summarize the  ingredients
included in the input file of the algorithm.

\begin{itemize}
\item One  of the main result obtained  studying virialized structures
  in numerical  simulations is that  haloes possess a  relatively flat
  density  profile  $\propto  r^{-1}$  near  the  center  and  steeper
  $\propto  r^{-3}$  outside   \citep{navarro97}.   The  scale  radius
  $r_{s}$  where  the slope  changes  defines  the halo  concentration
  $c_{vir}\equiv  R_{vir}/r_s$.   For  each  halo,   the  smooth  halo
  component is  modelled with  a NFW profile,  to which is  assigned a
  concentration $c_{sm}$ such that:
  \begin{eqnarray}
    & &\rho_{NFW,vir}(r|M_{vir},c_{vir}) =
    \rho_{NFW,s}(r|M_{sm},c_{sm}) + \nonumber \\
    & &\sum_{i=1}^{N_{tot}} \rho_{sub}(r-r_i|m_{sub,i})\,,
  \end{eqnarray}
  where the subscript  $vir$, $sm$ and $sub$ refer  to the total halo,
  the  smooth  component  and  the substructures  respectively,  $r_i$
  indicates the distance of the  subhalo from the host halo center and
  $\rho_{sub}(<r)$ is the subhalo density profile.  The equation above
  is solved numerically by minimizing:
\begin{eqnarray}
  & & \chi^2(c_{sm}) = \Bigg{\{} \rho_{NFW,vir}(r|M_{vir},c_{vir}) - \\
  \nonumber & & \left[  \rho_{NFW,s}(r|M_{sm},c_{sm})  + 
    \sum_{i=1}^{N_{tot}} \rho_{sub}(r-r_i|m_{sub,i}) \right] \Bigg{\}}^2\,.
\end{eqnarray}
  
\item In the standard  scenario of structure formation haloes collapse
  as consequence of gravitational  instability, with the small systems
  assembling  first,  in  a  denser  universe,  than  the  large  ones
  \citep{wechsler02}.  This translates in small haloes to be typically
  more  concentrated than the  large ones.   In our  model we  use the
  \citet{zhao09}  relation to  take  into account  this effect,  which
  relates the  concentration of the  halo with the time  $t_{0.04}$ at
  which the main  halo progenitor assembles $4\%$ of  its mass through
  the equation:
  \begin{equation}
    c_{vir}(M_{vir},z) = 4 \left\{ 1 +
      \left[ \frac{t(z)}{3.75 t_{0.04}}\right]^{8.4} \right\}^{1/8}\,.
    \label{eqzhao}
  \end{equation}
  According to this model we  estimate the scatter in concentration at
  fixed halo mass using the scatter in $t_{0.04}$ from the generalized
  formation redshift  distribution derived by  \citet{giocoli12b}.  In
  Figure  \ref{scattercFIG}  we show  the  redshift  evolution of  the
  scatter  in  concentration as  a  function  of  redshift.  From  the
  figure, we  notice that  at low redshift  the scatter  is consistent
  with       the      results      of       numerical      simulations
  \citep{jing00,jing02,wechsler02,dolag04,sheth04a,neto07}  to  be  of
  the order of  $0.23$, while at high redshift  $\sigma_{\ln c}$ drops
  down due to the contraction  of the typical halo merging time scale.
  However, the decrease of the  scatter at high redshift does not seem
  to   exist  in   numerical  simulation,   as  recently   noticed  by
  \citet{bahttacharya11}.  The difference can  be due to the fact that
  numerical simulations are limited in mass resolution going to higher
  $z$ or that the scatter in  $c_{vir}$ at a fixed mass is not totally
  due to the different assembly halo histories.
  
\begin{figure}
\includegraphics[width=\hsize]{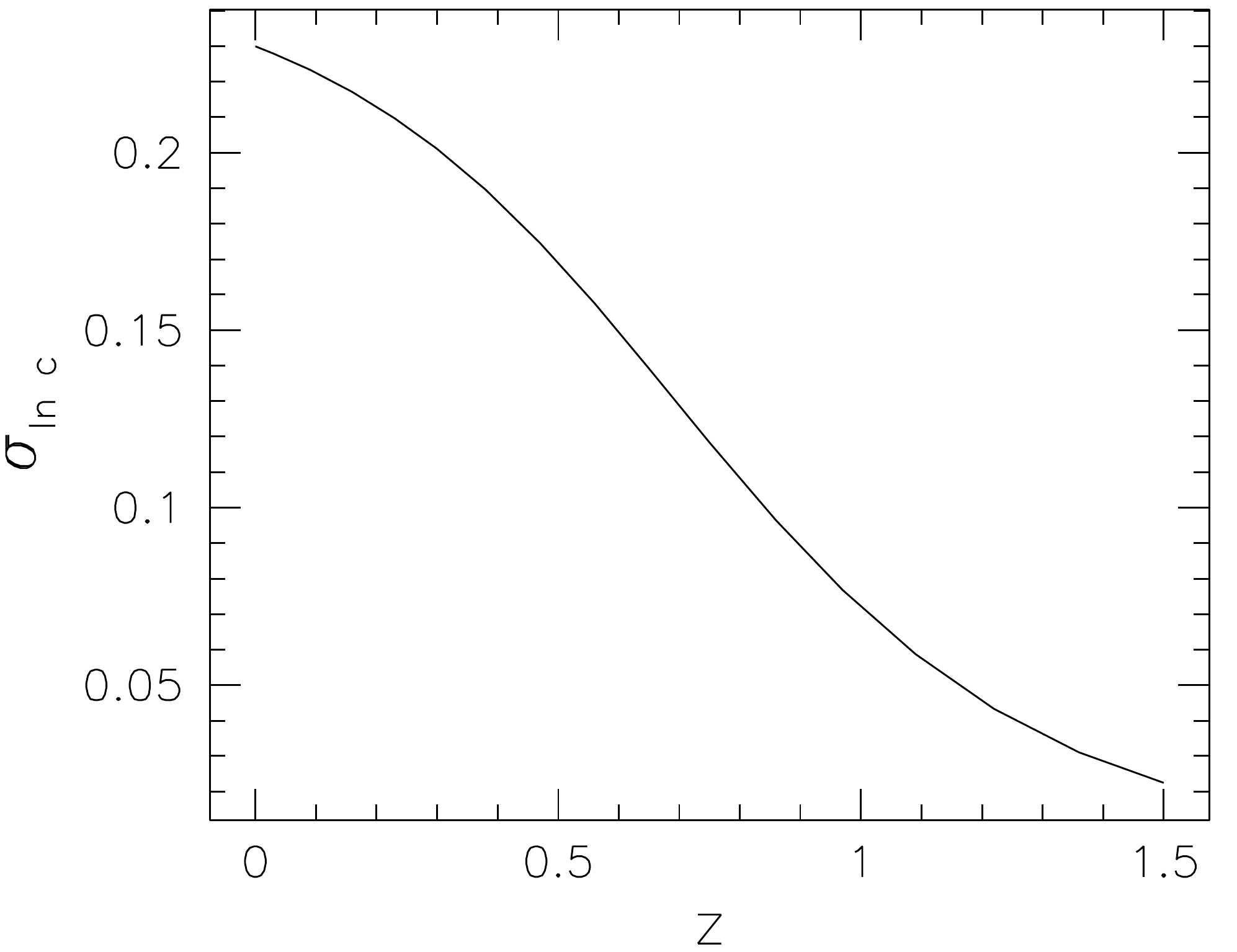}
\caption{\label{scattercFIG}Scatter in  concentration $\sigma_{\ln c}$
  at a fixed host halo mass as a function of $z$ due to the scatter in
  time when the main halo  progenitor assembles $4\%$ of its mass. The
  scatter has been  obtained mass averaging the scatter  at fixed halo
  mass  of a  sample  of haloes  with  mass larger  than $3.16  \times
  10^{13}M_{\odot}/h$.}
\end{figure}

\item  The improvement of  the computational  resources has  given the
  possibility  to  increase mass  and  force  resolution in  numerical
  simulations. One  of the first results revealed  by this improvement
  is that  haloes are not completely  smooth but made  up of different
  clumps. Since merging events are not totally efficient, these clumps
  represent the  core of progenitor haloes accreted  during the growth
  of           the            main           halo           progenitor
  \citep{springel01b,gao04,giocoli08b}. Different studies based on the
  analysis of  numerical simulations have  converged on the  fact that
  the subhalo mass function has  a power-law distribution with a slope
  $\alpha=-1.9$.  In modeling  the subhalo population in \texttt{MOKA}
  we will use the model proposed by \citet{giocoli10} which takes into
  account the dependence on redshift and  on the host halo mass of the
  distribution, plus  the scatter at  fixed mass due to  the different
  assembly history; this means  that the scatters of the concentration
  and  subhalo  mass  function,  for   a  fixed  halo  mass,  are  not
  independent  but are  correlated: haloes  more concentrated,  and so
  forming at higher  redshift, are more relaxed and  typically tend to
  possess   less  substructures   \citep{gao04,giocoli08b}.   We  will
  populate the host halo sampling  the subhalo mass function down to a
  mass resolution  of $10^{10}M_{\odot}/h$, which  is of the  order of
  the minimum subhalo mass  resolved in current cosmological numerical
  simulations \citep{springel05b,gottlober06,boylan-kolchin09}.
\item  The subhalo  density  distribution inside  the  hosts tends  to
  follow that of the host dark matter component. However, the combined
  effect of  gravitational heating and tidal stripping  does not allow
  subhaloes to be present in the central region of the cluster, mainly
  dominated  by  the density  profile  of  the  smooth component  that
  increases  like  $r^{-1}$.  This  translates  into  a clump  density
  distribution which  is less concentrated and shallower  than the NFW
  profile toward the centre.  To spatially locate the subhaloes within
  the  virial radius  of the  cluster we  will adopt  the  relation by
  \citet{gao04},  since  it takes  into  account  these effects  being
  extracted from simulated cluters.
\item Haloes trace  the dark matter density field  in the Universe and
  their  clustering   is  a  function   of  both  the  mass   and  the
  redshift. Their  mutual gravitational interaction tends  to pull and
  stretch  their matter  distribution making  the haloes  triaxial. We
  model the halo triaxiality  using the results by \citet{jing02} from
  numerical studies. This model  correlates the axial ratios $e_a/e_c$
  and $e_b/e_c$ ($e_a<e_b<e_c$) with the typical collapsed mass $M^*$,
  at the considered  redshift, and the matter content  of the universe
  $\Omega_{m}$; apart  from this no other  tighter correlation between
  the halo triaxiality and its formation time is assumed.
\item  Orbiting in  the potential  well of  the host,  subhalo density
  profile tends to  be modified since (i) the  particles are heated at
  the  local temperature  of the  host, (ii)  pushed toward  the tidal
  radius and  (iii) eventually stripped.   This phenomenology modifies
  the initial NFW  profile into something that is  well described by a
  singular  isothermal sphere (SIS)  \citep{metcalf01}. The  radius of
  the subhaloes  is computed  in such away  preserve the  subhalo mass
  $m_{sub}$ and is defined as:
\begin{equation}
R_{sub} = \frac{G m_{sub}}{2 \sigma_v^2}\,.
\end{equation}
As  discussed in \citet{giocoli12}  we notice  that typically,  if the
subhalo is  located at a distance  $r$ from the host  halo center, the
following condition  is valid  $R_{t}(r) \leq R_{sub}$,  with $R_t(r)$
defined as \citep{tormen98b}:
\begin{equation}
R_t = r \left\{ \frac{m_{sub}}{ [ 2 - \partial \ln M_{vir}(r)/\partial
    \ln r ] M_{vir}(r)} \right\}^{1/3} \,,
\end{equation}
where  $M_{vir}(r)$ represents the  mass density  profile of  the host
halo.   The SIS  profile  will be  used  to model  the matter  density
distribution around clumps.
\item  In   the  standard  scenario   \citep{white78},  the  structure
  formation process is driven  by the dark matter component.  Baryons,
  feeling the halo gravitational potential, shock, cool and eventually
  form stars.   This phenomenology drives  the formation of  a massive
  and bright  galaxy at  the center of  the main halo  progenitor. The
  correct  modeling of the  bright central  galaxy (BCG)  is important
  since  it influences the  lensing cross  section of  galaxy clusters
  \citep{meneghetti03}.  The  central galaxy population  is modeled in
  \texttt{MOKA}   using  the   halo   occupation  distribution   (HOD)
  approach. We use the relation  by \citet{wang06} to estimate the BCG
  stellar mass $M_{BCG}$, assuming  that $M_{BCG}$ correlates with the
  depth  of the potential  well of  the halo  within which  it formed,
  thus:
  \begin{equation}
    M_{BCG} = \dfrac{2 M_{star,0}}{(M_{vir}/M_0)^{-\alpha}+(M_{vir}/M_0)^{-\beta}}\,.
  \end{equation}
  In this  relation, we  include a Gaussian  scatter in  $M_{BCG}$ for
  given  host  halo mass  with  $\sigma_{M_{BCG}}=0.148$,  we set  the
  parameters $\alpha=0.39$, $\beta=1.96$, $\log(M_{star,0})=10.35$ and
  $M_0=3.16 \times  10^{11} M_{\odot}/h$ (see  \citet{wang06} for more
  detalis about the parameters).

\item  The BCG  stellar  density profile  is  described following  the
  \citet{hernquist90}  profile:
\begin{equation}
\rho_{star}(r) = \frac{\rho_g}{(r/r_f)(1+r/r_g)^3}\,,
\end{equation}
where  the  scale  radius  $r_g$  is  related  to  the  half-mass  (or
effective)   radius   $R_e$   by   $r_g=0.551  R_e$.    As   done   by
\citet{keeton01}    we   define   the    effective   radius    to   be
$R_e=0.003R_{vir}$. The scale density $\rho_g$ can be estimated by the
definition of the total mass of a Hernquist model:
\begin{equation}
\rho_g = \frac{M_{star}}{2 \pi r_g^3}\,.
\end{equation}
We  checked  the robustness  of  our  results  using a  \cite{jaffe83}
profile which keeps unchanged our results.
\end{itemize}

In Table  \ref{tabSim} we  summarize the whole  set of  simulations we
have   performed:   the    first   two   samples,   \texttt{SPH}   and
\texttt{SPHwBCG},  assume the  haloes to  be spherical,  while the
last two, \texttt{ELLwBCG}  and \texttt{ELLwBCGwADC}, consider the
haloes  to be  triaxial.   In the  latter  case the  axial ratios  are
obtained   randomly   sampling    the   distributions   suggested   by
\citet{jing02}.  We  also notice that  the last three  samples include
the presence  of a BCG.  In  all samples we  included the contribution
from the subhalo population.  We recall the reader that the total halo
mass $M_{vir}$ is defined as the sum of all components, so:
\begin{equation}
M_{vir} = M_{smooth} + \sum_{i=1}^{N_{tot}} m_{sub,i} \;\; [\;+
M_{BCG} ]\,.
\end{equation}
In   \texttt{SPHwBCG}  and   \texttt{ELLwBCG},  the   presence   of  a
dissipative  baryonic  component in  the  host  halo  centre does  not
influence the  dark matter density  distribution.  This phenomenology,
acting on  the dark matter component  mainly in proximity  of the host
halo     centre     and     known     as     adiabatic     contraction
\citep{blumenthal86,rix97,keeton01,gnedin11}.  The  effects of baryons
on  the total mass  profile are  generally found  to only  modify halo
concentrations at the $10\%$ level, although some studies suggest that
low  mass systems may  me significantly  affected by  baryonic cooling
\citep{fedeli11}.      \texttt{ELLwBCGwADC}     includes     adiabatic
contraction, and  in Appendix \ref{app1}  we discuss how the  mass and
concentration   estimates    are   biased   with    respect   to   the
\texttt{ELLwBCG} sample.

\begin{table*}
  \caption{\label{tabSim} List of Simulation sets performed with \texttt{MOKA 1.5}}
\begin{tabular}{ l |  c | c | c| c |}
  \hline                       
  Simulation set name & Smooth component & Substructures & BCG & Adiabatic Contraction  \\ \hline \hline
  \texttt{SPH} & spherical & yes & no & no \\ 
  \texttt{SPHwBCG} & spherical & yes & yes & no \\
  \texttt{ELLwBCG} & triaxial & yes & yes & no \\ 
  \texttt{ELLwBCGwADC} & triaxial & yes & yes & yes \\
  \hline  
\end{tabular}
\end{table*}

\begin{figure*}
\includegraphics[width=6.75cm]{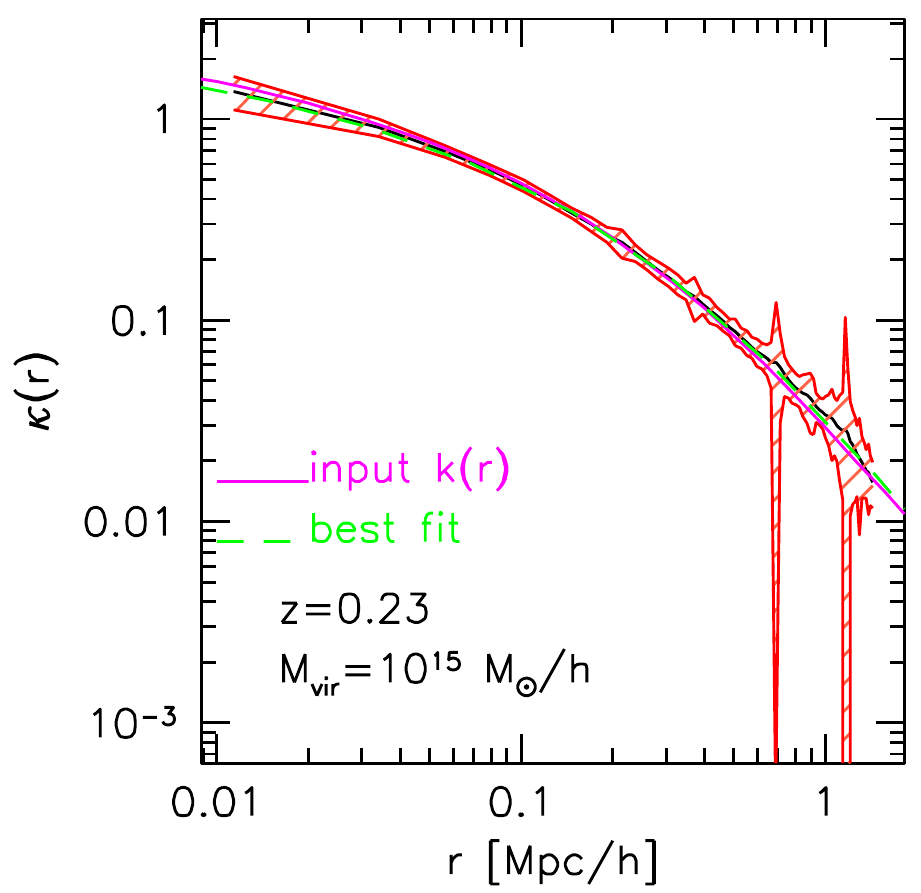}
\vspace{0.5cm}
\includegraphics[width=6.75cm]{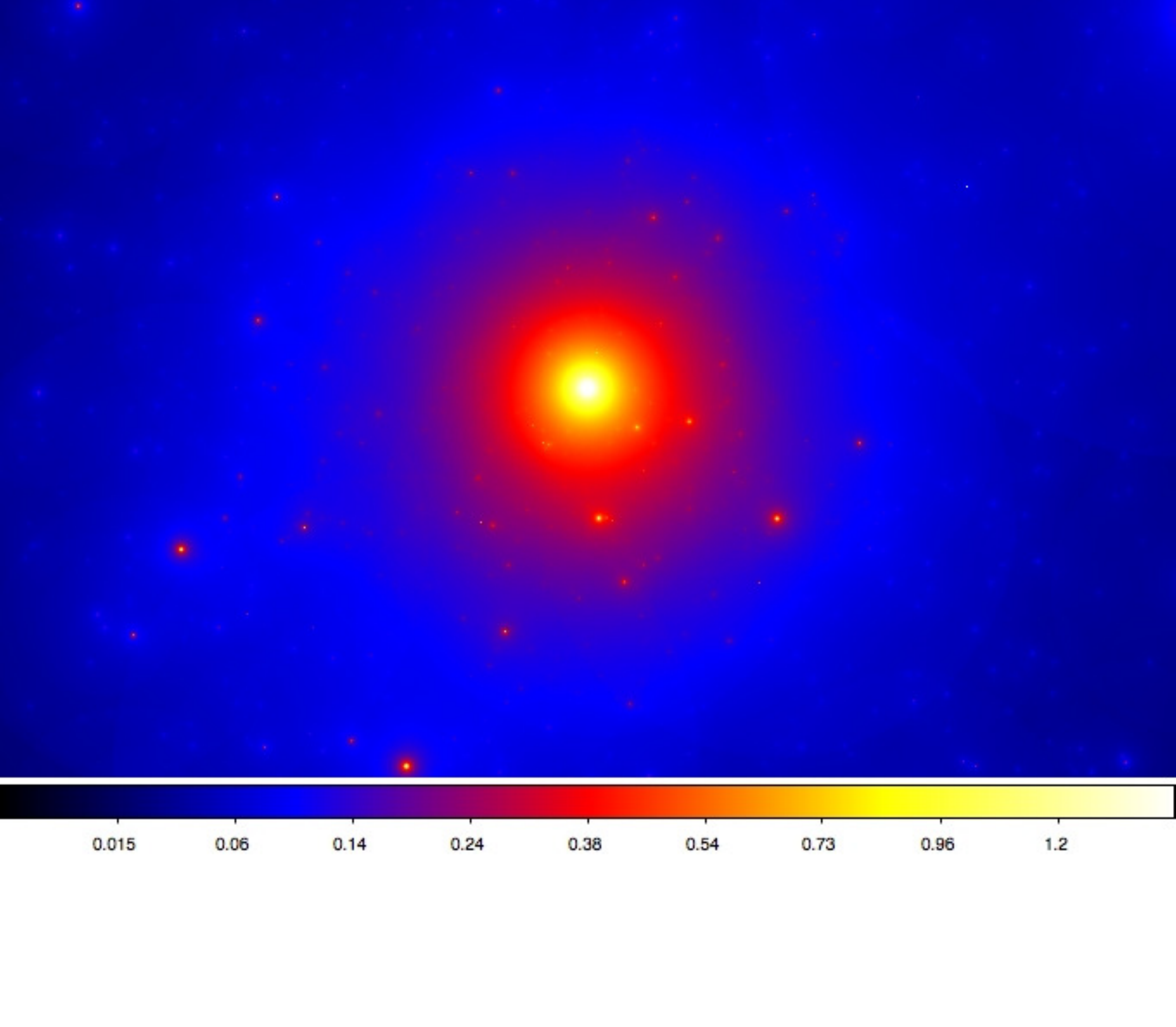}
\includegraphics[width=6.75cm]{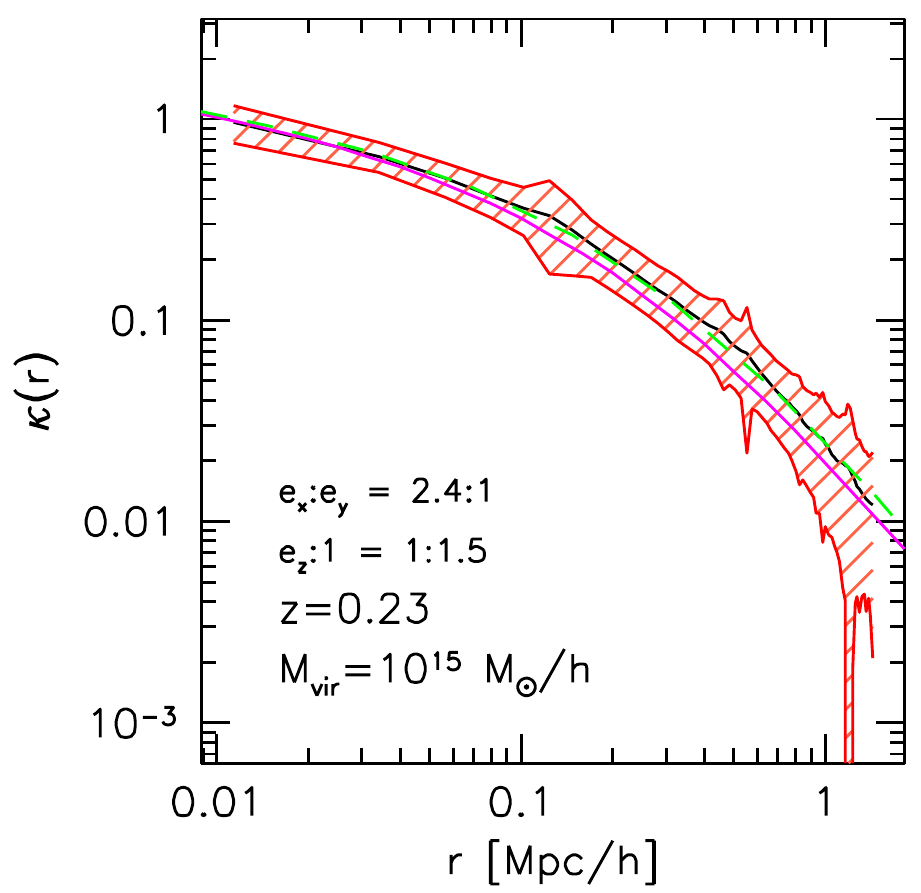}
\includegraphics[width=6.75cm]{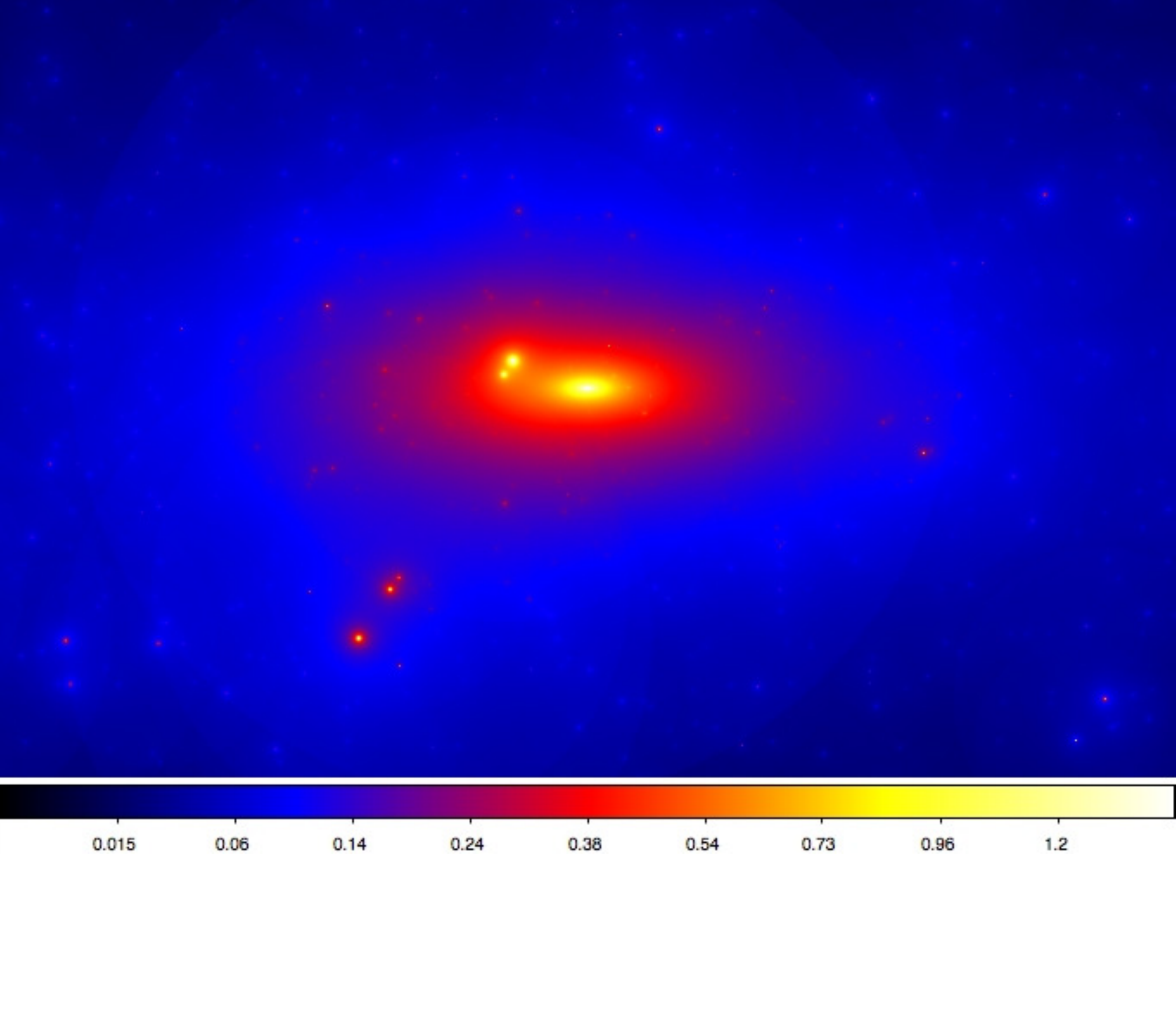}
\caption{\emph{Top panels}: spherical averaged convergence profile of a
  spherical   halo   with   substructures   and   BCG   created   with
  \texttt{MOKA}.   The shaded  region  shows the  rms  of the  average
  convergence  in each annulus  $8$ pixels  large. The  right subpanel
  shows  the  convergence map  of  the  central region.   \emph{Bottom
    panels}: spherical averaged convergence  profile of a triaxial halo
  with substructures and  BCG. In the label we show  the values of the
  components of  the axial ratios in  the plane of the  lens and along
  the line of sight. The solid  and dashed line show the input and the
  best fit converge profile respectively. The right subpanel shows the
  convergence  map   up  to  half   virial  radius  of   the  triaxial
  halo.\label{2clusterprof}}
\end{figure*}

As  example, in the  Figure \ref{2clusterprof}  we show  the spherical
averaged  convergence profiles  and  the maps  of  two clusters,  with
substructures and  BCG. In the  top panels, we  present the case  of a
spherical system  while in  the bottom panels  we show a  triaxial one
(the label  reports the  values of the  axial ratios projected  in the
plane of the lens and along  the line of sight).  We underline that in
all our simulations we do  not consider perturbations of the projected
matter density  due to  the large scale  structures along the  line of
sight \citep{hoekstra03}.  This is because we are mainly interested to
understand how  the halo properties their-selves  affect the recovered
mass and concentration  fitting a projected quantity. The  map of each
halo is resolved with a resolution  of $1024 \times 1024$ pixels up to
the  virial  radius $R_{vir}$  defined  as  the  radius enclosing  the
density   contrast   predicted  by   the   spherical  collapse   model
\citep{eke96}.  The profile is  built binning linearly the convergence
on annulus of $8$ pixels on which also the rms is estimated, showed by
the shaded region in the convergence profile of the figure.

\section{Fitting the Convergence Profile}
\label{section2}

The estimate of the total  galaxy cluster mass is important because it
is  by far  the most  accurate  predicted halo  property from  theory.
Ongoing and future surveys require an estimate of the mass with a bias
smaller than $10\%$  in order to allow the usage  of cluster counts as
cosmological probe.  However, \citet{okabe10b,oguri12} showed that the
combination  of  mass  and  concentration  estimates,  from  weak  and
weak+strong  lensing  measurements respectively,  can  be  used as  an
additional  cosmological  probe  through  the  $c-M$  relation.  These
results  have   revealed  a   good  agreement  with   the  theoretical
predictions  for massive  clusters  but a  steeper  dependence of  the
concentration on the mass wich what pure $N$-body simulations predict.
As discussed  by \citet{fedeli11} (see also  Appendix \ref{app1}), the
steepening of  the $c-M$  relation reflects the  response of  the dark
matter density distribution to the presence of the cold baryons in the
central regions of  haloes, that is stronger in  less massive, then in
more massive haloes, since baryon  cooling and star formation are more
efficient  in  those  systems  \citep{conroy07a,andreon10}.   In  this
direction it is very important  to understand the scatter and the bias
in  the  mass  and  in  the concentration  estimated  when  fitting  a
projected quantity: the convergence profile.  We are aware of the fact
that  the convergence is  not a  direct observable  and that  the most
standard way to measure the  cluster mass profile with weak lensing is
the tangential shear,  but our work is meant to  study the scatter and
the bias introduced in the  mass and in the concentration measurements
going  from  a  $3$D  to  a  $2$D quantity  of  the  measurement  halo
profiles. Since  our idea is  to illustrate the intrinsic  scatter and
the  bias  in the  mass  and  concentration  estimates, obtained  when
fitting  projected  quantities, we  present  and  discuss our  results
assuming  no observational  error  in the  cluster  converge map.  The
situation  would have  been different  if we  had used  the tangential
shear profile,  where the  dominant source of  noise is  the intrinsic
ellipticity distribution of background galaxies $\sigma_e \approx 0.3$
\citep{okabe10} that gives a Gaussian noise in each region of area $A$
where is measured given by:
\begin{equation}
\sigma_s^2 = \frac{\sigma_e^2}{n_{gal} A}\,,
\end{equation}
where $n_{gal}$ represents the background galaxy density distribution,
that goes  from $10$  gal/arcmin$^2$ for Dark  Energy Survey  (DES) or
similar to $40$ gal/arcmin$^2$ for deep ground based observations like
Large Synoptic Survey Telescope (LSST) or space based observation like
Euclid.   It is worth  mentioning that  for high  quality observations
with  large  source  densities,  the  effect of  shape  noise  becomes
comparable  or  subdominant  with  respect  to  the  intrinsic  source
scatters:  triaxiality, correlated (haloes  and subhaloes  around $3\,
Mpc/h$  from  the  cluster)  and uncorrelated  large-scale  structures
(systems more  distant than $3\,Mpc/h$  from the cluster).  To isolate
the effects of halo substructures and triaxiality we will describe our
results using the cluster  convergence profile. We also underline that
the results are  specific to the method chosen  for measuring the halo
properties (the converge profile).
\begin{figure*}
\includegraphics[width=5.5cm]{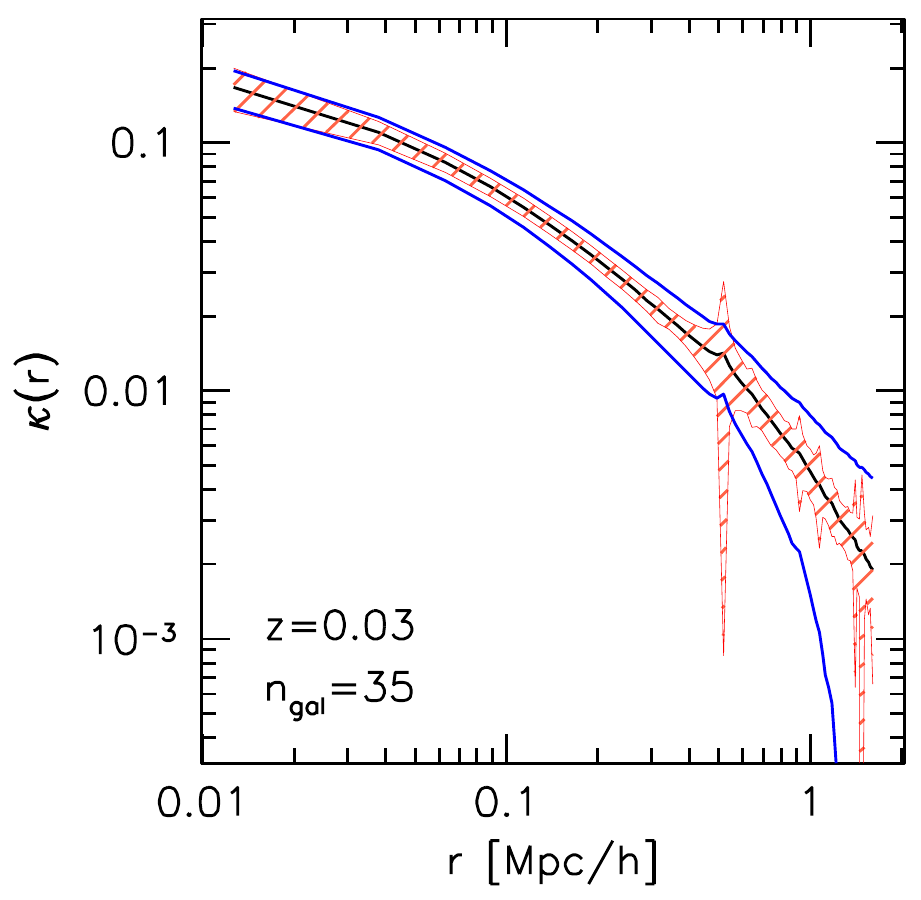}
\includegraphics[width=5.5cm]{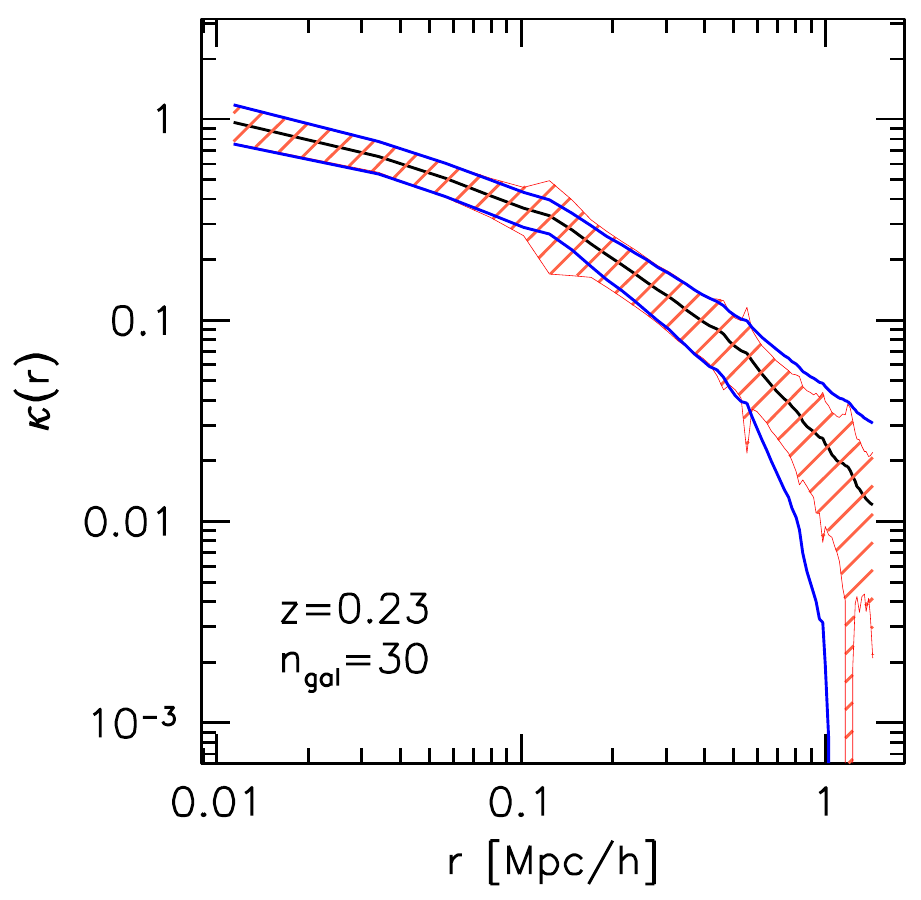}
\includegraphics[width=5.5cm]{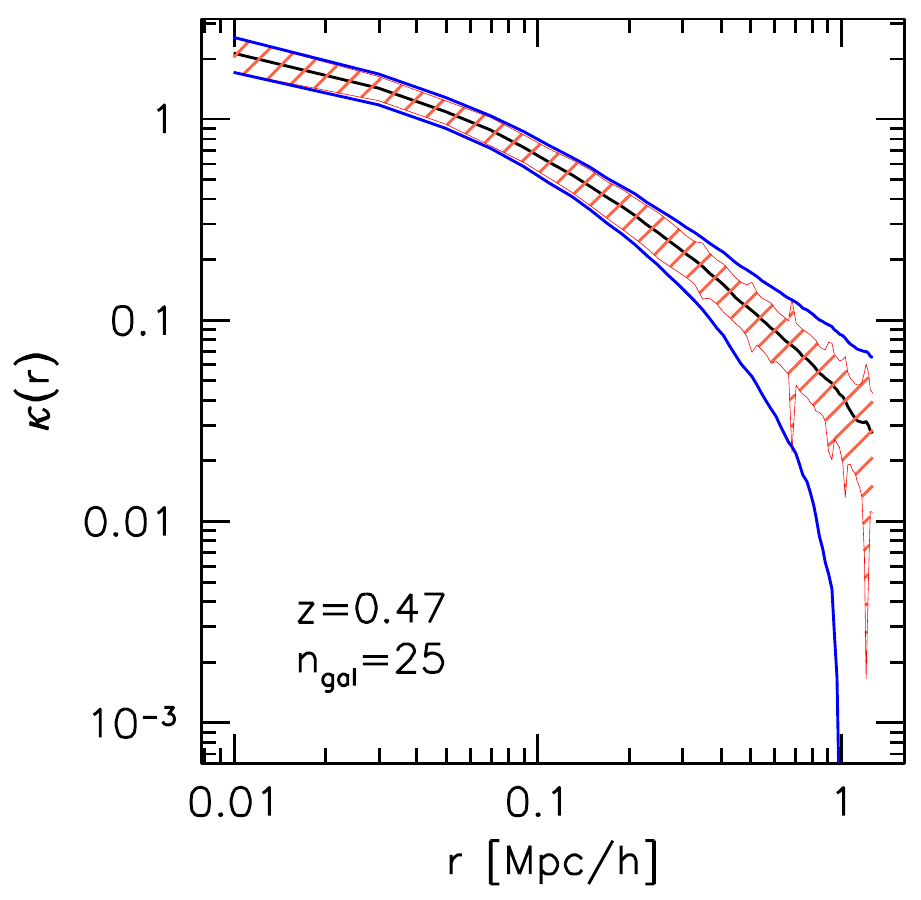}
\caption{Convergence profile of  three galaxy cluster-size haloes with
  $M_{vir}=10^{15}M_{\odot}/h$, located  at three different redshifts.
  In   each   panel,   the    shaded   region   represents   the   rms
  $\sigma_{\kappa}$  of  the  average  convergence  measured  in  each
  annulus, while  the solid curves  represent the noise  expected from
  the intrinsic ellipticities of source galaxies $\sigma_s$. \label{srefplot}}
\end{figure*}
In order  to compare the rms  of the convergence  with the statistical
observational noise, we show  in Figure \ref{srefplot} the convergence
profile  of three clusters  located at  redshift $z=0.03$,  $0.23$ and
$0.47$.  In each panel  the shaded region represents $\sigma_{\kappa}$
while  the solid  curves enclose  the statistical  observational noise
$\sigma_s$.  We have  considered 35, 30 and 25  gal/arcmin$^2$ for the
cluster located  at $z=0.03$,  $0.23$ and $0.47$,  respectively.  From
the figure  we notice that while  in the inner region  of the clusters
the two  errors are almost  identical, going toward the  virial radius
the noise expected from the intrinsic ellipticities of source galaxies
becomes larger than the rms of the convergence.

For  each halo  the  spherical averaged  convergence  profile is  best
fitted  using a single  component NFW  function in  order to  obtain a
simultaneous   estimate  for   both   the  mass   $M_{fit}$  and   the
concentration $c_{fit}$.   These are obtained considering  that, if we
define  $x$  and  $y$ as  the  coordinates on  the  lens  plane,  the
convergence  $\kappa(r)$   of  a  NFW   profile  can  be   written  as
\citep{bartelmann96,meneghetti03}:
\begin{equation}
  \Sigma_{NFW}(x,y)= \frac{2 \rho_s r_s}{r^2 -1 } F(r)\,, \\
\label{kappaNFW}
\end{equation}
where $r\equiv\sqrt{x^2+y^2}/r_s$, and $F(r)$:
\begin{displaymath}
  F(r) = \left\{ \begin{array}{ll}
    1- \frac{2}{\sqrt{r^2-1}} \arctan\sqrt{\dfrac{r-1}{r+1}}&r>1, \\
    1- \frac{2}{\sqrt{1-r^2}}
    \mathrm{arctanh}\sqrt{-\dfrac{r-1}{r+1}} &r<1, \\
    0& r=1;\\
  \end{array} \right.
\end{displaymath}
$r_s$  is the scale  radius, defining  the concentration  $c_{vir} \equiv
R_{vir}/r_s$,  and  $\rho_s$ is the  dark  matter  density  at the  scale
radius,
\begin{equation}
  \rho_s = \frac{M_{vir}}{4 \pi r_s^3}
  \left[ \ln(1+c_{vir}) - \frac{c_{vir}}{1+c_{vir}}\right]^{-1}\,,
\label{eqrhos}
\end{equation}
and so
\begin{equation}
\kappa_{NFW}(r) = \dfrac{\Sigma_{NFW}(r)}{\Sigma_{crit}}\,, \label{kappaNFW2}
\end{equation}
where $\Sigma_{crit}=c^{2}D_{s}/(4 \pi G D_{l}D_{ls})$ is the critical
surface  mass  density, depending  on  the angular-diameter  distances
$D_{l}$, $D_{s}$  and $D_{ls}$ from the  observer to the  lens, to the
source,  and from  the lens  to  the source,  respectively.  From  the
equations above, since $\kappa_{NFW}(r)$ depends both on halo mass and
concentration, $M_{fit}$ and $c_{fit}$  are obtained by minimizing the
$\chi^2$ function:
\begin{equation}
\chi^{2}(M_{fit},c_{fit}) = \sum_{i=1}^{N_{radial\,bins}} \dfrac{[\kappa_{NFW}(r_{i}|M_{fit},c_{fit})-\kappa(r_{i})]^{2}}{\sigma_{\kappa,i}^{2}}\,,
\label{chi2prof} 
\end{equation}
where $N_{radial\,bins}$ represents the number of radial bins on which
the  convergence profile  is estimated.  In the  default  setting wich
considers the  whole profile $[0,1]R_{vir}$ the number  of radial bins
is $64$,  since the map is made  by $1024 \times 1024$  pixels and the
profile is linearly built on annuli of $8$ pixels.

\subsection{Spherical Haloes}
\begin{figure*}
\begin{center}
\includegraphics[width=8.8cm]{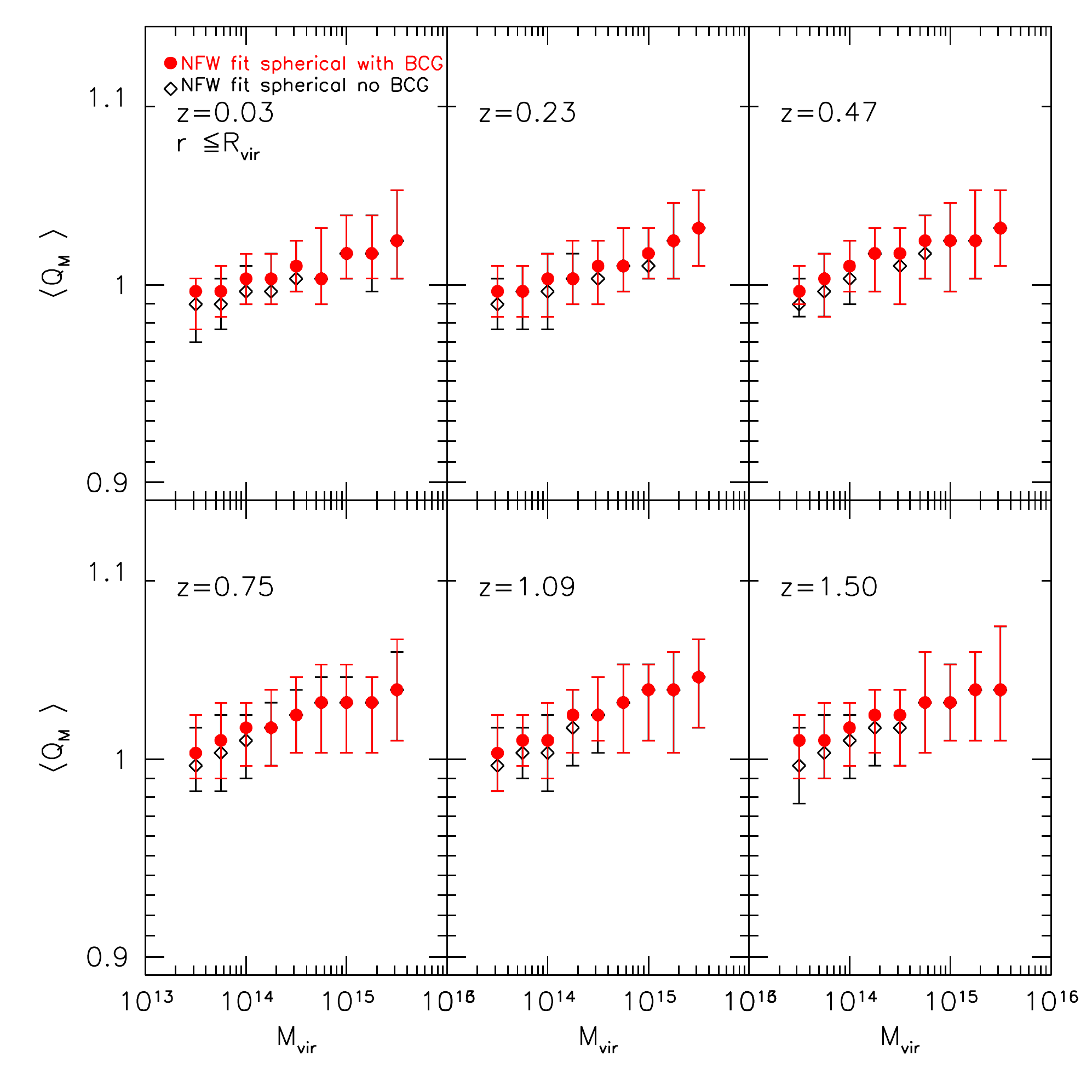}
\includegraphics[width=8.8cm]{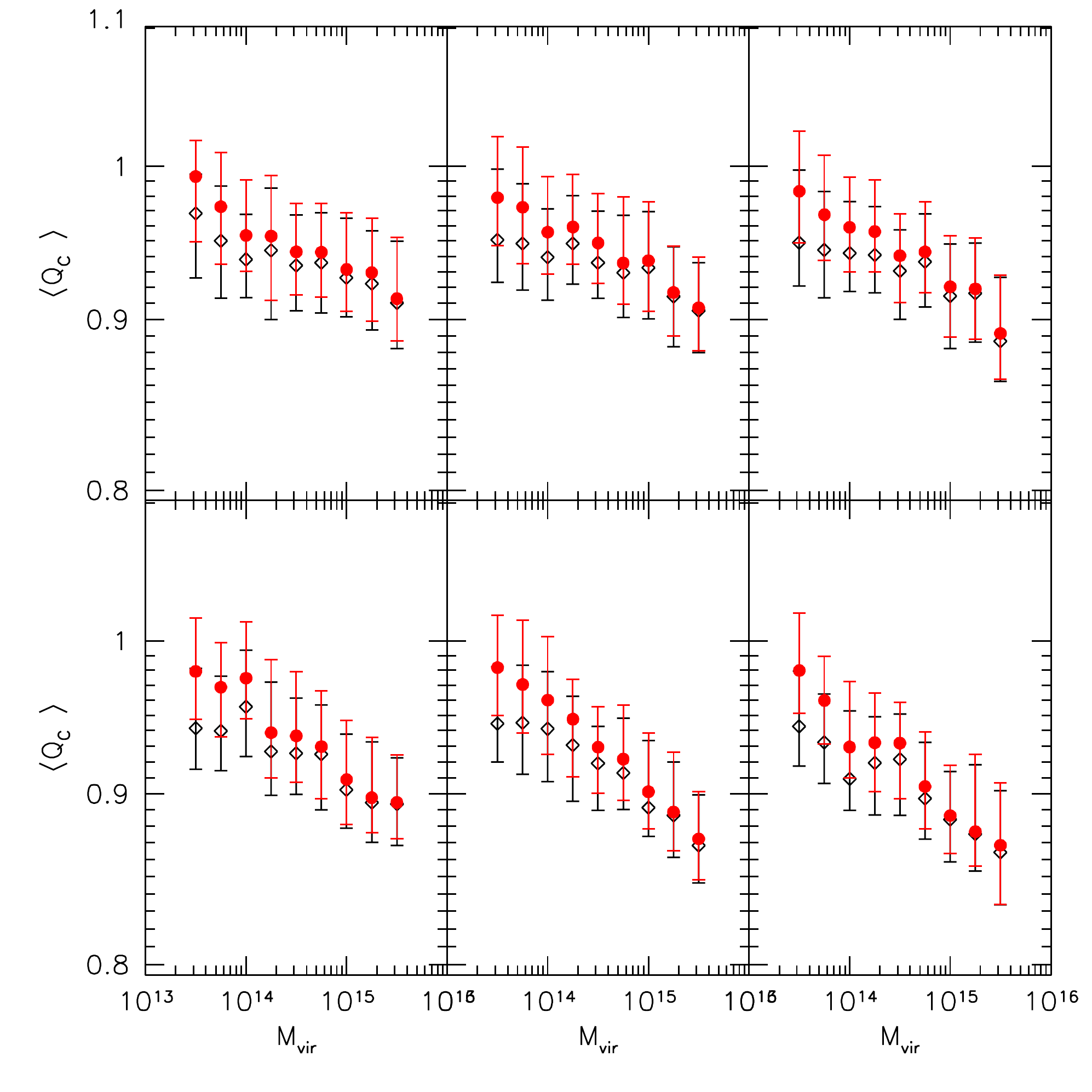}
\caption{Median rescaled  mass (\emph{left panels})  and concentration
  (\emph{right panels}) estimates as a  function of the halo mass, for
  spherical substructured  haloes at  six different redshifts.
  The  mass  of each  halo  has been  estimated  by  best fitting  the
  spherical averaged convergence profile of each halo up to the virial
  radius $R_{vir}$, with a NFW function varying both the mass and the
  concentration. The two different data points refer to haloes without
  (open  diamonds)  and  with  (filled  circles)  the  bright  central
  galaxy.   The  error   bars  enclose   the  first   and   the  third
  quartiles.\label{rescaledmassSPH}}
\end{center}
\end{figure*}

Figure  \ref{rescaledmassSPH}  shows   the  rescaled  mass  $Q_M\equiv
M_{fit}/M_{vir}$   and   concentration   $Q_C\equiv   c_{fit}/c_{vir}$
estimates  -- where $M_{vir}$  and $c_{vir}$  represent the  true halo
mass and  concentration - as a  function of $M_{vir}$  for the samples
\texttt{SPH} (open diamonds)  and \texttt{SPHwBCG} (filled circles) at
six different redshifts.  The  estimates have been obtained by fitting
the profile up to the halo  virial radius: this in practice would mean
that for a realistic cluster we  would need to combine both strong and
weak  lensing  data.   The  different  data  points  with  error  bars
represent  the  median and  the  two  quartiles,  respectively of  the
distribution in each mass bin.
From  the left  figure  we notice  that  the halo  mass  is very  well
recovered without any bias and with an error of only few percents. The
small trend with the halo mass is due to the presence of substructures
that  contaminate the  projected mass  distribution: the  most massive
haloes possess a larger fraction  of mass in substructures, because of
their later formation time, with  respect to the least massive systems
\citep{gao04,delucia04,vandenbosch05}.  Unlike the mass, the estimated
concentration (right  panels) shows a  negative bias and  a decreasing
behavior with the  host halo mass, and so  with the subhalo abundance.
From this  figure we  notice that the  concentration estimate  is more
influenced by  the presence  of substructures in  the 2D map  than the
mass estimate.  At  the smallest masses, the BCG seems  to play a more
relevant role, adding an additional bias of few percent.

\begin{figure*}
\begin{center}
\includegraphics[width=8.8cm]{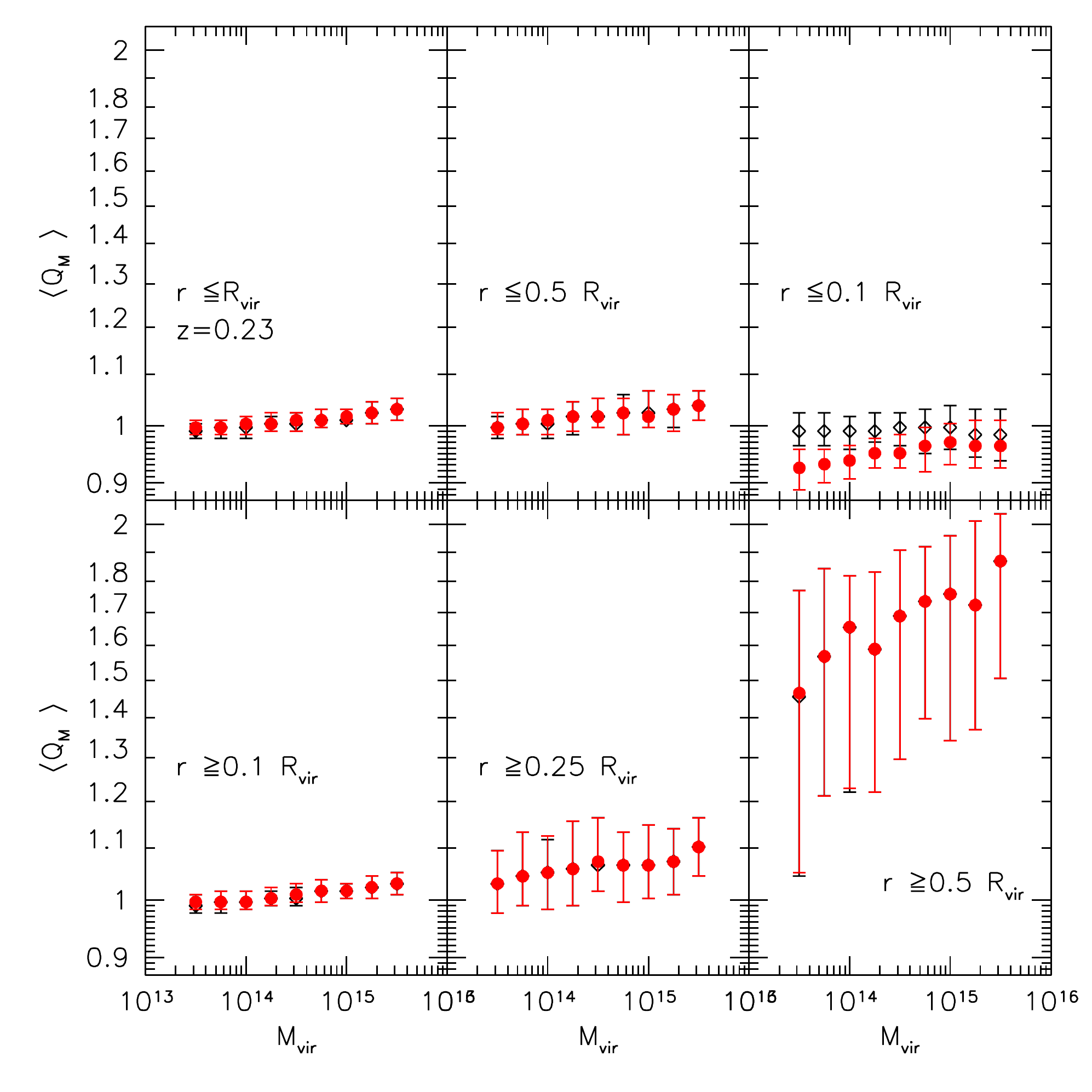}
\includegraphics[width=8.8cm]{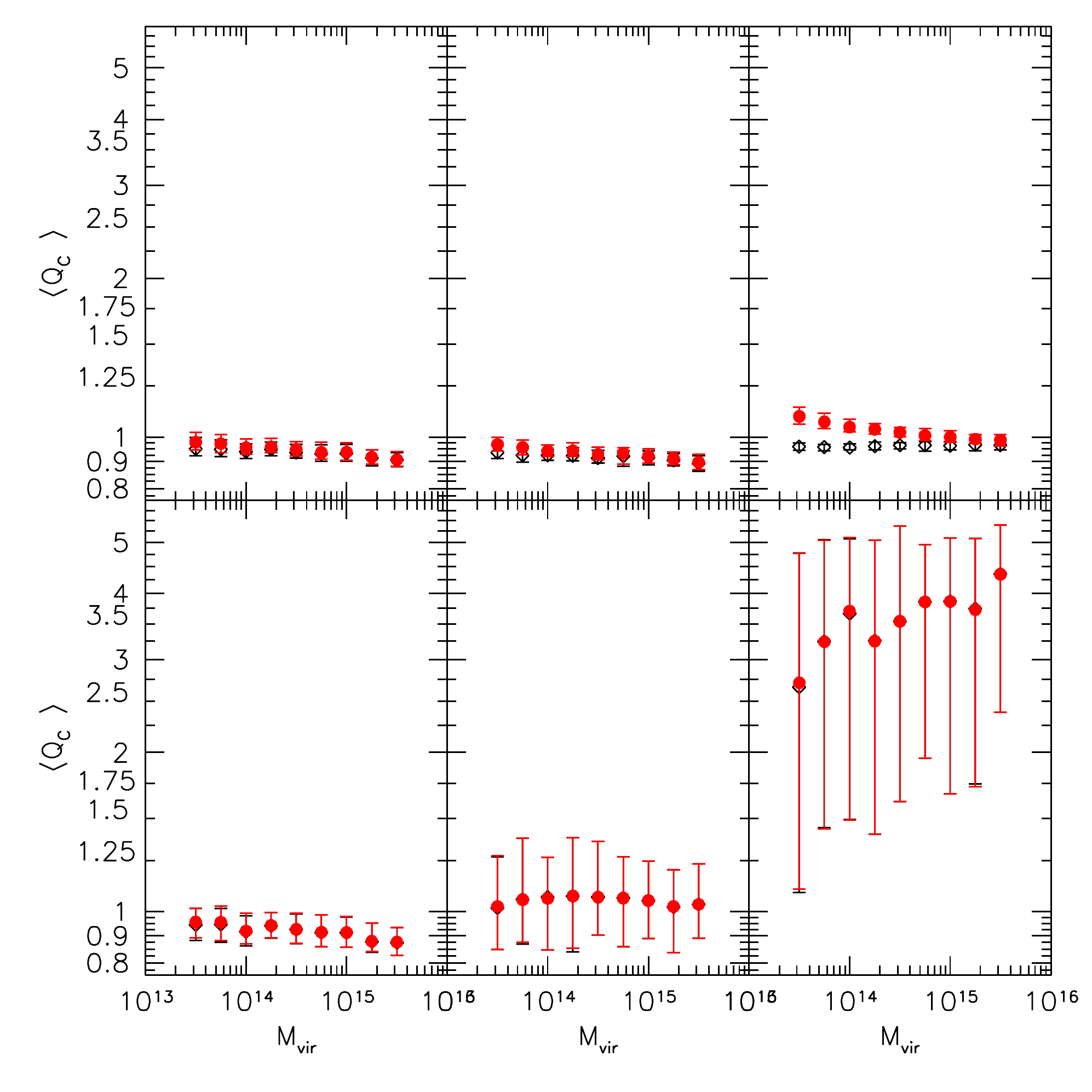}
\caption{Median rescaled  mass (\emph{left panels})  and concentration
  (\emph{right panels}) estimates as a  function of the halo mass, for
  spherical  and substructured  haloes at  redshift $z=0.23$.  The top
  panels  show the  relation obtained  fitting the  spherical averaged
  convergence profile up  to $R_{vir}$, $0.5R_{vir}$ and $0.1R_{vir}$;
  while  the   bottom  ones  from  $R_{vir}$   down  to  $0.1R_{vir}$,
  $0.25R_{vir}$  and $0.5R_{vir}$. Symbols  and error  bars are  as in
  Figure \ref{rescaledmassSPH}.\label{rescaledmassSPHDr}}
\end{center}
\end{figure*}

To discuss  the importance of the  radial range used  when fitting the
convergence profile,  in Figure \ref{rescaledmassSPHDr}  we again show
the estimated  mass and  the concentration as  a function of  the true
halo mass,  for the same two  halo samples discussed  before.  In this
case we show how the estimated masses and concentrations depend on the
radial range on  which the fit is performed.  In  the upper panels, we
fit the  profile between $[R_{min},  R_{max}]/R_{vir} = [0,  1]$, $[0,
0.5]$,  $[0, 0.1]$  (upper panels  of  Fig.  \ref{rescaledmassSPHDr}),
$[0.1, 1]$,  $[0.25, 1]$ and  $[0.5, 1]$ (lower panels).   Results are
shown only  for haloes  at redshift $z=0.23$,  since we  verified that
they do not  depend on redshift.  It is worth noticing  that we do not
observe significant differences apart  from the following cases: ($i$)
in  the fit  over  the radial  range  $[0.25, 1]  R_{vir}$, where  the
scatter is almost double; ($ii$) when the region $[0, 0.1] R_{vir}$ is
considered: it roughly corresponds  to the strong lensing region where
the contribution from the BCG is comparable to that of the dark matter
component,  while  in   presence  of  a  BCG  the   mass  is  slightly
underestimated almost the opposite is valid for the concentration (see
the different data  points); ($iii$) in the fit  over the radial range
$[0.5,   1]  R_{vir}$,   where   both  mass   and  concentration   are
overestimated by a factor of $1.5$ and $3.5$, respectively.

\begin{figure*}
\begin{center}
\includegraphics[width=8.8cm]{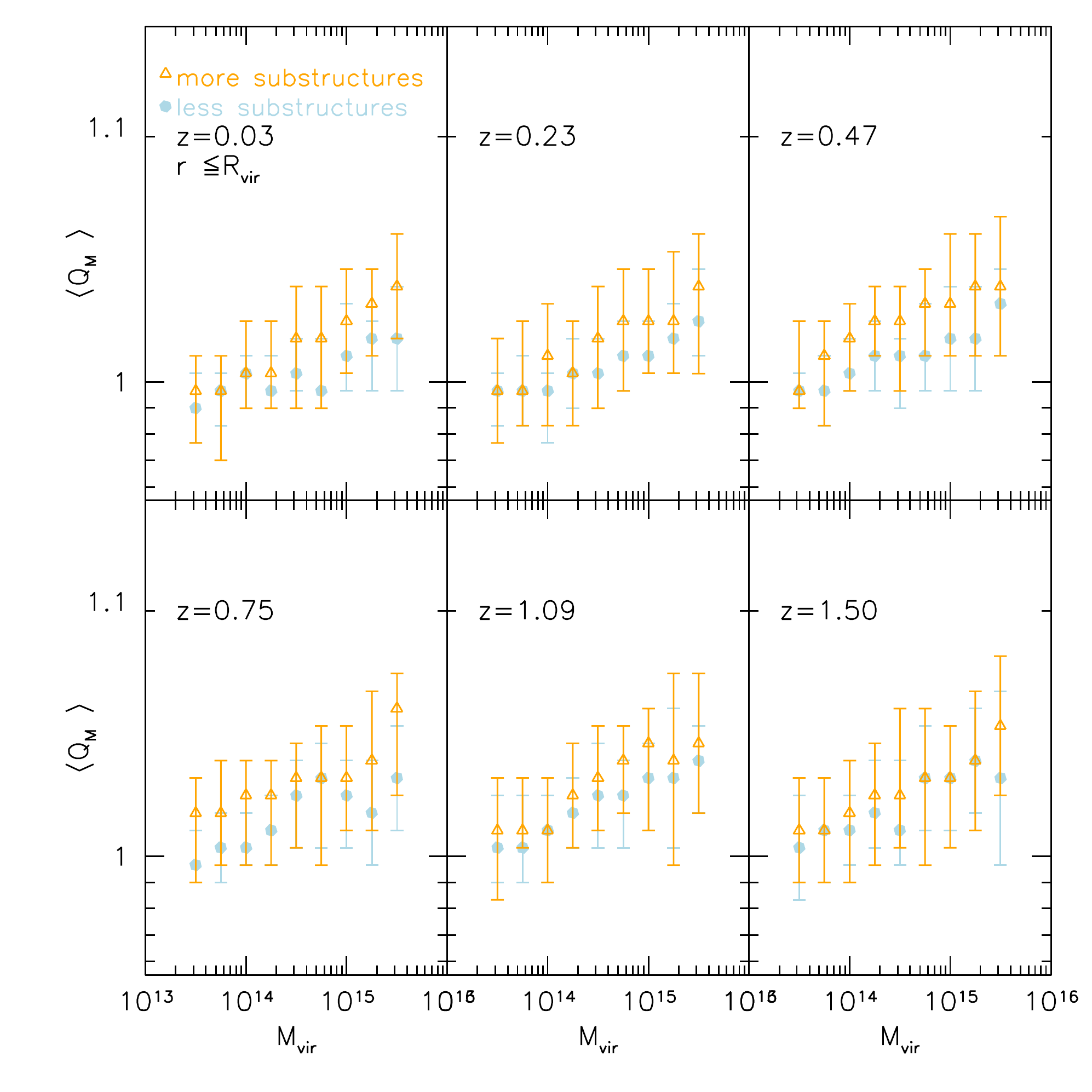}
\includegraphics[width=8.8cm]{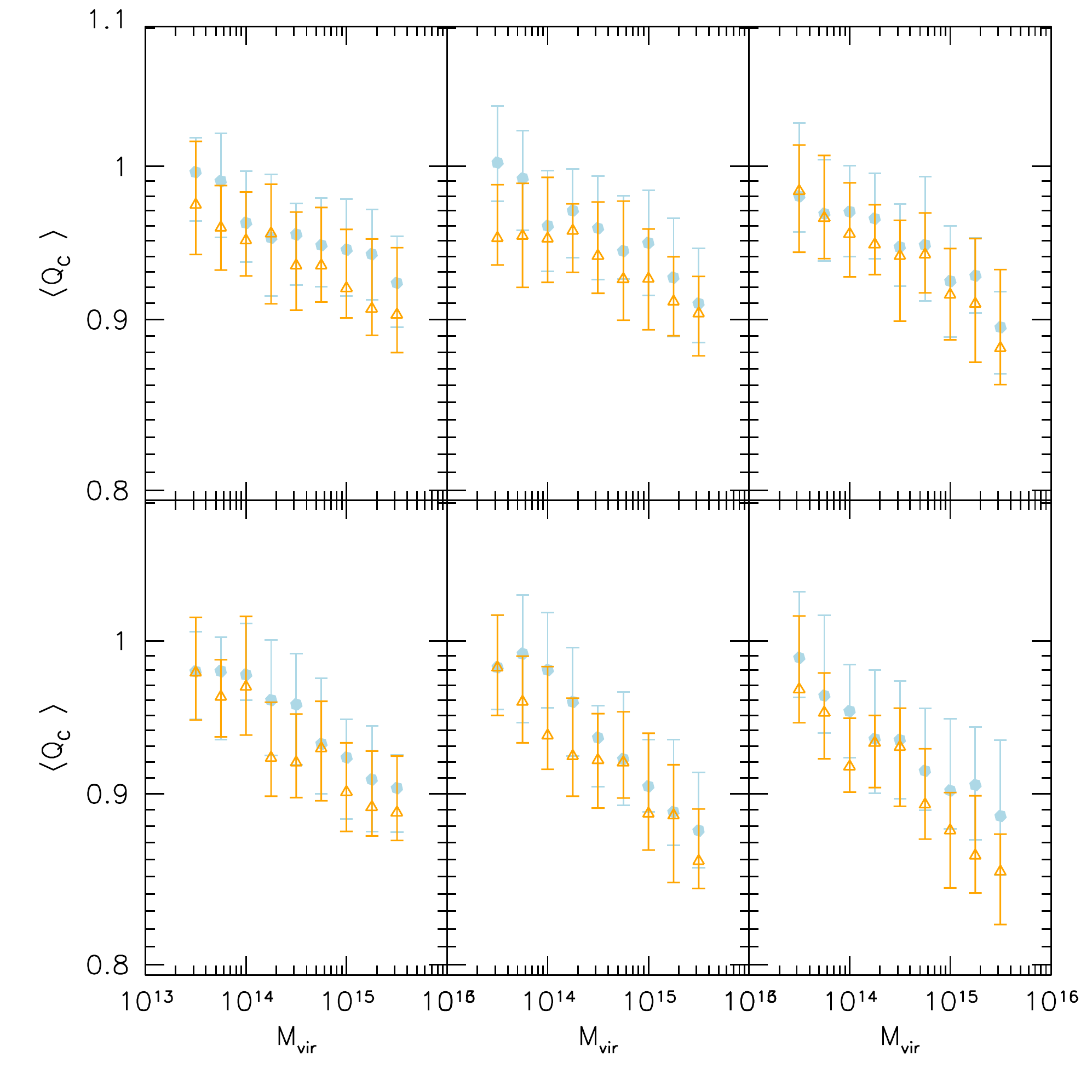}
\caption{Median rescaled  mass (\emph{left panels})  and concentration
  (\emph{right panels}) estimates  as a function of the  halo mass for
  the \texttt{SPHwBCG} subsamples with  more (open triangles) and less
  (filled circles) substructures.  The error bars correspond the first
  and the third quartile.\label{rescaledmassSPHsub}}
\end{center}
\end{figure*}

In order  to quantify how much  the presence of  substructures in host
haloes  biases  the mass  and  the  concentration  estimates, we  have
divided our  sample in  haloes with more  and less  substructures.  In
each  mass bin  we  estimate the  median  host halo  mass fraction  in
substructures  $\langle  f_{sub} \rangle$  and  we  define  a halo  as
``substructured'' if its $f_{sub}$ is larger than the median, $f_{sub}
>  \langle  f_{sub} \rangle$.   On  the contrary,  ``unsubstructured''
haloes  have   $f_{sub}  <   \langle  f_{sub}  \rangle$.    In  Figure
\ref{rescaledmassSPHsub} we show  the estimated mass and concentration
at different  redshifts (as  in Figure \ref{rescaledmassSPH})  for the
samples of  substructured (open triangles)  and unsubstructured haloes
(filled circles).  The  haloes are assumed to be  spherical and with a
BCG  at their  centres.  From  the figure  we notice  that,  while the
estimated masses are  larger than the true ones  for the substructured
sample than  for the  unsubstructured, the opposite  is valid  for the
estimated concentrations.

\subsection{Triaxial Haloes}
The spherical  approximation used to describe the  halo matter density
distribution is far from  being realistic.  Different studies based on
numerical  simulations   showed  that  the  halo   shape  is  triaxial
\citep{sheth01b,jing02,kazantzidis04}.   This  is  because during  the
collapse phase  a halo  is influenced by  its surrounding  tidal field
which   stretches  and  shears   its  matter   distribution.   Various
observations of galaxy  clusters, both in optical and  in X-rays, have
confirmed this  picture and  have revealed that  a discrete  number of
X-ray-selected clusters present also  their major axis elongated along
the line  of sight.  This  condition is typically sufficient,  but not
necessary, to  cause, in  their optical images,  the presence  of very
distorted  gravitational arcs  and large  Einstein rings,  and  also a
discrepancy   between  galaxy   cluster   masses  and   concentrations
determined   from  X-ray   and  gravitational   lensing  observations.
Combining    X-ray,   weak    and    strong   lensing    observations,
\citet{morandi10}  have  proposed a  method  to  determine the  galaxy
cluster elongation along the line of sight (with an error of the order
of  $5\%$)  and  reconcile  the   discrepancy  of  the  mass  and  the
concentration estimates. The  observational estimate of the elongation
of the cluster halo along of the line of sight is very challenging and
only possibile for  clusters that are very massive  and located at low
redshifts.

Let us now consider the case of a triaxial prolate or oblate halo, and
define its axes as $e_a$, $e_b$ and $e_c$; if the halo is spherical in
the plane of  the sky ($e_a=e_b=1$) and $e_c$ is  parallel of the line
of sight, the convergence of the halo will be defined as:
\begin{equation}
  \Sigma(x,y) = e_c \int_{-\infty}^{\infty} \rho(x,y,\zeta')
  \mathrm{d} \zeta'\,,  
  \label{elongcor} 
\end{equation}
with $\zeta'\equiv \zeta/e_c$. 

In \texttt{MOKA} we have included the model proposed by \citet{jing02}
to describe the main halo  ellipticity, according to which the stellar
component of  the BCG  and the subhalo  spatial distribution  are also
perturbed. Using the axial ratio probability distributions we estimate
$e_a$, $e_b$  and $e_c$, requiring that  $e_a e_b e_c =  1$.  Once the
axial ratios are  known, we randomly orient the  halo choosing a point
on  a  sphere  identified  by  its  azimuthal  and  elevation  angles;
according  to  this  we  rotate  the halo  ellipsoid  and  deform  the
convergence map. The elongation $e_z$ represents the largest component
of the ellipsoid axes projected along the line of sight.

\begin{figure*}
\begin{center}
\includegraphics[width=8.8cm]{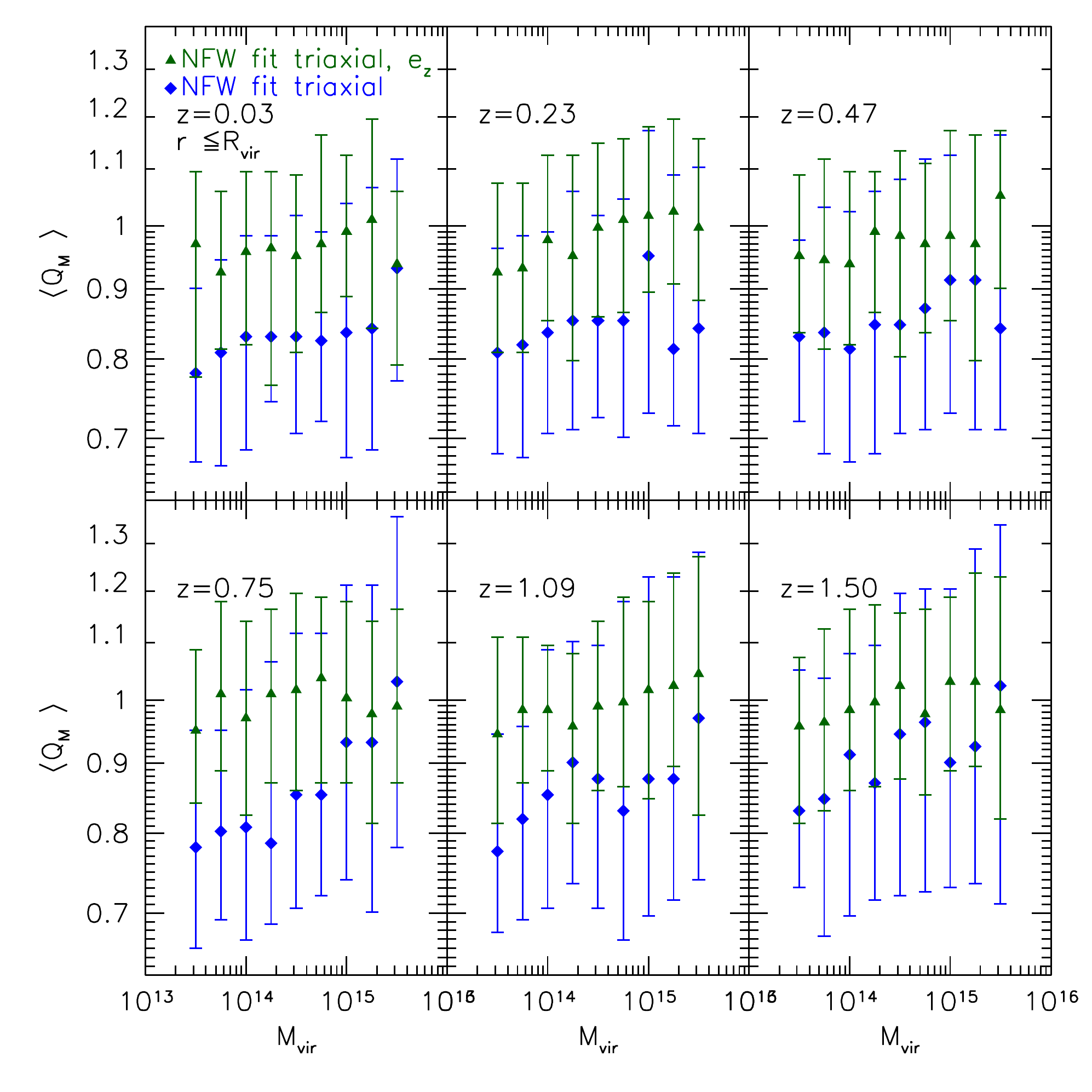}
\includegraphics[width=8.8cm]{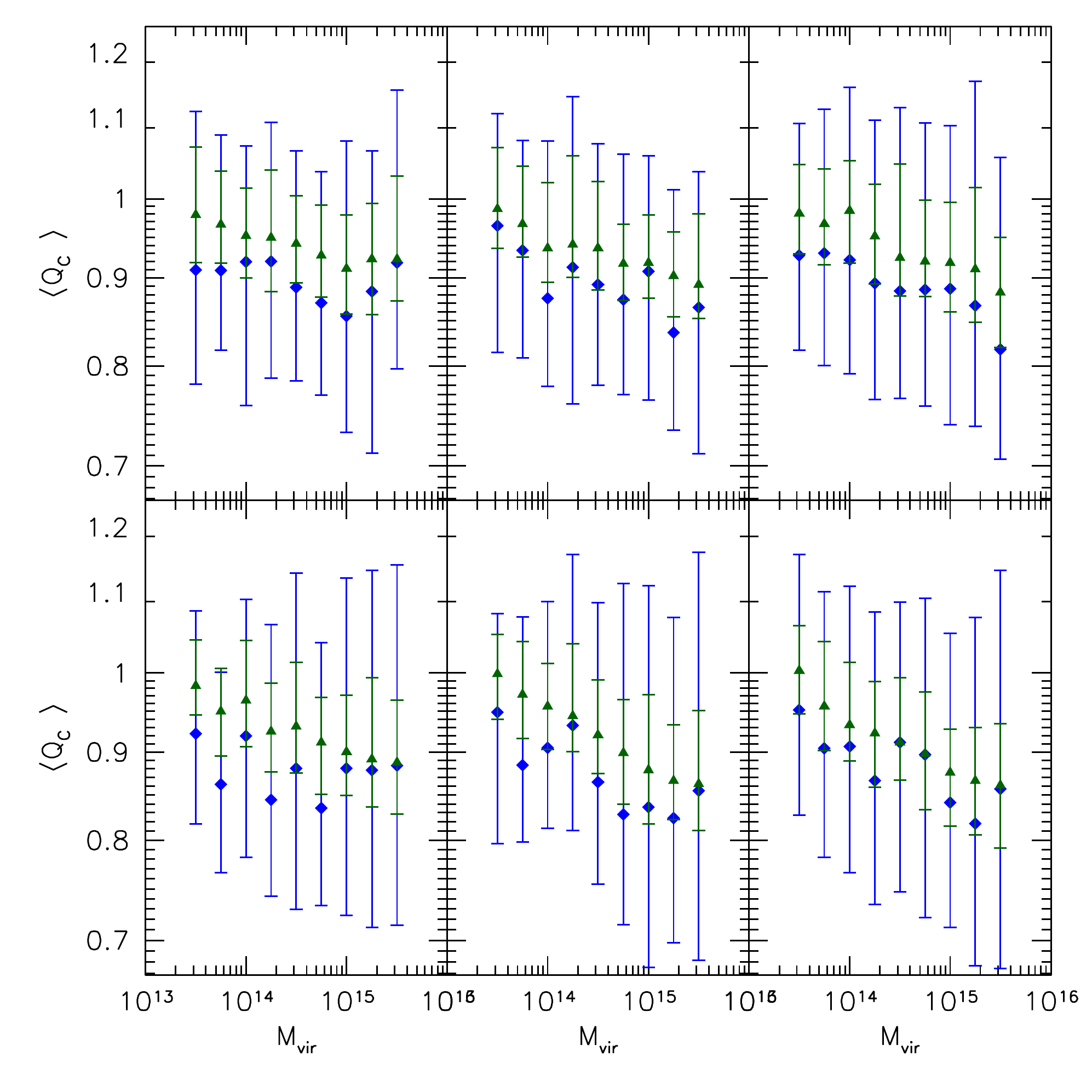}
\caption{Median rescaled  mass (\emph{left panels})  and concentration
  (\emph{right panels}) estimates as a  function of the halo mass, for
  triaxial  substructured haloes  with BCG,  located at  six different
  redshifts.  The  mass of each  halo has been estimated  best fitting
  the spherical averaged  convergence profile of each halo,  up to the
  virial radius $R_{vir}$, with a  NFW function varying both the mass
  and the  concentration. The two  different data points refer  to two
  cases  in which  the  fit  has been  performed  not knowing  (filled
  diamonds) or  knowing (filled  triangles) the halo  elongation along
  the line of  sight.  The error bars enclose the  first and the third
  quartiles     of      the     distribution     in      each     mass
  bin.\label{rescaledmassELL}}
\end{center}
\end{figure*}

In Figure \ref{rescaledmassELL}, we show $Q_M$ and $Q_c$ as a function
of the host halo mass, obtained by fitting the convergence profiles of
the  \texttt{ELLwBCG} sample  up  to $R_{vir}$  and  at six  different
redshifts.   In  fitting  the  convergence  profile  of  the  triaxial
clusters, and in estimating mass  and concentration, we proceed in two
ways:   ($i$)  in   the   first   case  we   make   use  directly   of
$\kappa_{NFW}(r)$  from equations  \ref{kappaNFW}  and \ref{kappaNFW2}
and  minimize $\chi^2(M_{fir},c_{fit})$ (eq.   \ref{chi2prof}); ($ii$)
in  the  second  case  we  multiply $\kappa_{NFW}(r)$  by  $e_z$,  the
elongation  of the  halo  along  the line  of  sight (assumed  without
error),  and   estimate  $M_{fit}$  and   $c_{fit}$  again  minimizing
$\chi^2$.   We refer  to  this  last case  as  mass and  concentration
estimates corrected for elongation.

In the left  panels of Fig. \ref{rescaledmassELL}, we  notice that the
mass bias,  found also  by \citet{meneghetti11} and  \citet{rasia12} -
who fitted  the tangential shear profile  using simulated observations
of galaxy clusters - almost disappears when we correct for elongation.
In the right  panels we show the results  for the concentration.  When
we do not  include the elongation in the  calculations, both $Q_M$ and
$Q_c$ are underestimated of $10-20\%$. Conversely, when we correct for
elongation,  the  mass  estimate  is  unbiased,  while  the  resulting
concentration is still slightly  underestimated, since $Q_M$ and $Q_c$
are  not independent  of each  other, considering  that  $c_{vir} \sim
M_{vir}^{1/3}$ they do not scale in the same way but as: $\Delta Q_c =
\left( \Delta Q_M \right)^{1/3}$.

\begin{figure*}
\begin{center}
\includegraphics[width=8.8cm]{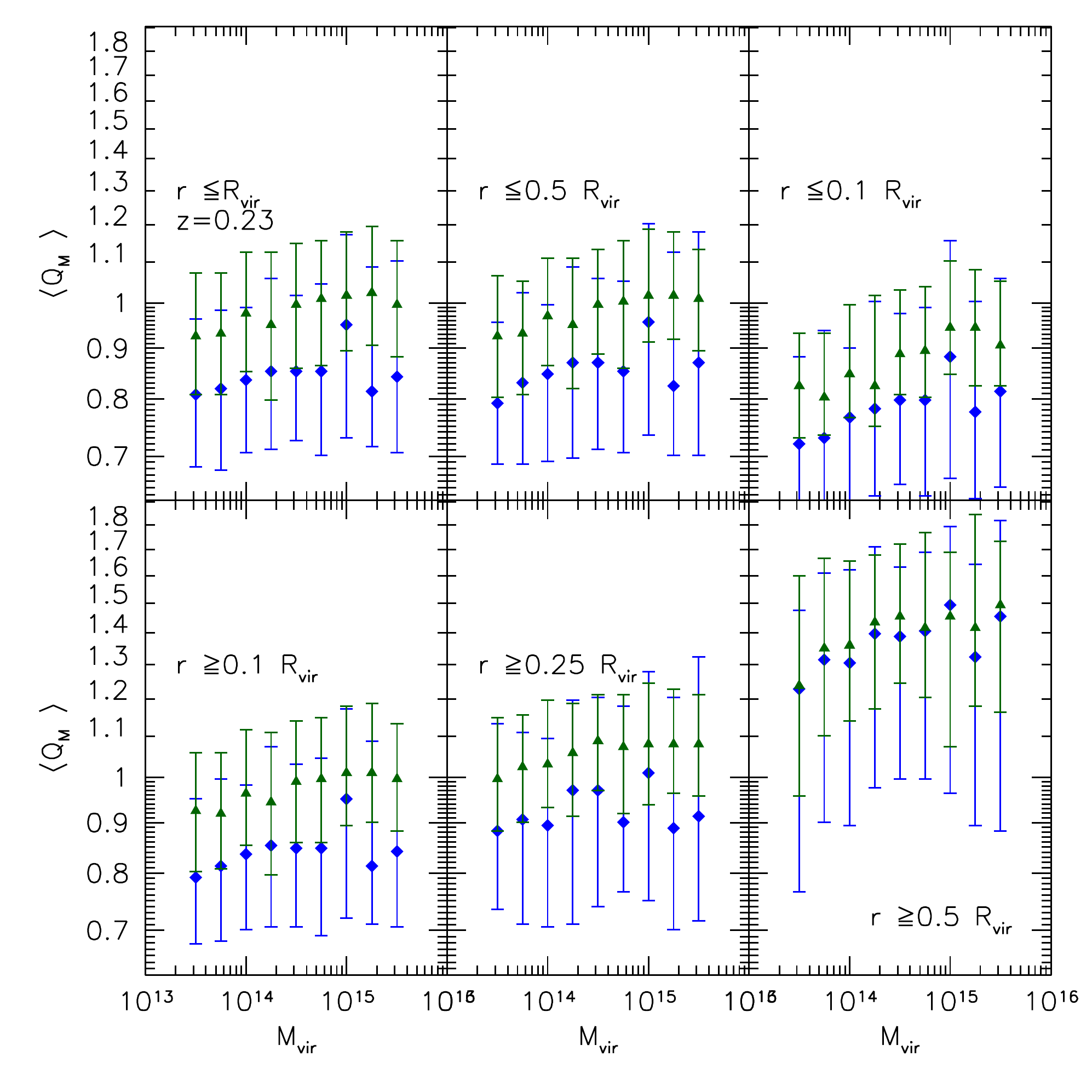}
\includegraphics[width=8.8cm]{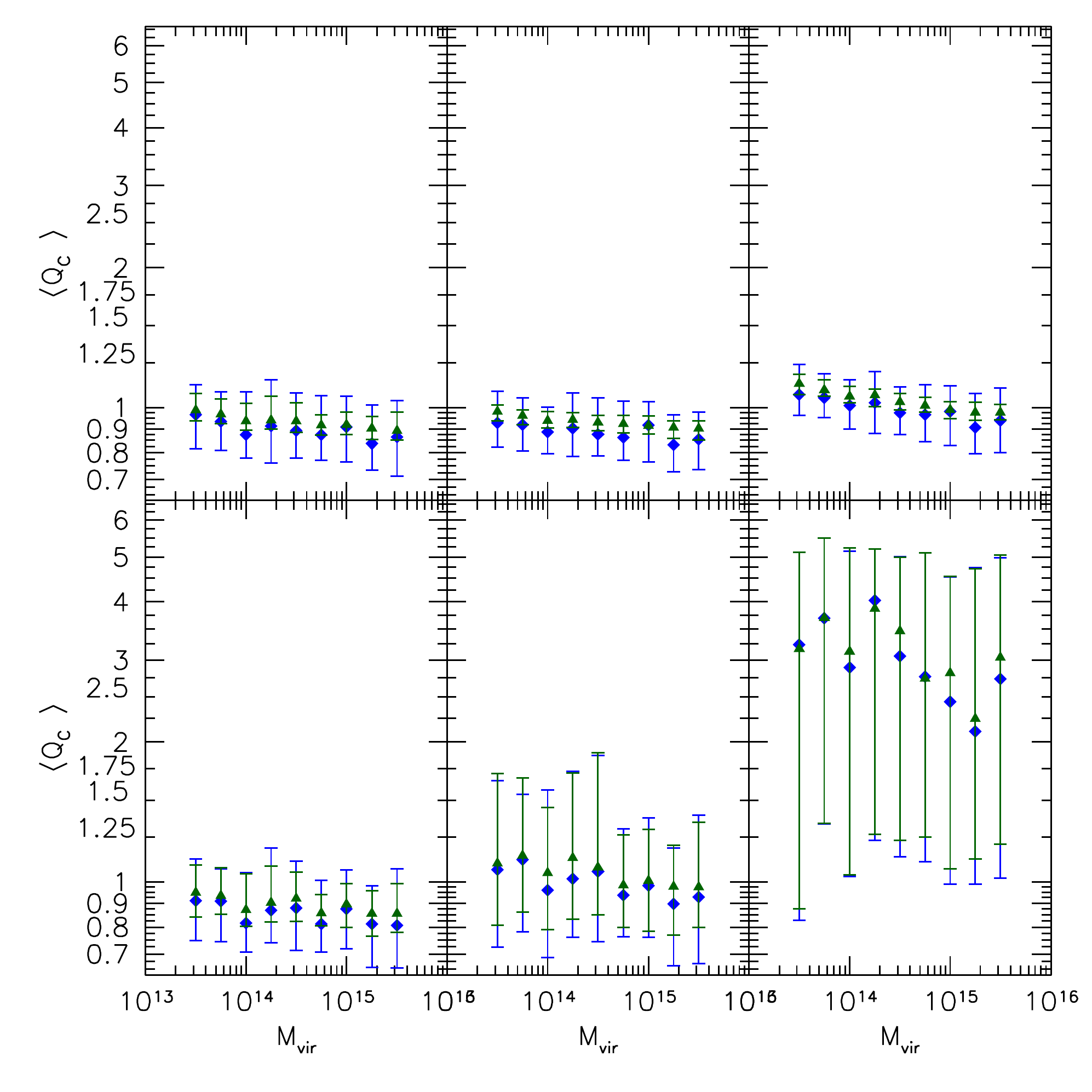}
\caption{Median rescaled  mass (\emph{left panels})  and concentration
  (\emph{right panels}) estimates as a  function of the halo mass, for
  triaxial  and  substructured haloes  with  BCG  located at  redshift
  $z=0.23$.   The top panels  show the  relation obtained  fitting the
  profile without any assumption on  the value of the elongation up to
  $R_{vir}$, $0.5R_{vir}$ and $0.1R_{vir}$; while the bottom ones from
  $R_{vir}$ down to $0.1R_{vir}$, $0.25R_{vir}$ and $0.5R_{vir}$.  The
  different data points  and error bars are the same  as in the Figure
  \ref{rescaledmassELL}.\label{rescaledmassELLDr}}
\end{center}
\end{figure*}

To discuss  the importance  of the range  used to fit  the convergence
profile, in Figure \ref{rescaledmassELLDr} we show the median rescaled
mass and concentration  as a function of the true  halo mass, when the
fit of  the convergence  profile is done  on different  radial ranges.
These   are  the   same   chosen  for   the   analysis  presented   in
Fig. \ref{rescaledmassSPHDr}. Again we note that the error bars become
larger when fitting  on a narrower radial range.  In  the figure we do
not  observe significant  differences  in the  radial  ranges $[0,  1]
R_{vir}$,  $[0,  0.5] R_{vir}$,  $[0.1,  1]  R_{vir}$  and $[0.25,  1]
R_{vir}$. Fitting the most central region, where the BCG dominates the
convergence profile,  the masses  and the concentrations  are slightly
underestimated  and  overestimated,  respectively.   As show  for  the
spherical  case,   fitting  the  profile  outside   $0.5  R_{vir}$  we
overestimate  both  the halo  mass  and  concentration  with not  much
difference between the two cases in  which we do or do not correct the
convergence profile for elongation.

\begin{figure*}
\begin{center}
\includegraphics[width=8.8cm]{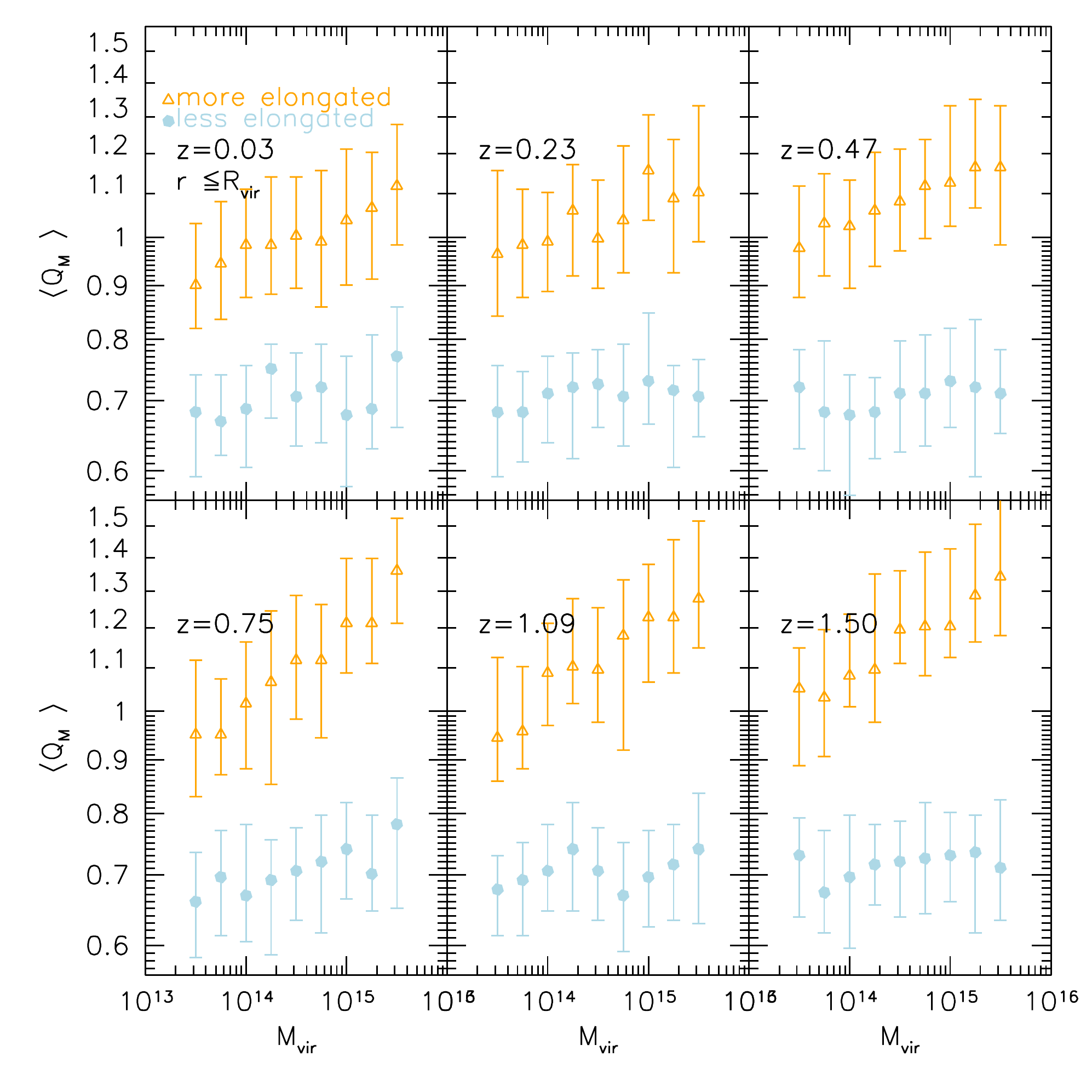}
\includegraphics[width=8.8cm]{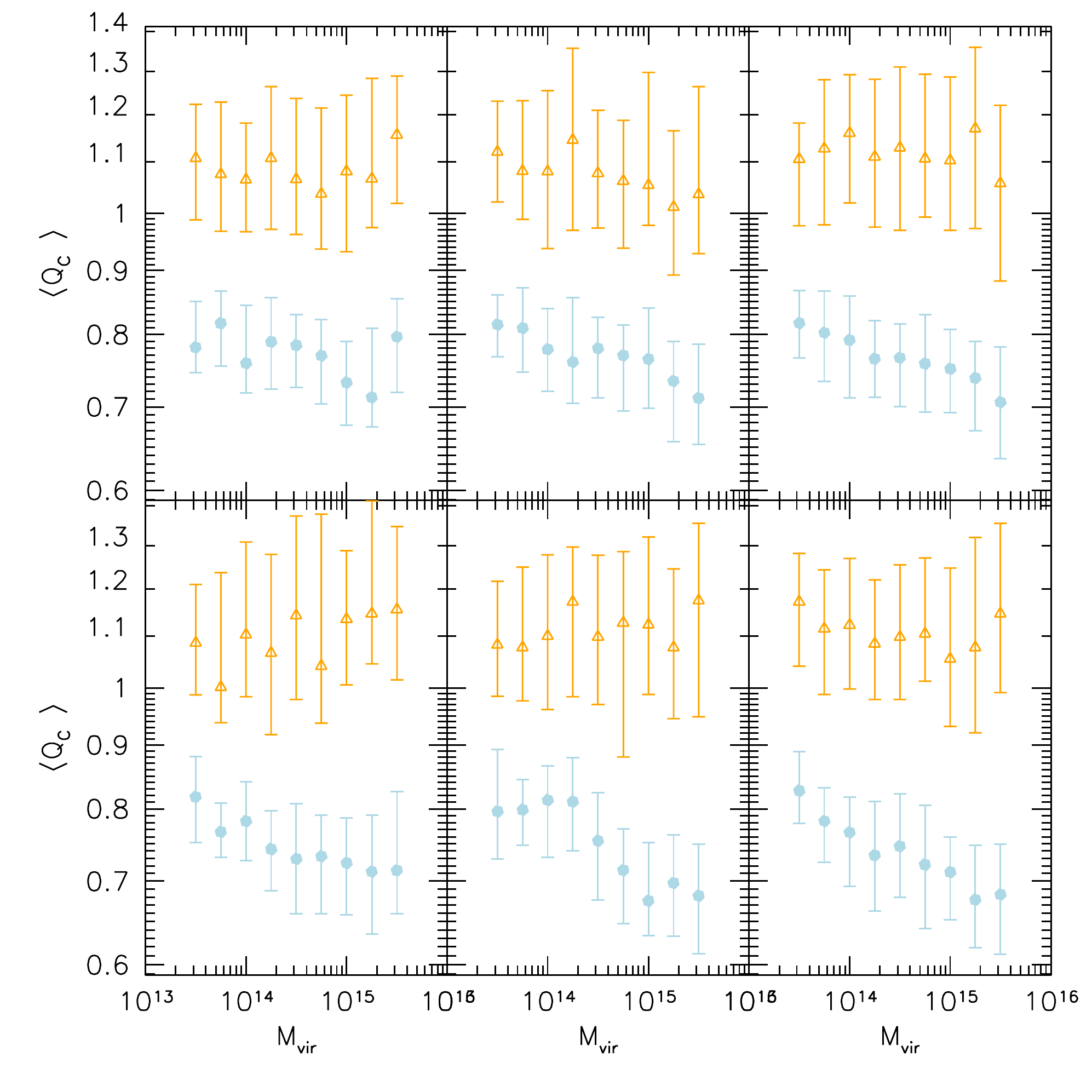}
\caption{Median rescaled  mass (\emph{left panels})  and concentration
  (\emph{right panels}) estimates as a function of the hist halo mass,
  for the  \texttt{ELLwBCG} sample by fitting the  profile without any
  assumption on  the value of  the elongation using a  NFW-funtion. In
  each mass  bin we  divided the haloes  in more (open  triangles) and
  less  (filled  circles) elongated  along  the  line  of sight.   The
  various    panels    show   the    relation    at   six    different
  redshifts.\label{rescaledmassELLell}}
\end{center}
\end{figure*}

Fitting  the  profile without  any  assumption  on  the value  of  the
elongation  creates a strong  bias in  the mass  estimates due  to the
elongation  along the  line of  sight. This  can be  better quantified
dividing the halo  sample, in each mass bin according  to the level of
elongation,   in  more   and  less   elongated  systems.    In  Figure
\ref{rescaledmassELLell}   we  again  show   the  median   masses  and
concentrations  as a  function of  the true  mass for  the  samples of
haloes with large  and small elongation.  The separation  of haloes in
these two classes was made by measuring the median elongation $\langle
e_z \rangle$ in each mass bin.  We define as ``less elongated'' haloes
those with $e_z < \langle e_z \rangle$, and vice-versa.  This analysis
shows that the two samples have a large differences both in mass (left
panel) and in concentration (right  panel) of about $35-40\%$.  We can
summarize  that  while  more  and  less substructured  haloes  have  a
difference in the estimates of the order of few percent, the estimates
between  more and less  elongated haloes  can differ  up to  more than
$50\%$, in the case of very massive clusters.

This can  originate very important systematics when  trying to recover
cosmological  information from  the mass  and  concentration estimates
based  on  lensing  analysis  without  taking  into  account  possible
projection effects \citep{sereno12,coe12}.

\section{Cosmology from the $c-M$ relation of a wide field survey}
\label{section3}
Several  studies  aimed   at  constraining  cosmological  parameters  by
measuring  the  $c-M$  relation   of  galaxy  clusters.  For  example,
\citet{ettori10}  used  a  sample  of  $44$  X-ray  luminous  clusters
observed   with  XMM-Newton.   They  measured   the  masses   and  the
concentrations  by assuming spherically  symmetric X-ray  emitting gas
distributions   in   hydrostatic    equilibrium   with   the   cluster
gravitational potentials,  which are  modeled by NFW  functionals.  By
comparing  with  theoretical   $c-M$  relations,  they  constrain  the
degeneracy   between   $\Omega_m$   and  $\sigma_8$.    Checking   the
consistency  of   the  observed   cluster  $c-M$  relation   with  the
theoretical  expectations   in  the  framework   of  the  $\Lambda$CDM
cosmology   is  also   one  of   the  goals   of  the   ongoing  CLASH
Multi-Cycle-Treasury   program   of   the   Hubble   Space   Telescope
\citep{postman11}.  Interesting   results  are  emerging   from  these
analyses,  whose significance  is  however limited  by the  relatively
small sizes of the cluster samples.
 
The situation  is going to  significantly improve in the  near future.
Indeed, several  optical surveys are planned, which  will deliver high
quality data suitable  for the weak lensing analysis  of large samples
of           galaxy           clusters          (e.g.            KiDS,
DES\footnote{\texttt{http://www.darkenergysurvey.org}},          LSST),
\citet{kuijken10,anderson01}).  In particular, in the wide survey that
will be operated by the  Euclid mission, several dozen of thousands of
galaxy clusters will be detectable up to redshift $2$. Now, we attempt
to  quantify the robustness  of the  cosmological constraints  that it
will be  possible to  derive from the  analysis of such  large cluster
samples.

The  first step  towards  our goal  is  to model  a cluster  selection
function.  In  this work, we  make the simple assumption  that cluster
detections will  be not  contaminated above a  constant mass  limit of
$5\times   10^{13}M_{\odot}/h$    for   $z<0.4$   and    $1.5   \times
10^{14}M_{\odot}/h$ for $z \geq 0.4$.   This reflects the fact that at
high redshift we  tend to select only massive  clusters. More detailed
selection functions should be  derived for specific surveys, depending
on their  depth, observing  bands, etc.  The  mass limit used  in this
work  is  consistent with  the  analytical  estimates  for the  EUCLID
photometric  selection  function  \citep{euclidredbook}.  We  use  the
selection   function   in   combination  with   the   \citet{sheth99b}
mass-function to  generate a  population of cluster-sized  dark matter
halos reproducing the mass  and redshift distributions expected in the
context of a corresponding $\Lambda$CDM cosmology (see Sect.~1 for the
exact choice  of the cosmological  parameters). We use these  halos to
populate  a  light-cone corresponding  to  a  survey  area of  $15000$
sq. degrees.  The sampled volume extends from $z=0.03$ and $z=1.5$ and
is subdivided  in slices equally  spaced in d$\log(1+z)$.   We further
use the \citet{zhao09} $c-M$  relation, including a redshift dependent
log-normal scatter  as shown in Figure~\ref{scattercFIG},  to assign a
concentration to each halo.

As we discussed in the previous sections, projection effects introduce
substantial    biases   in    the   measurements    of    masses   and
concentrations. Such  biases can  be quantified using  the simulations
done with the  {\tt MOKA} code. More specifically,  for any input mass
and  redshift, we  can  use  the simulations  to  derive the  expected
distribution of the measured  masses and concentrations, assuming they
are obtained by fitting  the two-dimensional mass profiles of triaxial
halos.
\begin{figure*}
\begin{center}
\includegraphics[width=8.8cm]{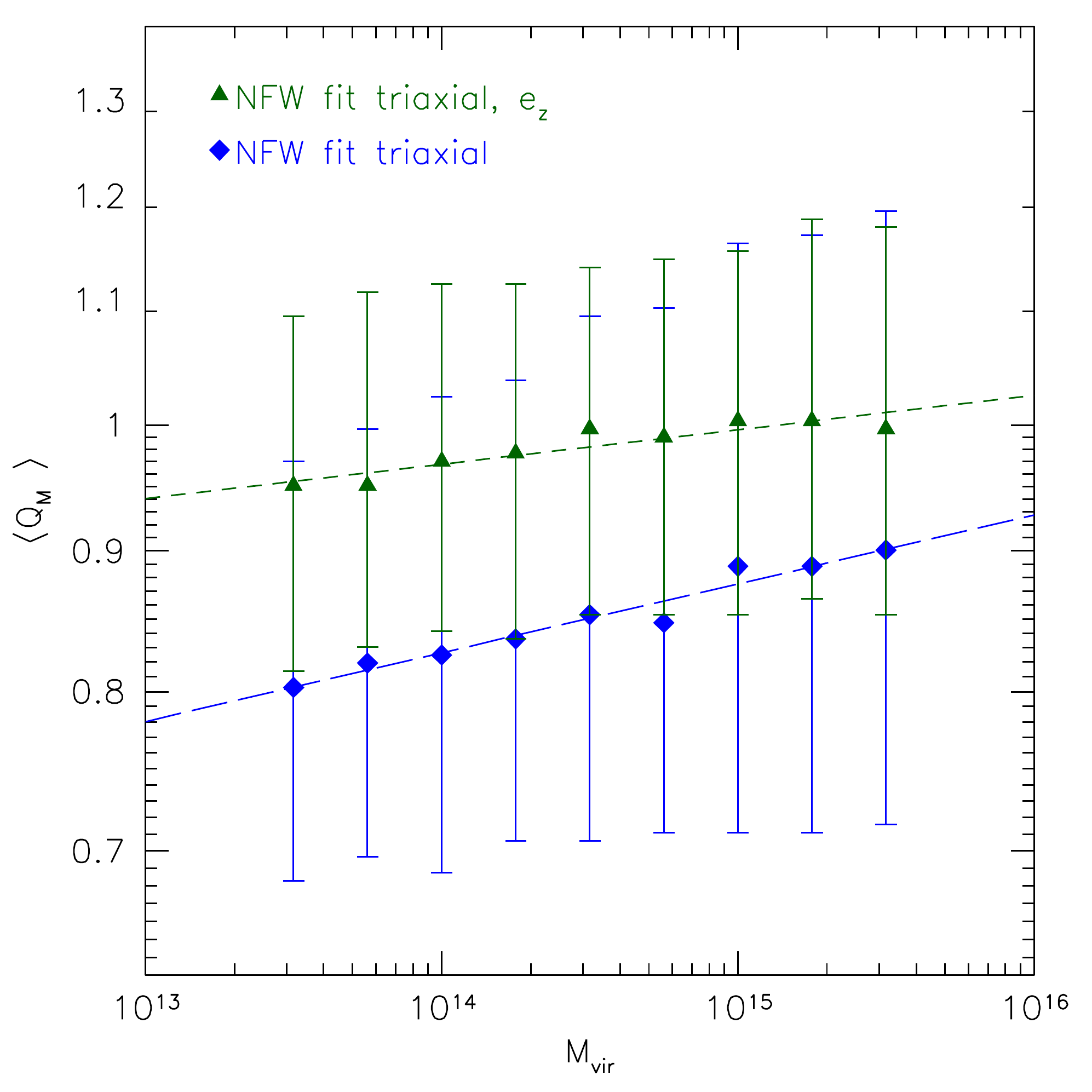}
\includegraphics[width=8.8cm]{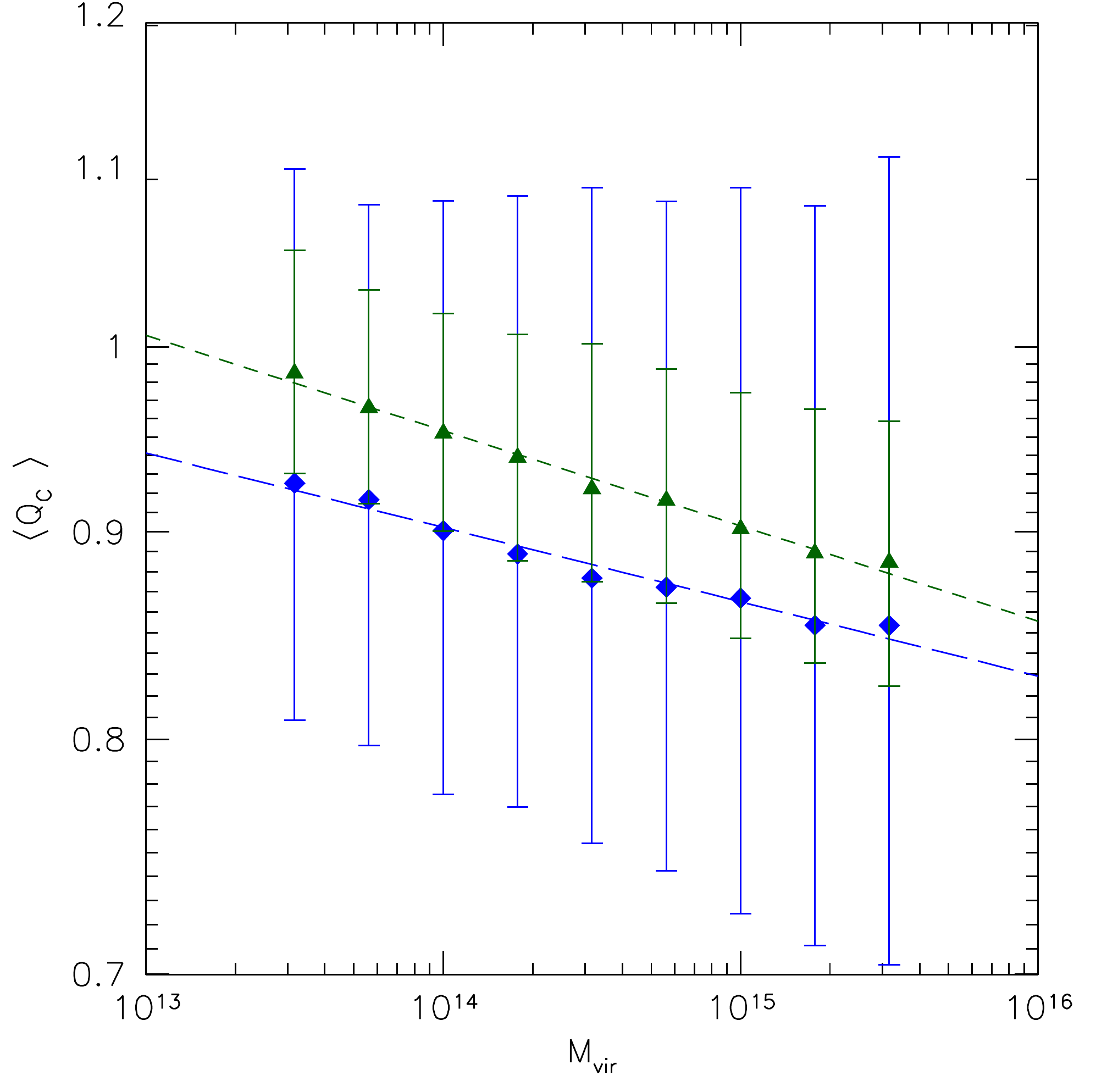}
\caption{Median  rescaled mass  (\emph{left panel})  and concentration
  (\emph{right panel}) as a function  of the true halo mass, combining
  the results of $16$ redshift  slices between $z=0.03$ up to $z=1.5$,
  for  the \texttt{ELLwBCG}  sample. Diamonds  and triangles  show the
  results for the NFW fits without and with correction for elongation,
  respectively.  The error bars show  the first and the third quartile
  of  the distribution.   The  long and  short-dashed  lines show  the
  least-squares fits to the data.\label{MestandCestintf}}
\end{center}
\end{figure*}
In the left and right  panels of Figure \ref{MestandCestintf}, we show
how the rescaled masses and concentrations are expected to change as a
function of the halo virial  mass $M_{vir}$. The results were obtained
by combining the \texttt{ELLwBCG} samples at $16$ different redshifts.
Triangles and  diamonds refer  to the cases  where the  projected mass
profiles were  fitted with and  without correcting for  the elongation
along  the line  of  sight (Eq.~\ref{elongcor} with $e_c=e_z$).  They
indicate the  median rescaled masses  and concentrations in  each mass
bin.  The  error-bars correspond to  the inter-quartile ranges  of the
distributions in the mass bins. We performed linear least-squares fits
to the data  points and we measured the relations  given by short- and
the long-dashed lines  in each panel. We can  explicit these relations
as
\begin{eqnarray}
\log (Q_M) &=& 0.025 \log (M_{vir}) - 0.433\,;\,rms=0.094 \label{QM1}\\
\log (Q_C) &=& -0.018 \log (M_{vir}) + 0.212\,;\,rms=0.081  \label{QC1} 
\end{eqnarray}
when no  correction for  elongation is applied.  When the fit  is done
correcting for $e_z$, we obtain
\begin{eqnarray}
\log (Q_M) &=& 0.013 \log (M_{vir}) - 0.190\,;\,rms=0.066 
\label{QM2} \\
\log (Q_C) &=& -0.023 \log (M_{vir}) + 0.308\,;\,rms=0.029 \label{QC2} 
\end{eqnarray}
We use these equations and their  r.m.s. to assign to each halo in the
light-cone  values  of $Q_M$  and  $Q_c$.  Therefore,  we compute  the
``biased" masses and concentrations as
\begin{eqnarray}
	M_{fit}&=&M_{true}\times Q_M \\
	c_{fit}&=&c_{true}\times Q_c \;.
\end{eqnarray} 

\begin{figure}
\begin{center}
\includegraphics[width=\hsize]{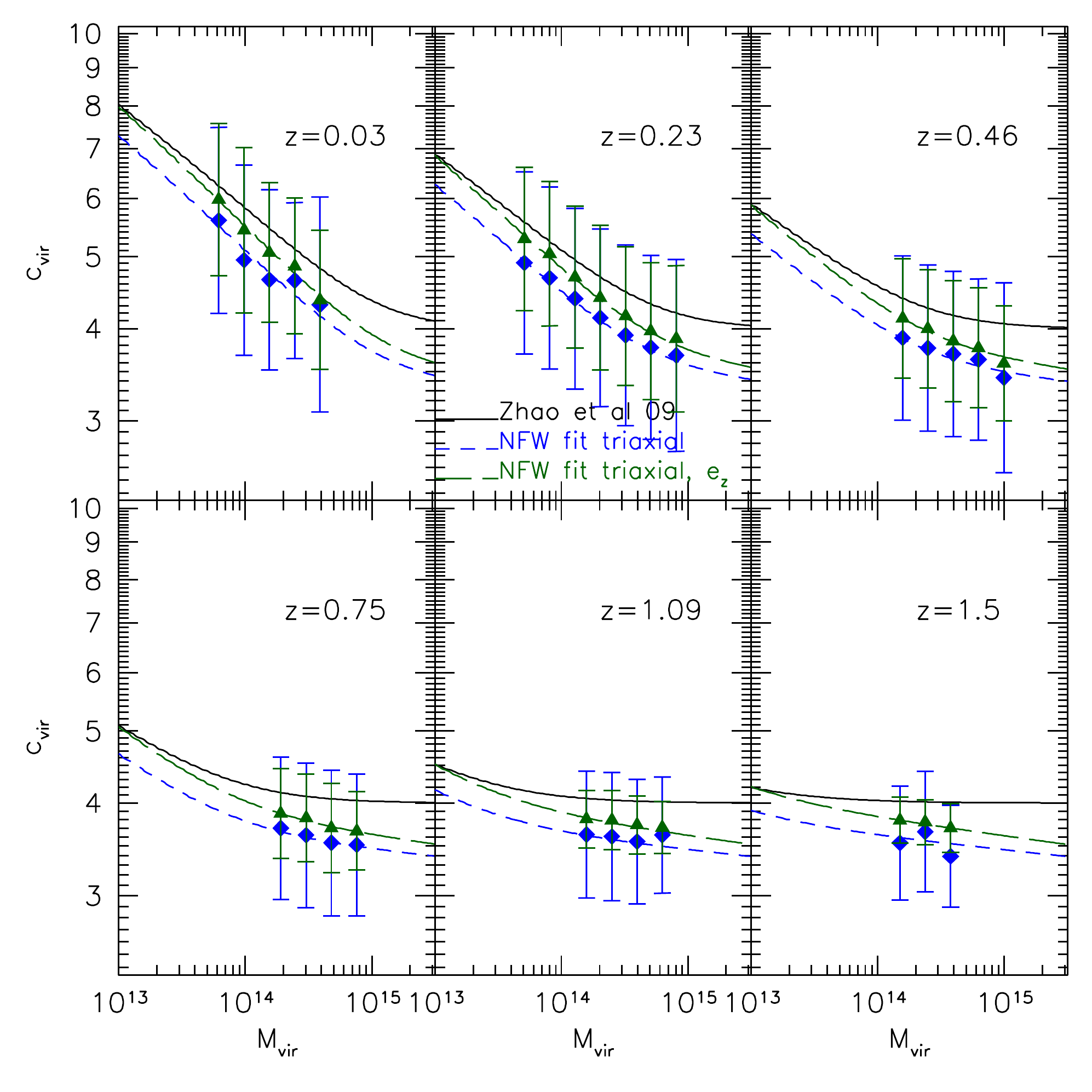}
\end{center}
\caption{\label{fcmrel}Concentration-mass  relations at  six different
  redshifts.  The different points show the average concentrations and
  r.m.s.in different  mass bins.  The  solid curve indicate  the input
  $c-M$  relation of  Zhao et  al.   Diamonds and  triangles show  the
  results obtained by fitting the profiles without and with correction
  for  elongation.  The  short  and  the  long-dashed  lines  are  the
  functionals that best fit the data (see text for more details).}
\end{figure}

Once we  have realistically included  the biases caused  by projection
effects, we can proceed to measure the halo $c_{fit}-M_{fit}$ relation
and use it to constrain the cosmological parameters. In particular, in
this section we focus on $\sigma_8$, $\Omega_m$, and $w$.

We show in  Figure \ref{fcmrel} the results obtained  in six different
redshift bins ranging between  $z=0.03$ (upper-left panel) and $z=1.5$
(bottom-right  panel). The  solid  black curve  show  the input  $c-M$
relation by \citet{zhao09}. As usual, the green triangles and the blue
diamonds  indicate the  concentrations and  masses recovered  with and
without correcting  for elongation.  As  we can see, the  diamonds are
bised lower then the triangles with  respect to the Zhao et al.  $c-M$
relation.      These     reflect      the     results     shown     in
Figure~\ref{MestandCestintf}, where  we see that  correcting for $e_z$
generally allows to recover the correct  input mass, but it leads to a
still  slightly lower  estimate of  the cluster  concentration.  These
results seem to indicate that a correction for elongation, as given in
Eq.~\ref{elongcor} with  $e_c=e_z$, should  be applied to  the cluster
mass. As  explained before  when we correct  for elongation,  the mass
estimate is  unbiased, while the resulting  concentration is corrected
but  not enough, since  $Q_M$ and  $Q_c$ are  not independent  of each
other, considering that $c_{vir} \sim M_{vir}^{1/3}$ they do not scale
in  the   same  way   but  as:  $\Delta   Q_c  =  \left(   \Delta  Q_M
\right)^{1/3}$.

We fit the triangles and the diamonds with functions of the kind
\begin{equation}
	c_{2D}(M)=c_{3D}(M)\times N\;M^{\lambda} \;,
\end{equation}
where $c_{3D}$ is  the input $c-M$ relation (in our  case the model by
\citet{zhao09}) and we find that they are well fitted by the following
functional forms:
\begin{equation}
c_{2D}(M) = c_{3D}(M) \times \\ \left\{ \begin{array}{ll}
      1.630 \; M^{-0.018},\;\mathrm{(short-dashed)} \\
      \\
      2.033 \; M^{-0.023},\;\mathrm{(long-dashed),}\\      
  \end{array} \right. \label{cmcorr}
\end{equation}
respectively.   Finally,   we  use  the   recovered  $c_{fit}-M_{fit}$
relations  to constrain  the cosmological  parameters.  To  do  so, we
compare  them  to  corresponding   $c-M$  relations  of  Zhao  et  al.
recalculated  by   varying  the  cosmological   parameters.   In  this
experiment,  we  vary  only   two  of  the  investigated  cosmological
parameters simultaneously, while we keep  the third fixed and equal to
value used in  our simulations.  The purpose of  this experiment is to
quantify  the   degeneracies  between  cosmological   parameters  when
constraining  them  using  the  $c-M$  relation.   If  the  couple  of
parameters which  is changed is ($\Phi,\Psi$), we  define the $\chi^2$
variable as
\begin{eqnarray}
  \chi^2(\Phi,\Psi) &&= \sum_{j}^{redshifts}\sum_i^{N_{mass\,bin}} \\ 
     && \dfrac{\left\{
     \log[c_{fit,i}(z_j)] - \log[c_{model}(M_{fit,i},z_j|\Phi,\Psi)]
    \right\}^2}{\sigma_{\log c,i}^2}
    \nonumber
\end{eqnarray}
where  $\sigma_{\log c,i}$  represents  the concentration  $rms$ in  a
given mass bin. Then, we  find the couples of cosmological parameters,
$(\bar\Phi,\bar\Psi)$, that minimize $\chi^2$.

\begin{figure*}
\begin{flushleft}
\hspace{1.4cm} no bias: only log-normal scatter in $c_{vir}$ \hspace{0.8cm}
bias: NFW fit triaxial \hspace{1.6cm}
bias: NFW fit triaxial, knowing $e_z$
\end{flushleft}
\begin{center}
\includegraphics[width=5cm]{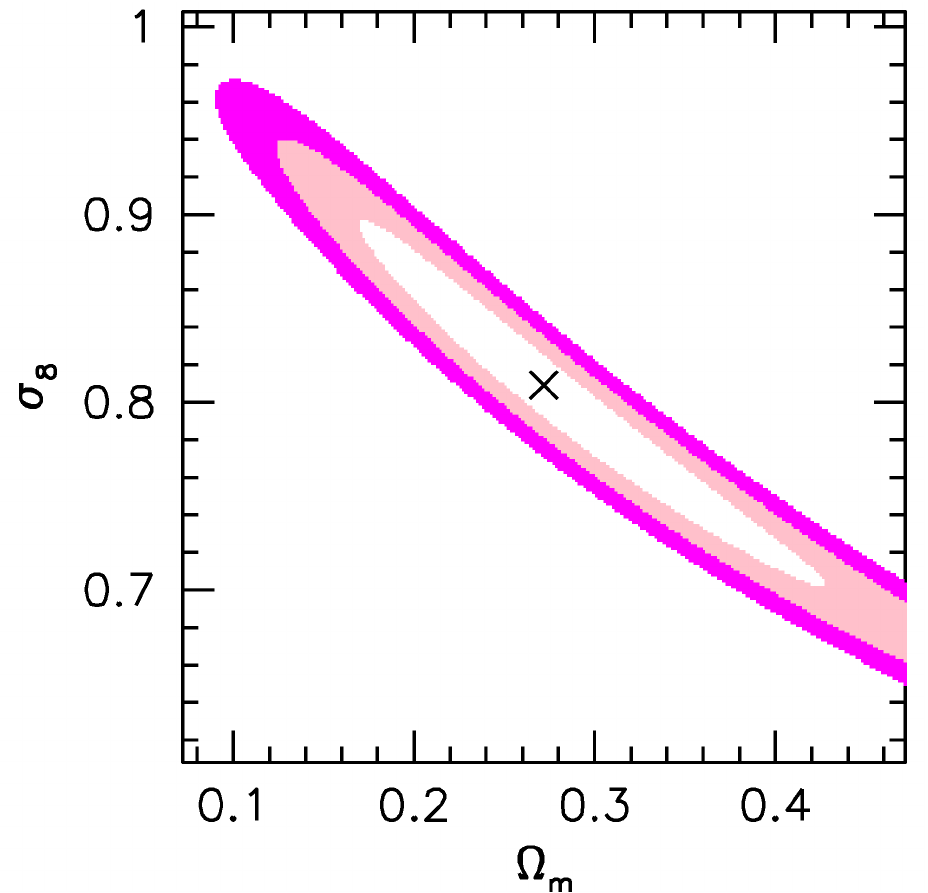}
\includegraphics[width=5cm]{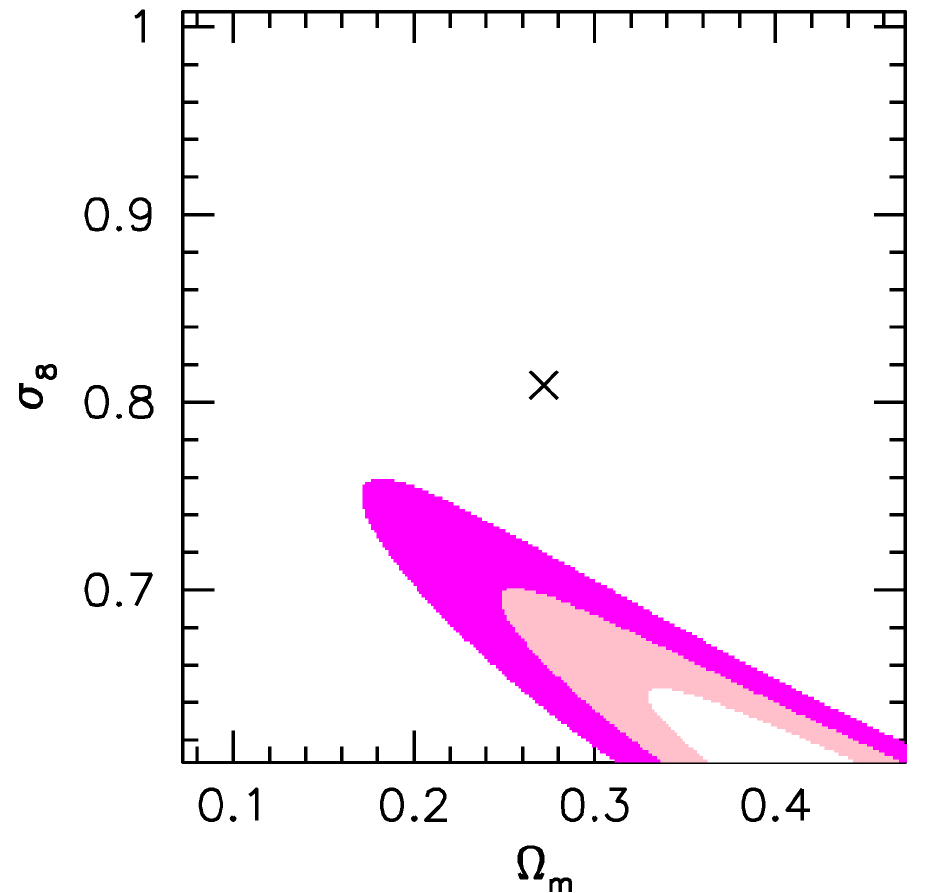}
\includegraphics[width=5cm]{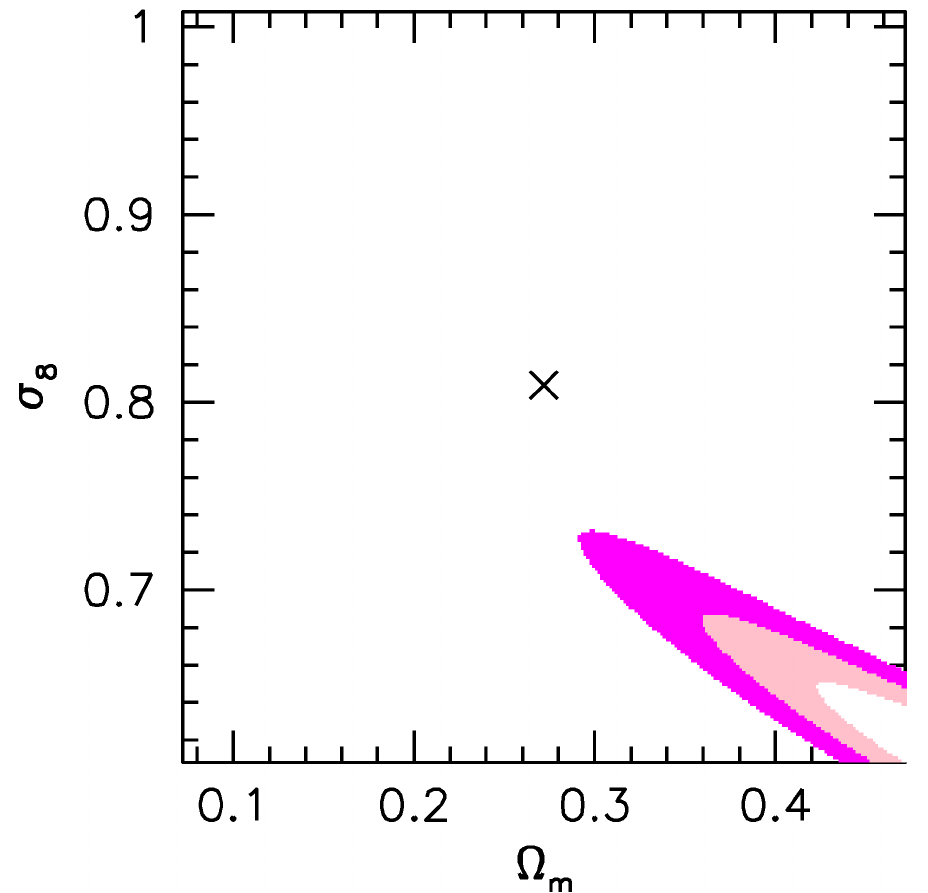}
\includegraphics[width=5cm]{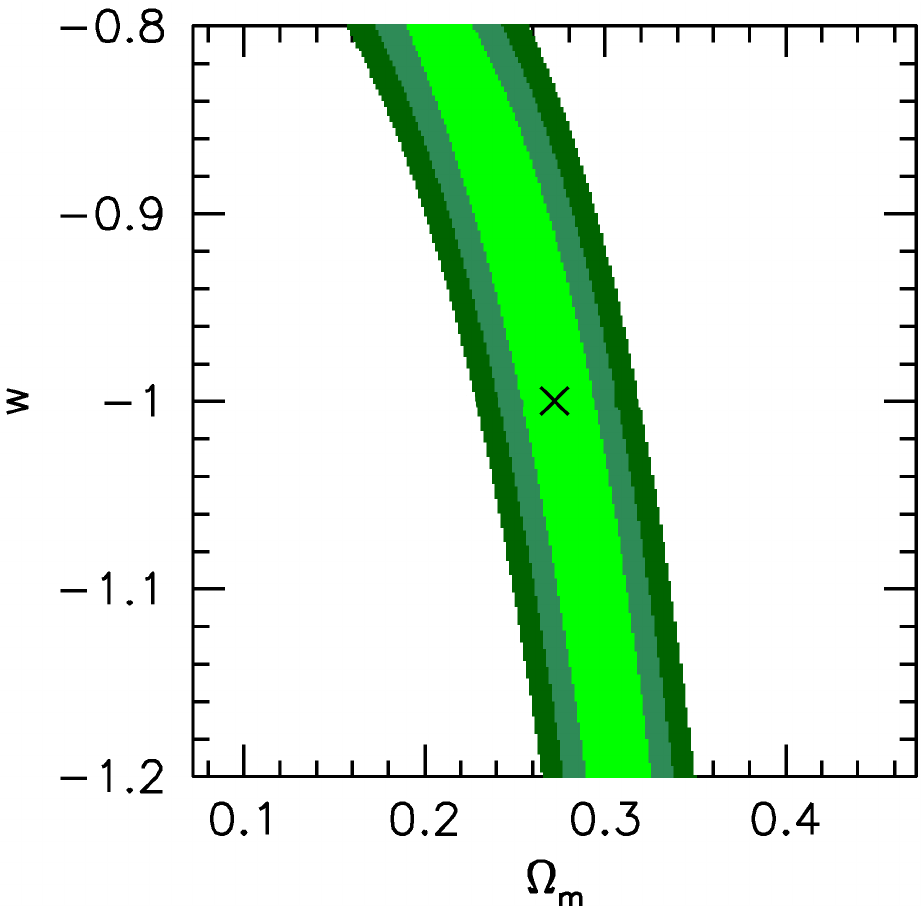}
\includegraphics[width=5cm]{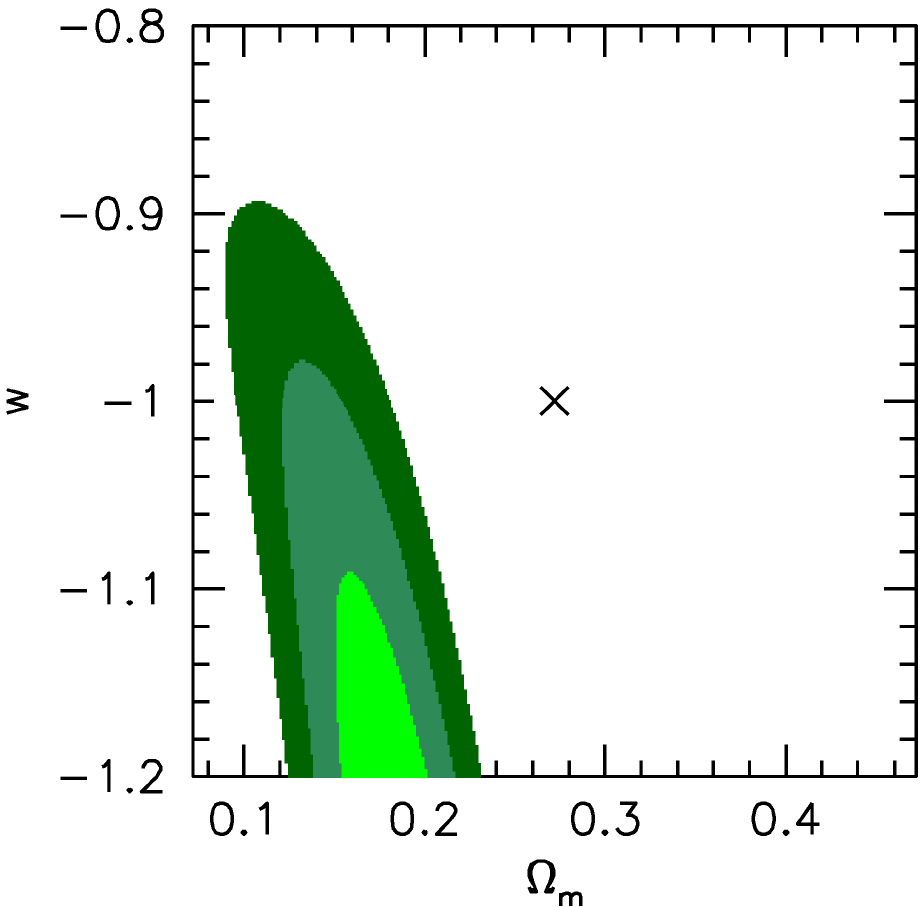}
\includegraphics[width=5cm]{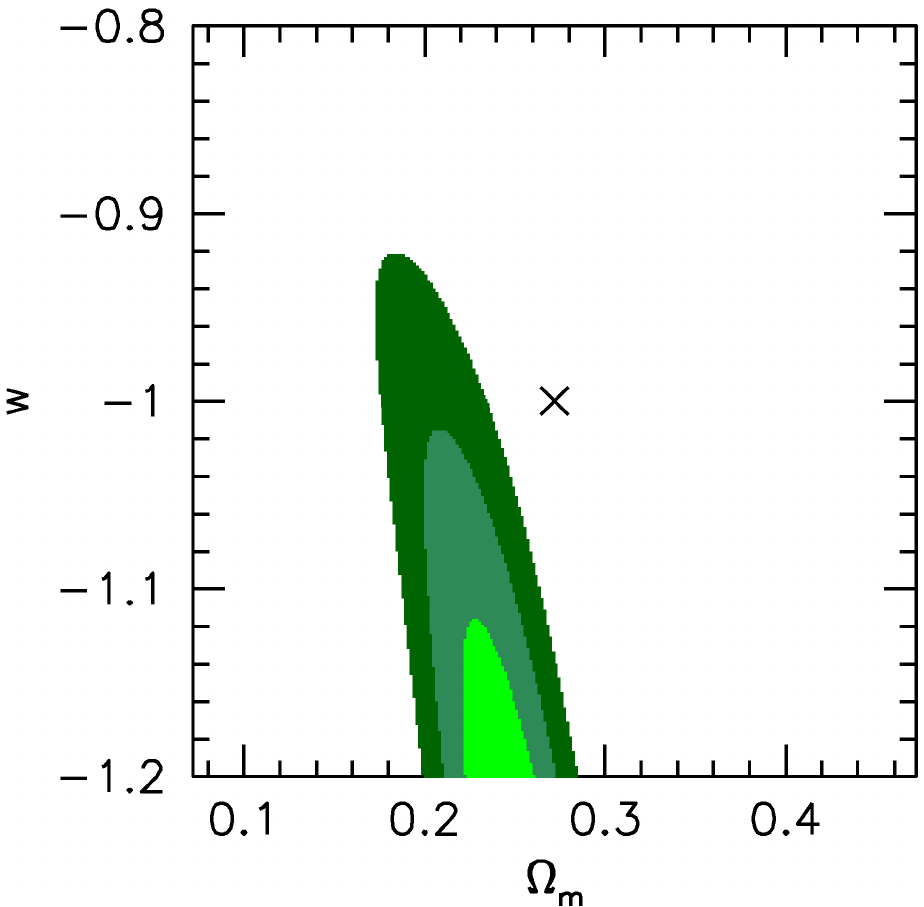}
\includegraphics[width=5cm]{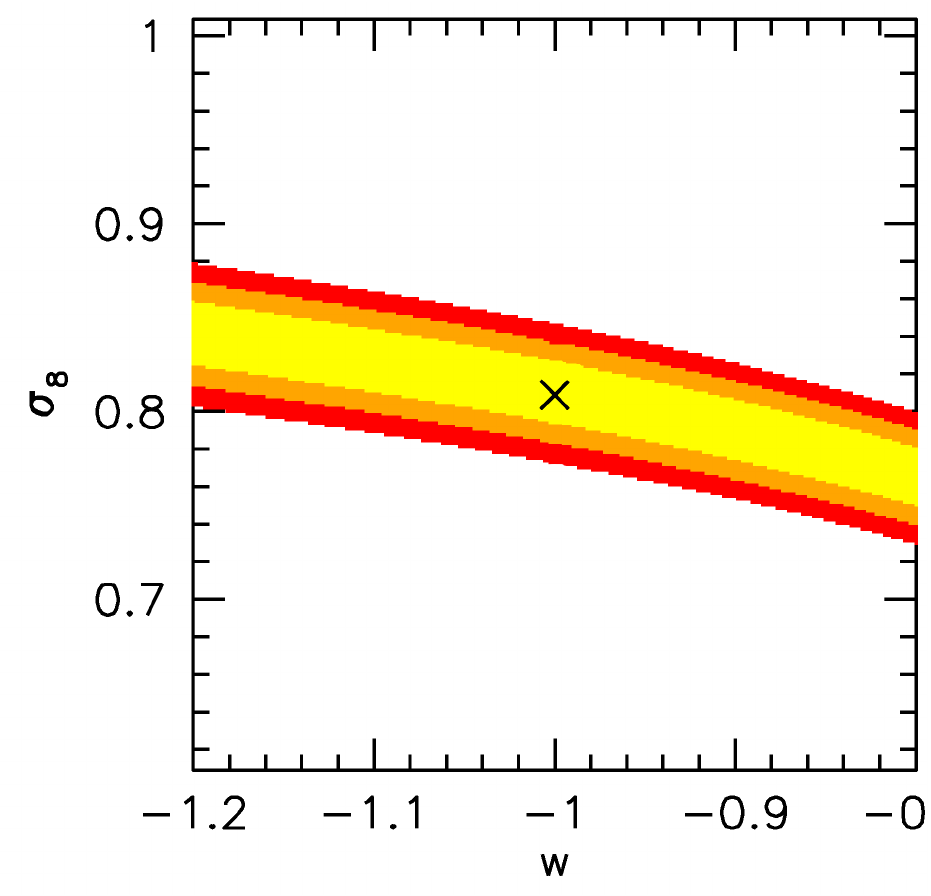}
\includegraphics[width=5cm]{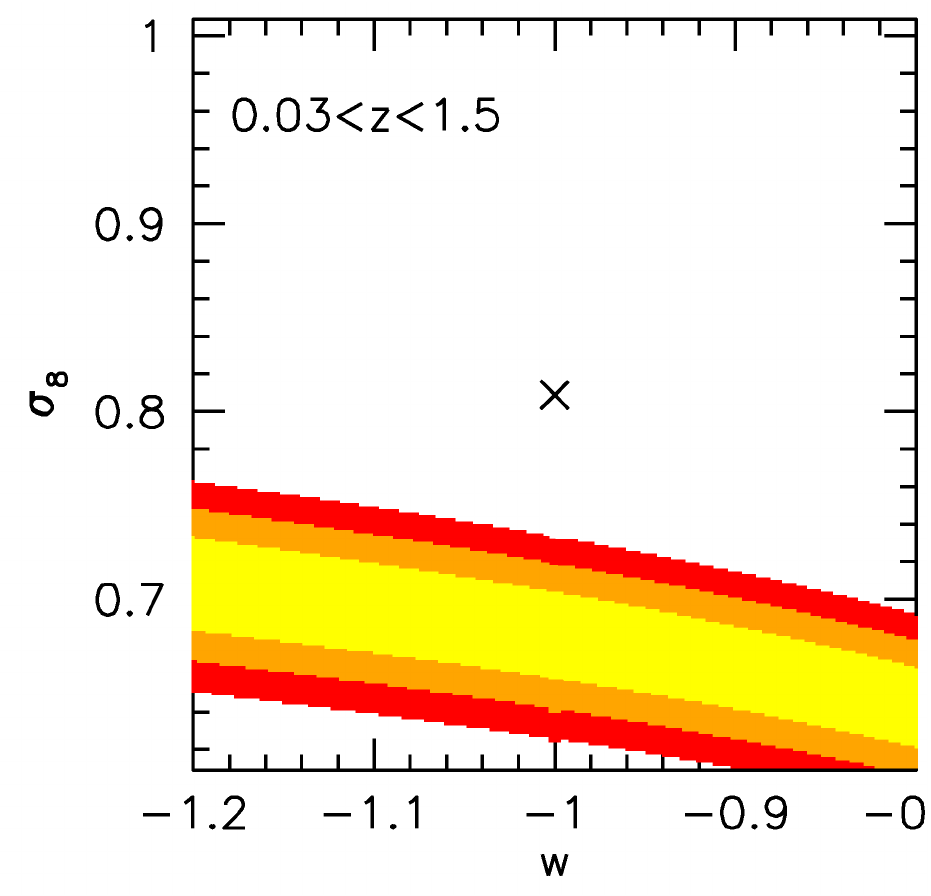}
\includegraphics[width=5cm]{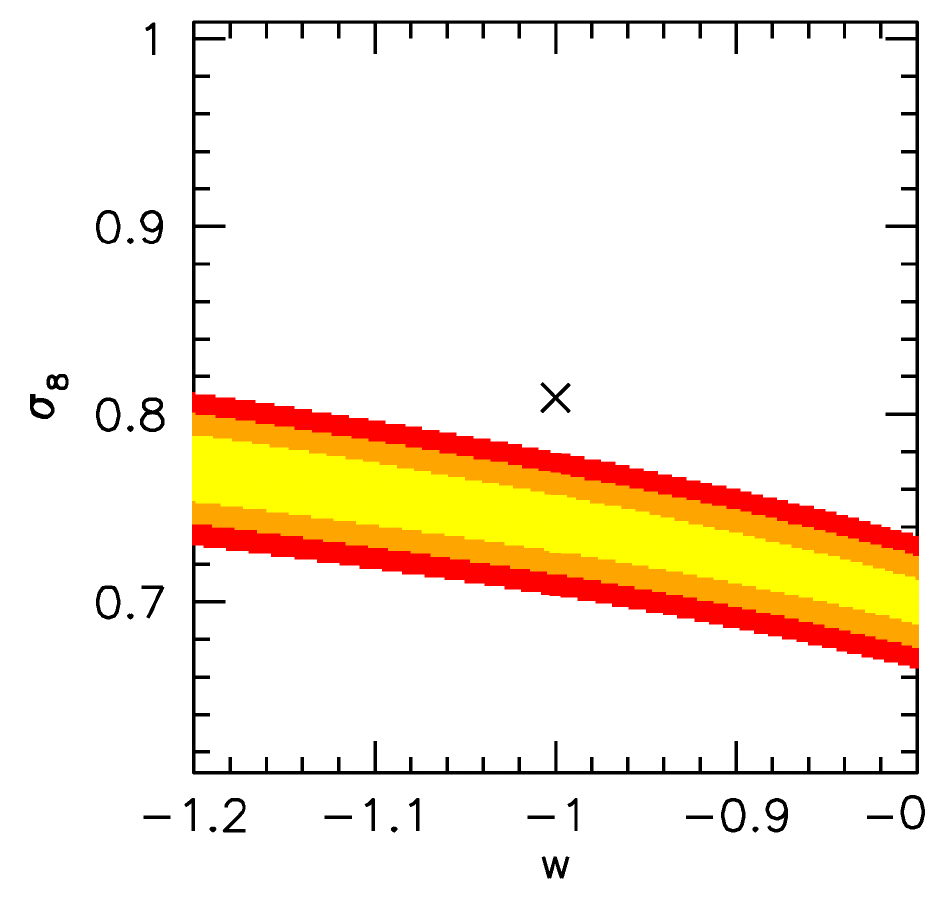}
\caption{$1$,  $2$ and $3$  $\sigma$ confidence  levels in  the planes
  $\sigma_8,\Omega_m$    (\emph{top    panels})    and    $\sigma_8,w$
  (\emph{bottom panels}).  The results  refer to a $15000$ sq. degrees
  survey, combining measurements of  the $c-M$ relation in 16 redshift
  bins between $z=0.03$  and $z=1.5$.  In the left  panels, we fit the
  input $c-M$ relation, adding only a log-normal scatter in $c_{vir}$.
  In the middle column, we use the $c_{fit}-M_{fit}$ relation measured
  without applying  any correction for elongation.  Finally, the right
  panels  refer  to the  case  where,  concentrations  and masses  are
  recovered  by  correcting  for   the  true  halo  elongations  using
  Eq.~\ref{elongcor}  with $e_c=e_z$.  The  true sets  of cosmological
  parameters  are   indicated  by  the   black  crosses  ($\Lambda$CDM
  cosmological model).\label{cont1}}
\end{center}
\end{figure*}

In  Figure \ref{cont1},  we show  the 1,  2, and  3$\sigma$ confidence
levels for the couples  of cosmological parameters $(\sigma_8,w)$ (top
panels),  $(w,\Omega_m)$ (middle  panels), and  $(\sigma_8,w)$ (bottom
panels).  The results refer to a $15000$ sq. degrees survey, combining
measurements  of  the  $c-M$  relation  in 16  redshift  bins  between
$z=0.03$ and  $z=1.5$.  In the  left panels ,  we fit the  input $c-M$
relation, adding only a log-normal scatter in $c_{vir}$.  As expected,
we recover the input sets of cosmological parameters, corresponding to
the $\Lambda$CDM cosmological model. In  the middle column, we use the
$c_{fit}-M_{fit}$  relation measured  without applying  any correction
for elongation.   In this case, the  $c-M$ relation is  best fitted by
models with lower  $\sigma_8$ and lower $\Omega_m$ when  we fix $w=-1$
(upper panel).   Keeping $\sigma_8=0.809$ (middle  panel), the results
are sensitive  to $\Omega_m$,  and values of  $w$ lower than  $-1$ are
preferred  (phantom dark  energy). In  both upper  and  middle panels,
however, the  input cosmological parameters are  outside the $3\sigma$
of the  recovered sets of parameters.  Fixing $\Omega_m=0.272$ (bottom
panel),  the  input values  of  $\sigma_8$  and  $w$ are  outside  the
measurements.    The   right  panels   refer   to   the  case   where
concentrations  and masses are  recovered by  correcting for  the true
halo elongations using  Eq.~\ref{elongcor} with $e_c=e_z$. The results
of the  fit in this case  are qualitatively similar to  those shown in
the panels in the middle  column, but the offset between recovered and
input cosmological parameters is slightly weaker.

In  Table  \ref{tabcosmo},  we  quantify the  degeneracy  between  the
couples    of    cosmological    parameters   estimated    from    the
concentration-mass relation.  Results are reported  for  $3-\sigma$
and $1-\sigma$ (in parentheses) confidence limits.

\begin{table*}
  \caption{Degeneracy between cosmological parameters estimated from the concentration-mass
    relation. The subscript ($0$) indicates the input parameters, which correspond to a $\Lambda$CDM cosmological model ($\Omega_{m,0}=0.272$, $w_{0}=-1$ and $\sigma_{8,0}=0.809$)
    \label{tabcosmo}}
\begin{tabular}{c|c|c|}
 parameters & $\alpha$ & $\beta$ \\ \hline \hline
 $\dfrac{\sigma_{8}}{\sigma_{8,0}}=\beta \left( \dfrac{\Omega_{m}}{\Omega_{m,0}}\right)^{\alpha}$ & $-0.235\;(-0.265)$ & $0.983\;(0.994)$ \\ \hline 
 $\dfrac{w}{w_{0}}=\beta \left( \dfrac{\Omega_m}{\Omega_{m,0}}\right)^{\alpha}$ & $1.082\;(1.112)$ & $-1.019\;(-1.013)$ \\ \hline 
 $\dfrac{\sigma_{8}}{\sigma_{8,0}}=\beta \left( \dfrac{w}{w_{0}}\right)^{\alpha}$ & $0.231\;(0.231)$ & $0.999\;(1.000)$ \\ \hline \hline
\end{tabular}
\end{table*}

\begin{figure}
\begin{flushleft}
\hspace{0.5cm}
bias: NFW fit triaxial \hspace{0.4cm}
bias: NFW fit triaxial, knowing $e_z$
\end{flushleft}
\begin{center}
\includegraphics[width=4cm]{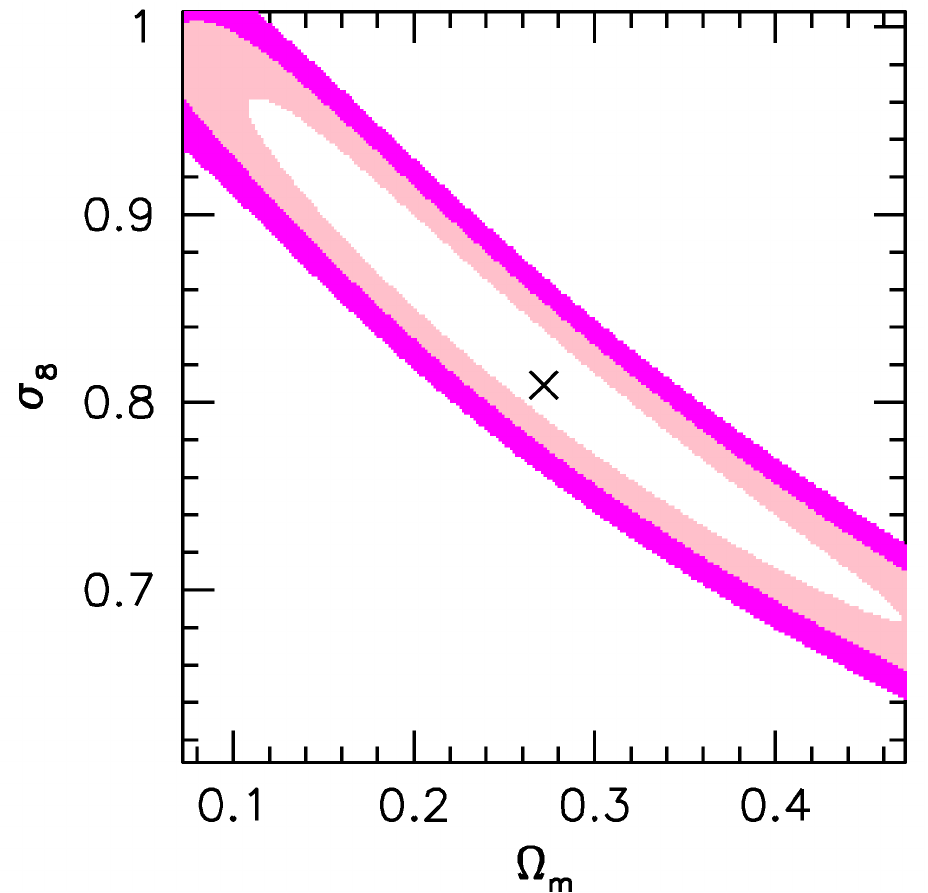}
\includegraphics[width=4cm]{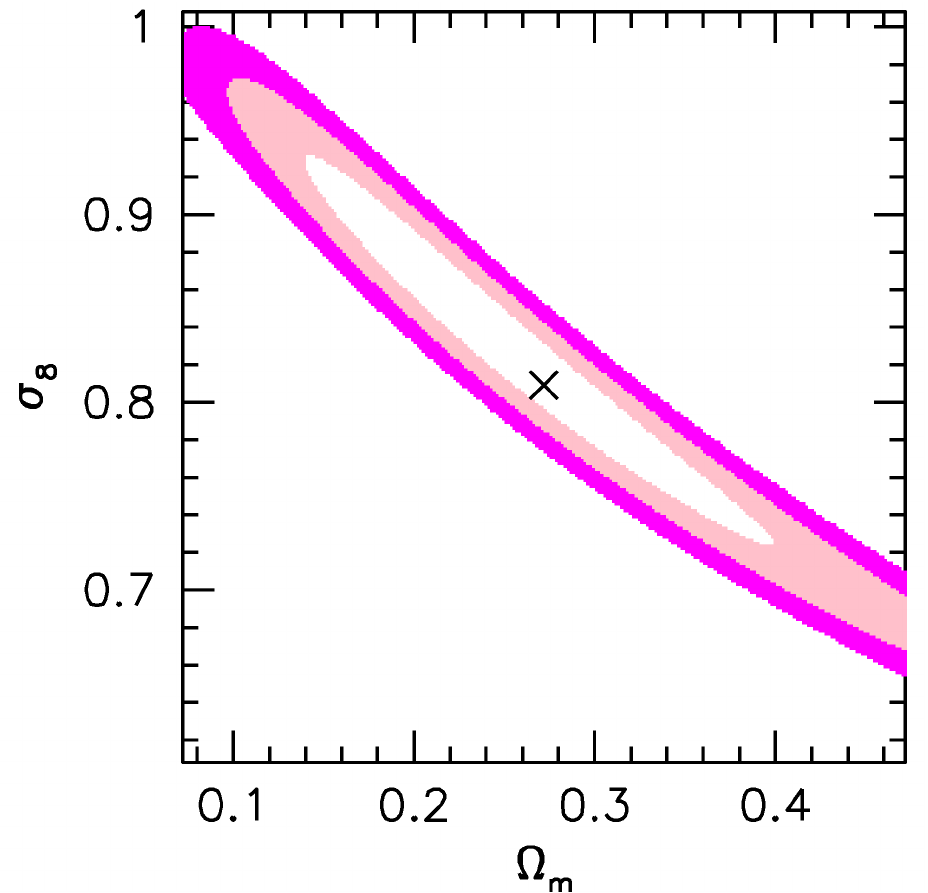}
\includegraphics[width=4cm]{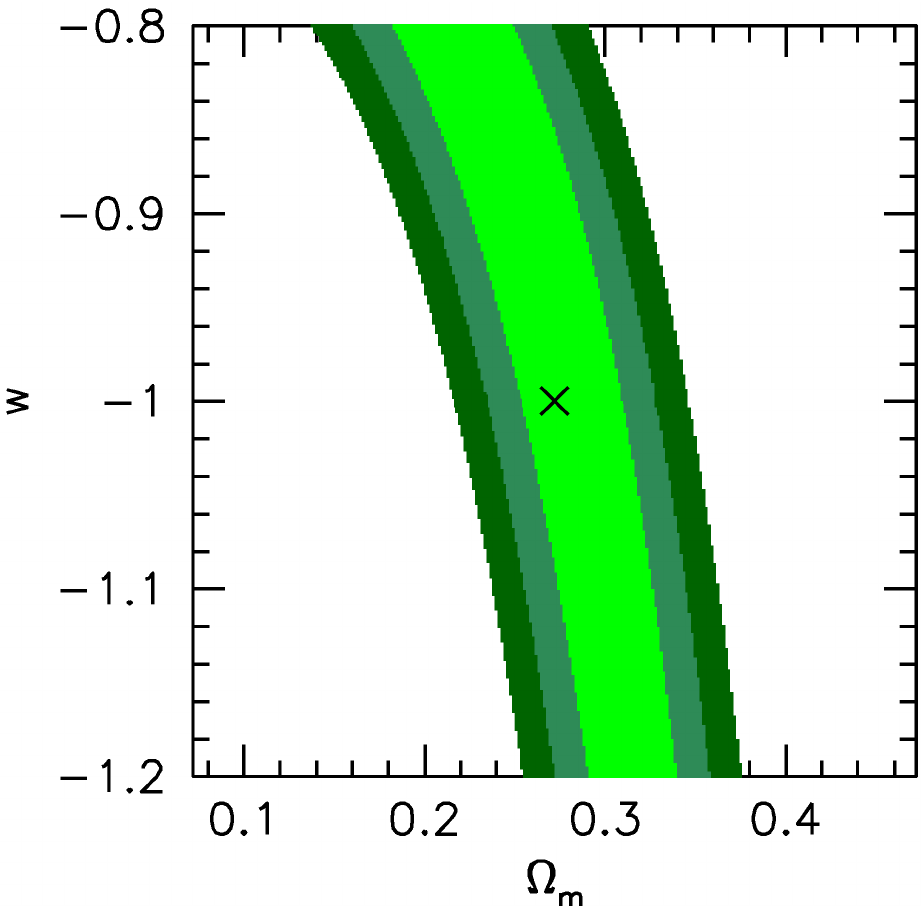}
\includegraphics[width=4cm]{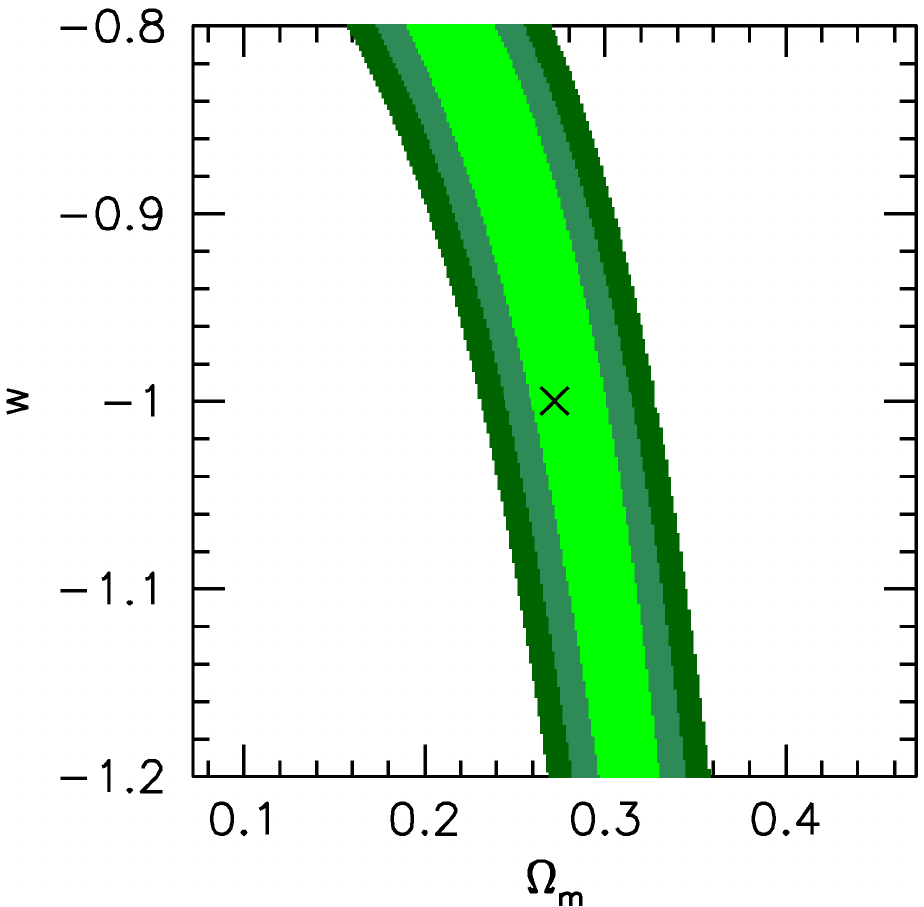}
\includegraphics[width=4cm]{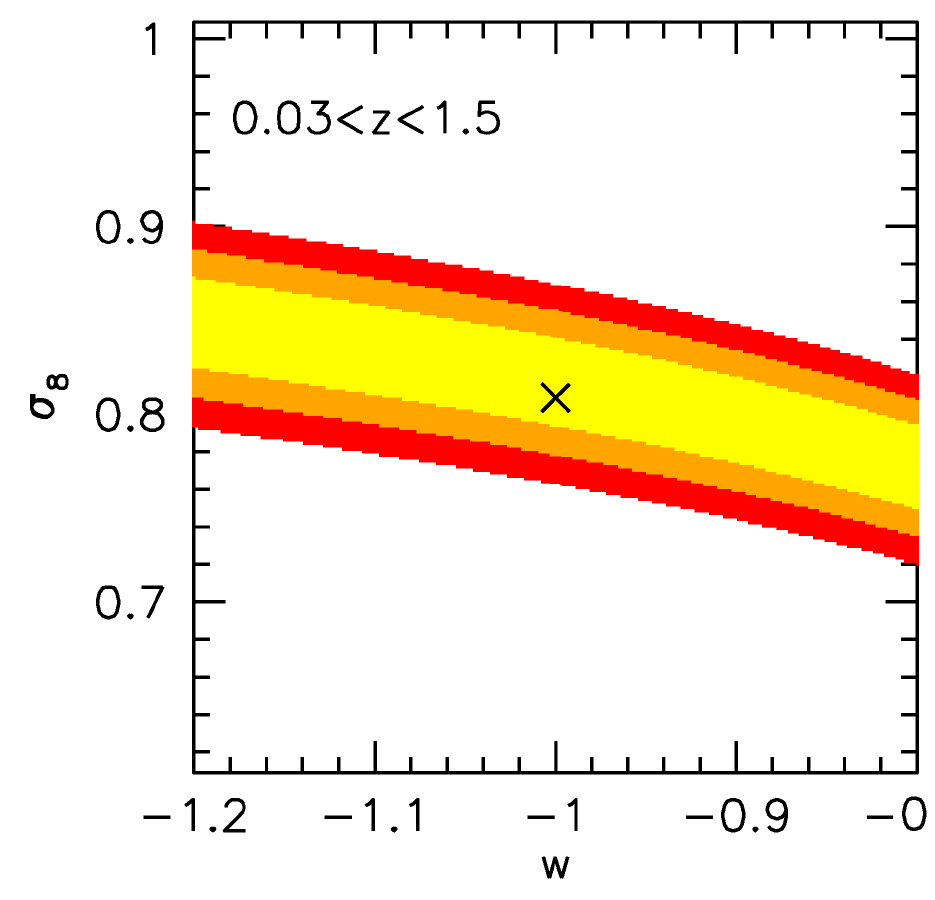}
\includegraphics[width=4cm]{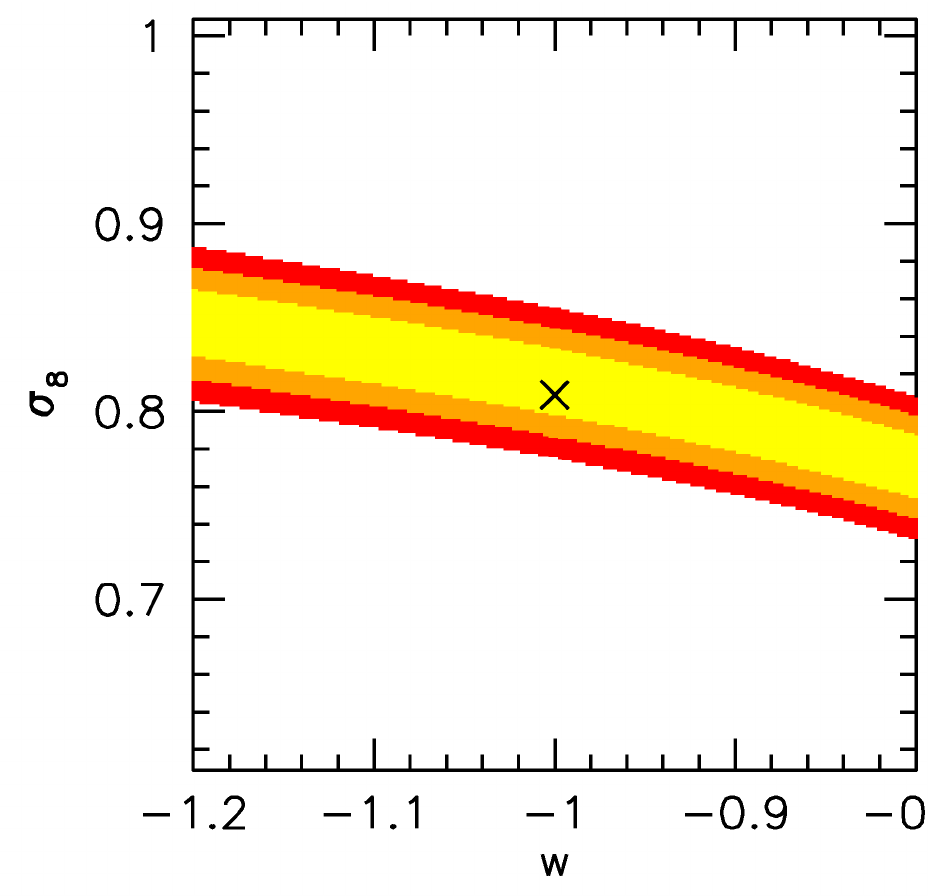}
\caption{$1$,  $2$  and  $3\sigma$  confidence levels  in  the  planes
  ($\sigma_8,\Omega_m$  (\emph{top panels}),  $\sigma_8-\Omega_m$, and
  $\sigma_8-w$  (\emph{bottom panels}). Here,  the relations  given in
  Eq.~\ref{cmcorr}  are  used  to  correct  the biases  on  the  $c-M$
  relations.\label{cont3}.}
\end{center}
\end{figure}

In Figure \ref{cont3}  we present again the confidence  levels for the
cosmological parameters  obtained when fitting the  $c-M$ relations in
the cases without  and with correction for elongation  (left and right
panels, respectively). In these  cases, however, we used the relations
in  Eq.~\ref{cmcorr}   to  correct  for   the  biases  in   the  $c-M$
relations. As expected the input cosmological models are now perfectly
recovered within $1\sigma$.

\section{Summary and Conclusions}
\label{lastsection}
In this  work, we  studied how well  the three-dimensional  masses and
concentrations of  galaxy clusters are recovered  from their projected
mass distributions. We did this by fitting the convergence profiles of
a  large  sample  cluster-sized  halos  generated  with  the  publicly
available    code   \texttt{MOKA}   adopting    NFW-functionals.    We
investigated   several    sources   of   possible    biases   in   the
measurements. In  particular, we studied the biases  introduced by the
presence of 1) substructures of  different masses, 2) a massive galaxy
at the center of  clusters, and 3) cluster triaxiality.  Additionally,
we  considered the  systematics related  to the  choice of  the radial
range  where   the  projected  mass   profile  is  fitted.    We  also
investigated  how the  mass and  concentration measurements  change by
adding  a  correction  factor  that  takes into  account  the  cluster
elongation  along  the line-of-sight.   Finally,  we investigated  the
robustness    of   cosmological    constraints   derived    from   the
concentration-mass  relation  measured  in  large  samples  of  galaxy
clusters.

Our findings can be summarized as follows:
\begin{itemize}

\item by  distributing substructures in  spherically symmetric cluster
  halos,  we  found that  the  concentrations  recovered from  cluster
  convergence profiles are on  average biased low by $\sim5-6\%$. This
  bias  tends to  increase as  a function  of mass,  because  the most
  massive haloes  are expected to  possess a larger fraction  of their
  mass  in substructures.   These substructures  have an  even smaller
  effect on  the cluster mass  estimates, which differ from  the input
  masses by just a few percent;

\item the presence of massive galaxies at the cluster center generally
  causes tiny over-estimate of  the mass, when the convergence profile
  is  fitted  with  simple  NFW functionals.   Such  over-estimate  is
  slightly more evident for small-mass halos, where it is of order few
  percent;

\item the above mentioned  results have little dependence on redshift,
  while they depend  more significantly on the radial  range chosen to
  perform  the fit  of the  convergence profile.  More  precisely, the
  effects of  massive central  galaxies are remarkable  if the  fit is
  restricted  to   the  very  inner   region  of  the   clusters  ($r<
  0.1R_{vir}$). In  this case, the cluster mass  can be under-estimate
  by up  to $10\%$, while  the concentration can be  over-estimated by
  the same amount;

\item  modeling  the  halo  triaxiality  following  the  \cite{jing02}
  prescription   and   assuming    random   halo   orientations,   the
  three-dimensional masses  recovered from the fit  of the convergence
  profiles are on  average biased low by $\sim  10-20\%$. However, for
  more massive  clusters the mass bias  is not larger  than $10\%$, as
  also  found  by  \citet{meneghetti11,rasia12} using  the  tangential
  shear profile of simulated galaxy clusters. This bias is larger than
  what  found   by  \citet{corless09}  whose   triaxial  clusters  are
  constructed without substructures and with a different model for the
  axial  ratios $e_a/e_c$ and  $e_b/e_c$ \citep{shaw06}  imposing $e_a
  e_b e_c =1$, as we  do, but with $e_a<e_b<e_c=1$.  We explained such
  under-estimate of the mass as due to a larger probability to observe
  clusters  elongated  on  the  plane   of  the  sky  than  along  the
  line-of-sight  and  to  the  combined  effects  of  triaxiality  and
  presence  of  substructures  --   as  quantified  and  discussed  in
  Section~\ref{section2}.   We  notice  that  in this  case,  we  also
  measure low  concentrations, this bias in  halo properties estimates
  has  also  been  studied  and  discussed is  several  other  studies
  \citep{clowe04,oguri05,gavazzi05,corless07};

\item   introducing  a  correction   for  elongation   multipling  the
  convergence by $e_z$, we found that the bias in mass can be removed.
  This correction  also moves $Q_c$  toward unity but still  keeping a
  small negative bias;

\item  by simulating  a cluster  survey  covering an  area of  $15000$
  sq. degrees, we estimated  that triaxiality and substructure induced
  biases on the 3D $c-M$ relation, that would translate into biases on
  the values  of cosmological parameters  like $\sigma_8$, $\Omega_m$,
  and $w$,  if no  correction for elongation  is applied.   The biases
  decrease when correcting for elongation.
\end{itemize} 
  
We conclude that, in order to use the cluster $c-M$ relation, measured
from  the  cluster  projected   density  profiles  as  they  might  be
determined  by   means  of  the  cluster   lensing  signal  \citep[see
e.g.][]{coe12,sereno12}, to constrain  the cosmological parameters, it
is necessary to take into account possible biases due to substructures
and halo triaxiality. This can be done introducing a 2D $c-M$ relation
correcting the 3D one for projection effects.

\section*{Acknowledgements}
We thank the  anonymus referee for his useful  comments that helped to
improve the presentation and the discussion of our results.

We also  thank Giulia Despali, Ravi  K. Sheth and  Giuseppe Tormen for
useful comments and interesting suggestions.

We  acknowledge  financial  contributions from  contracts  I/009/10/0,
EUCLID-IC  phase A/B1, PRIN-INAF  2009, ASI-INAF  I/023/05/0, ASI-INAF
I/088/06/0, ASI I/016/07/0  COFIS, ASI Euclid-DUNE I/064/08/0, ASI-Uni
Bologna-Astronomy Department Euclid-NIS  I/039/10/0 and PRIN MIUR dark
energy  and cosmology with  large galaxy  surveys.  CG's  research has
been  partially supported  by  the project  GLENCO,  funded under  the
Seventh Framework Programme, Ideas, Grant Agreement n. 259349.
\appendix
\section{Adiabatic Contraction}
\label{app1}
The presence  of a dissipative baryonic component  influences the dark
matter distribution. This  is because cold baryons that  settle at the
host    halo    centre    act    contracting    the    dark    matter.
\citet{blumenthal86} described the adiabatic contraction analytically,
finding a good agreement with numerical simulations.  However the halo
contraction, as  measured in  numerical simulations with  gas physics,
may depend on  the parameters that model the  star formation processes
and the  gas physics.   In these cases,  the evolution of  the central
density  may also  be affected  by spurious  numerical effects  due to
two-body  scattering of  massive particles.   Also the  differences in
some  models  can  be due  to  a  higher  density threshold  for  star
formation,  for example  in the  case  in which  the supernova  energy
release is more concentrated  and creates rapid potential fluctuations
near  the center.  The  model by  \citet{blumenthal86} assumes  that a
spherically  symmetric  halo  can  be  thought of  as  a  sequence  of
concentric  shells,  made  of  particles  on  circular  orbits,  which
homogeneously  contract while  conserving the  angular  momentum.  The
initial  and final  density profiles  -- characterized  by  an initial
radius  $r_i$ and  a  final radius  $r_f$,  when a  central galaxy  is
present -- are related by
\begin{equation}
  r \left[ M_{BCG}(r) + M_{DM,f} \right] = r_{i} M_{DM,i}(r_{i})\,,
\label{ad1}
\end{equation}
where
\begin{equation}
  M_{DM,f} = M_{DM,i} \left( 1-f_{cool}\right)\,,
\end{equation}
and $f_{cool}$ is  the baryon fraction in the halo  that cools to form
the central  galaxy. To  solve the adiabatic-contraction  equation, we
need to derive $r$ from equation (\ref{ad1}). When the stellar density
distribution is described by a Hernquist model we can read:
\begin{equation}
  f_{cool} r^{3} + (r + r_{h})^{2}
  \left[(1-f_{cool}) r - r_{i} \right] m_{i}(r_{i}) = 0\,,
\label{eqadcontraction}
\end{equation}
while for a Jaffe model we have:
\begin{eqnarray}
\left[ f_{cool} + m_i(r_i) ( 1- f_{cool}) \right] r^2 &+& \nonumber \\
m_i(r_i) (r_j - f_c r_j + r_i) r&+& r_i r_j m_i(r_i) = 0
\;
\end{eqnarray}
where $r_h$ and  $r_j$ represent respectively the scale  radius of the
two models.   The two equations have  a single positive  real root. We
recall that $m_{i}$ in the previous equations defines the initial mass
profile normalized  by the halo  virial mass. From the  equation above
results  that  the  presence   of  a  dissipative  baryonic  component
contracts the dark matter distribution.

The  model we  described  above, and  used  to model  the dark  matter
contraction  in our  haloes, is  often defined  as  standard adiabatic
contraction (SAC). \citet{gnedin11}  have defined a modified adiabatic
contraction  (MAC) model  that  takes into  account  the results  from
numerical  simulations in  which gas  physics and  merging  events are
included. They argue that the MAC model gives excellent predictions to
the data,  while the SAC model  tends to overestimate  the dark matter
content toward  the halo  centre.  We also  stress that  the adiabatic
contraction observed in numerical  simulations depends directly on the
adopted  parameters. Cosmological  simulations performed  by different
authors, with very different codes  and physics inputs, agree that the
contraction effect is present, but at a weaker level than suggested by
the SAC model  and with a significant variation  from system to system
and from input model to input model.

\begin{figure}
\begin{center}
\includegraphics[width=8.8cm]{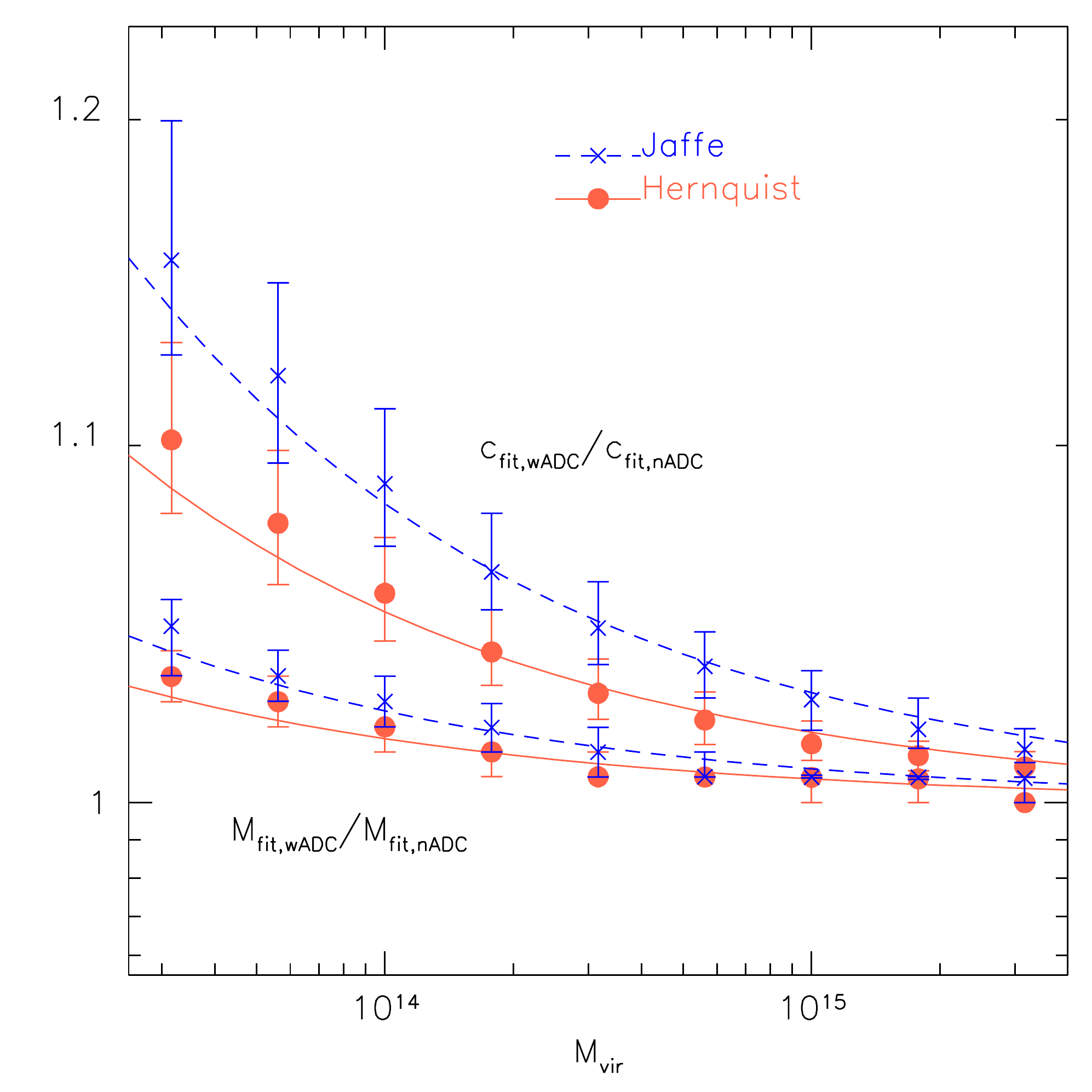}
\caption{Median  ratio between  the concentration  and  mass estimates
  obtained from the simulations performed at $z=0.03$, $0.23$, $0.47$,
  $0.75$,  $1.09$ and  $1.50$ with  and without  the inclusion  of the
  adiabatic contraction, as a function of the host halo mass.  We show
  the  results  for the  BCG  following  a  \citet{hernquist90} and  a
  \citet{jaffe83} profile. The error bars show the first and the third
  quartiles.\label{fcontration} The  curves show  the best fit  to the
  data   --  see   equations~(\ref{aeq1})   (\ref{aeq2})  (\ref{aeq3})
  (\ref{aeq4}).}
\end{center}
\end{figure}

Figure  \ref{fcontration}  shows the  median  ratio  of  the best  fit
estimate of the mass  (crosses) and the concentration (filled circles)
between  the  run with  (\texttt{ELLwBCGwADC})  and without  adiabatic
(\texttt{ELLwBCGwADC})  contraction as  a  function of  the host  halo
mass.  We  notice that in systems  where the ratio  between the BCG
mass and  the virial  mass is higher  (small systems)  the contraction
tends to increase  the concentration up to a  $10\%$, while for haloes
where $M_{BCG}/M_{vir}$ is lower (larger systems) the ratio approaches
to unity.  From the figure we also notice that the ratio between the two
mass estimates is influenced by  the adiabatic contraction by only few
percents.

In the Figure, the two  solid curves,  that consider  the case of  a BCG  following a
\citet{hernquist90} profile, correspond 
\begin{eqnarray}
\label{aeq2}
\log \left(\frac{M_{fit,wADC}}{M_{fit,nADC}}\right) = 
\exp\left[-\left(\log M_{vir} - 13.58 \right) \right]/88.80\,, \\
\log \left(\frac{c_{fit,wADC}}{c_{fit,nADC}}\right) = 
\exp\left[-\left(\log M_{vir} - 14.12 \right) \right]/51.04\,, 
\label{aeq1}
\end{eqnarray}
respectively, while the dashed curves, that consider the case of a BCG
following a \citet{jaffe83} model correspond:
\begin{eqnarray}
\label{aeq4}
\log \left(\frac{M_{fit,wADC}}{M_{fit,nADC}}\right) = 
\exp\left[-\left(\log M_{vir} - 13.95 \right) \right]/89.72\,, \\
\log \left(\frac{c_{fit,wADC}}{c_{fit,nADC}}\right) = 
\exp\left[-\left(\log M_{vir} - 14.53 \right) \right]/49.04\,, 
\label{aeq3}
\end{eqnarray}
which best fit the data points.  These equations can be used to modify
the   concentration-mass  relation  measured   in  pure   dark  matter
simulation to  include the effects of a  dissipative stellar component
at the host halo centre.

\bibliographystyle{mn2e}
\bibliography{cgiocoli}

\begin{thebibliography}{}

\bibitem[\protect\citeauthoryear{{Andreon}}{{Andreon}}{2010}]{andreon10}
{Andreon} S.,  2010, \mnras, 407, 263

\bibitem[\protect\citeauthoryear{{Bah{\'e}}, {McCarthy} \& {King}}{{Bah{\'e}}
  et~al.}{2012}]{bahe12}
{Bah{\'e}} Y.~M.,  {McCarthy} I.~G.,    {King} L.~J.,  2012, \mnras, 421, 1073

\bibitem[\protect\citeauthoryear{{Bartelmann}}{{Bartelmann}}{1996}]{bartelmann96}
{Bartelmann} M.,  1996, \aap, 313, 697

\bibitem[\protect\citeauthoryear{{Becker} \& {Kravtsov}}{{Becker} \&
  {Kravtsov}}{2011}]{becker11}
{Becker} M.~R.,  {Kravtsov} A.~V.,  2011, \apj, 740, 25

\bibitem[\protect\citeauthoryear{{Becker}, {McKay}, {Koester}, {Wechsler},
  {Rozo}, {Evrard}, {Johnston}, {Sheldon}, {Annis}, {Lau}, {Nichol} \&
  {Miller}}{{Becker} et~al.}{2007}]{becker07}
{Becker} M.~R.,  {McKay} T.~A.,  {Koester} B.,  {Wechsler} R.~H.,  {Rozo} E.,
  {Evrard} A.,  {Johnston} D.,  {Sheldon} E.,  {Annis} J.,  {Lau} E.,  {Nichol}
  R.,    {Miller} C.,  2007, \apj, 669, 905

\bibitem[\protect\citeauthoryear{{Bhattacharya}, {Habib} \&
  {Heitmann}}{{Bhattacharya} et~al.}{2011}]{bahttacharya11}
{Bhattacharya} S.,  {Habib} S.,    {Heitmann} K.,  2011, ArXiv e-prints

\bibitem[\protect\citeauthoryear{{Blumenthal}, {Faber}, {Flores} \&
  {Primack}}{{Blumenthal} et~al.}{1986}]{blumenthal86}
{Blumenthal} G.~R.,  {Faber} S.~M.,  {Flores} R.,    {Primack} J.~R.,  1986,
  \apj, 301, 27

\bibitem[\protect\citeauthoryear{{Boylan-Kolchin}, {Springel}, {White},
  {Jenkins} \& {Lemson}}{{Boylan-Kolchin} et~al.}{2009}]{boylan-kolchin09}
{Boylan-Kolchin} M.,  {Springel} V.,  {White} S.~D.~M.,  {Jenkins} A.,
  {Lemson} G.,  2009, \mnras, 398, 1150

\bibitem[\protect\citeauthoryear{{Clowe}, {De Lucia} \& {King}}{{Clowe}
  et~al.}{2004}]{clowe04}
{Clowe} D.,  {De Lucia} G.,    {King} L.,  2004, \mnras, 350, 1038

\bibitem[\protect\citeauthoryear{{Coe}, {Umetsu}, {Zitrin}, {Donahue},
  {Medezinski}, {Postman}, {Carrasco}, {Anguita}, {Geller}, {Rines},
  {Diaferio}, {Kurtz} et~al.,}{{Coe} et~al.}{2012}]{coe12}
{Coe} D.,  {Umetsu} K.,  {Zitrin} A.,  {Donahue} M.,  {Medezinski} E.,
  {Postman} M.,  {Carrasco} M.,  {Anguita} T.,  {Geller} M.~J.,  {Rines} K.~J.,
   {Diaferio} A.,  {Kurtz} M.~J.,    et~al., 2012, ArXiv e-prints

\bibitem[\protect\citeauthoryear{{Conroy}, {Prada}, {Newman}, {Croton}, {Coil},
  {Conselice}, {Cooper}, {Davis}, {Faber}, {Gerke}, {Guhathakurta}, {Klypin},
  {Koo} \& {Yan}}{{Conroy} et~al.}{2007a}]{conroy07a}
{Conroy} C.,  {Prada} F.,  {Newman} J.~A.,  {Croton} D.,  {Coil} A.~L.,
  {Conselice} C.~J.,  {Cooper} M.~C.,  {Davis} M.,  {Faber} S.~M.,  {Gerke}
  B.~F.,  {Guhathakurta} P.,  {Klypin} A.,  {Koo} D.~C.,    {Yan} R.,  2007a,
  \apj, 654, 153

\bibitem[\protect\citeauthoryear{{Corless} \& {King}}{{Corless} \&
  {King}}{2007}]{corless07}
{Corless} V.~L.,  {King} L.~J.,  2007, \mnras, 380, 149

\bibitem[\protect\citeauthoryear{{Corless} \& {King}}{{Corless} \&
  {King}}{2009}]{corless09}
{Corless} V.~L.,  {King} L.~J.,  2009, \mnras, 396, 315

\bibitem[\protect\citeauthoryear{{De Lucia}, {Kauffmann}, {Springel}, {White},
  {Lanzoni}, {Stoehr}, {Tormen} \& {Yoshida}}{{De Lucia}
  et~al.}{2004}]{delucia04}
{De Lucia} G.,  {Kauffmann} G.,  {Springel} V.,  {White} S.~D.~M.,  {Lanzoni}
  B.,  {Stoehr} F.,  {Tormen} G.,    {Yoshida} N.,  2004, \mnras, 348, 333

\bibitem[\protect\citeauthoryear{{Dolag}, {Bartelmann}, {Perrotta},
  {Baccigalupi}, {Moscardini}, {Meneghetti} \& {Tormen}}{{Dolag}
  et~al.}{2004}]{dolag04}
{Dolag} K.,  {Bartelmann} M.,  {Perrotta} F.,  {Baccigalupi} C.,  {Moscardini}
  L.,  {Meneghetti} M.,    {Tormen} G.,  2004, \aap, 416, 853

\bibitem[\protect\citeauthoryear{{Eke}, {Cole} \& {Frenk}}{{Eke}
  et~al.}{1996}]{eke96}
{Eke} V.~R.,  {Cole} S.,    {Frenk} C.~S.,  1996, \mnras, 282, 263

\bibitem[\protect\citeauthoryear{{Ettori}, {Gastaldello}, {Leccardi},
  {Molendi}, {Rossetti}, {Buote} \& {Meneghetti}}{{Ettori}
  et~al.}{2010}]{ettori10}
{Ettori} S.,  {Gastaldello} F.,  {Leccardi} A.,  {Molendi} S.,  {Rossetti} M.,
  {Buote} D.,    {Meneghetti} M.,  2010, \aap, 524, A68+

\bibitem[\protect\citeauthoryear{{Ettori}, {Morandi}, {Tozzi}, {Balestra},
  {Borgani}, {Rosati}, {Lovisari} \& {Terenziani}}{{Ettori}
  et~al.}{2009}]{ettori09}
{Ettori} S.,  {Morandi} A.,  {Tozzi} P.,  {Balestra} I.,  {Borgani} S.,
  {Rosati} P.,  {Lovisari} L.,    {Terenziani} F.,  2009, \aap, 501, 61

\bibitem[\protect\citeauthoryear{{Fedeli}}{{Fedeli}}{2011}]{fedeli11}
{Fedeli} C.,  2011, ArXiv e-prints

\bibitem[\protect\citeauthoryear{{Gao}, {Navarro}, {Cole}, {Frenk}, {White},
  {Springel}, {Jenkins} \& {Neto}}{{Gao} et~al.}{2008}]{gao08}
{Gao} L.,  {Navarro} J.~F.,  {Cole} S.,  {Frenk} C.~S.,  {White} S.~D.~M.,
  {Springel} V.,  {Jenkins} A.,    {Neto} A.~F.,  2008, \mnras, 387, 536

\bibitem[\protect\citeauthoryear{{Gao}, {White}, {Jenkins}, {Stoehr} \&
  {Springel}}{{Gao} et~al.}{2004}]{gao04}
{Gao} L.,  {White} S.~D.~M.,  {Jenkins} A.,  {Stoehr} F.,    {Springel} V.,
  2004, \mnras, 355, 819

\bibitem[\protect\citeauthoryear{{Gavazzi}}{{Gavazzi}}{2005}]{gavazzi05}
{Gavazzi} R.,  2005, \aap, 443, 793

\bibitem[\protect\citeauthoryear{{Giocoli}, {Meneghetti}, {Bartelmann},
  {Moscardini} \& {Boldrin}}{{Giocoli} et~al.}{2012a}]{giocoli12}
{Giocoli} C.,  {Meneghetti} M.,  {Bartelmann} M.,  {Moscardini} L.,
  {Boldrin} M.,  2012a, \mnras, 421, 3343

\bibitem[\protect\citeauthoryear{{Giocoli}, {Moreno}, {Sheth} \&
  {Tormen}}{{Giocoli} et~al.}{2007}]{giocoli07a}
{Giocoli} C.,  {Moreno} J.,  {Sheth} R.~K.,    {Tormen} G.,  2007, \mnras, 376,
  977

\bibitem[\protect\citeauthoryear{{Giocoli}, {Tormen} \& {Sheth}}{{Giocoli}
  et~al.}{2012b}]{giocoli12b}
{Giocoli} C.,  {Tormen} G.,    {Sheth} R.~K.,  2012b, \mnras, 422, 185

\bibitem[\protect\citeauthoryear{{Giocoli}, {Tormen}, {Sheth} \& {van den
  Bosch}}{{Giocoli} et~al.}{2010a}]{giocoli10}
{Giocoli} C.,  {Tormen} G.,  {Sheth} R.~K.,    {van den Bosch} F.~C.,  2010a,
  \mnras, 404, 502

\bibitem[\protect\citeauthoryear{{Giocoli}, {Tormen} \& {van den
  Bosch}}{{Giocoli} et~al.}{2008}]{giocoli08b}
{Giocoli} C.,  {Tormen} G.,    {van den Bosch} F.~C.,  2008, \mnras, 386, 2135

\bibitem[\protect\citeauthoryear{{Gnedin}, {Ceverino}, {Gnedin}, {Klypin},
  {Kravtsov}, {Levine}, {Nagai} \& {Yepes}}{{Gnedin} et~al.}{2011}]{gnedin11}
{Gnedin} O.~Y.,  {Ceverino} D.,  {Gnedin} N.~Y.,  {Klypin} A.~A.,  {Kravtsov}
  A.~V.,  {Levine} R.,  {Nagai} D.,    {Yepes} G.,  2011, ArXiv e-prints

\bibitem[\protect\citeauthoryear{{Gottloeber}, {Yepes}, {Wagner} \&
  {Sevilla}}{{Gottloeber} et~al.}{2006}]{gottlober06}
{Gottloeber} S.,  {Yepes} G.,  {Wagner} C.,    {Sevilla} R.,  2006, ArXiv
  Astrophysics e-prints

\bibitem[\protect\citeauthoryear{{Hernquist}}{{Hernquist}}{1990}]{hernquist90}
{Hernquist} L.,  1990, \apj, 356, 359

\bibitem[\protect\citeauthoryear{{Hoekstra}}{{Hoekstra}}{2003}]{hoekstra03}
{Hoekstra} H.,  2003, \mnras, 339, 1155

\bibitem[\protect\citeauthoryear{{Hoekstra}, {Hartlap}, {Hilbert} \& {van
  Uitert}}{{Hoekstra} et~al.}{2011}]{hoekstra11}
{Hoekstra} H.,  {Hartlap} J.,  {Hilbert} S.,    {van Uitert} E.,  2011, \mnras,
  412, 2095

\bibitem[\protect\citeauthoryear{{Jaffe}}{{Jaffe}}{1983}]{jaffe83}
{Jaffe} W.,  1983, \mnras, 202, 995

\bibitem[\protect\citeauthoryear{{Jing}}{{Jing}}{2000}]{jing00}
{Jing} Y.~P.,  2000, \apj, 535, 30

\bibitem[\protect\citeauthoryear{{Jing} \& {Suto}}{{Jing} \&
  {Suto}}{2002}]{jing02}
{Jing} Y.~P.,  {Suto} Y.,  2002, \apj, 574, 538

\bibitem[\protect\citeauthoryear{{Kazantzidis}, {Kravtsov}, {Zentner},
  {Allgood}, {Nagai} \& {Moore}}{{Kazantzidis} et~al.}{2004}]{kazantzidis04}
{Kazantzidis} S.,  {Kravtsov} A.~V.,  {Zentner} A.~R.,  {Allgood} B.,  {Nagai}
  D.,    {Moore} B.,  2004, \apjl, 611, L73

\bibitem[\protect\citeauthoryear{{Keeton}}{{Keeton}}{2001}]{keeton01}
{Keeton} C.~R.,  2001, \apj, 561, 46

\bibitem[\protect\citeauthoryear{{Komatsu}, {Smith}, {Dunkley}, {Bennett},
  {Gold}, {Hinshaw}, {Jarosik}, {Larson}, {Nolta} \& {et~al.}}{{Komatsu}
  et~al.}{2011}]{komatsu11}
{Komatsu} E.,  {Smith} K.~M.,  {Dunkley} J.,  {Bennett} C.~L.,  {Gold} B.,
  {Hinshaw} G.,  {Jarosik} N.,  {Larson} D.,  {Nolta} M.~R.,    {et~al.} 2011,
  \apjs, 192, 18

\bibitem[\protect\citeauthoryear{{Kuijken}}{{Kuijken}}{2010}]{kuijken10}
{Kuijken} K.,  2010, in {Block} D.~L.,  {Freeman} K.~C.,   {Puerari} I.,  eds,
  Galaxies and their Masks {Dark Haloes as Seen with Gravitational Lensing}.
p.~361

\bibitem[\protect\citeauthoryear{{Lacey} \& {Cole}}{{Lacey} \&
  {Cole}}{1993}]{lacey93}
{Lacey} C.,  {Cole} S.,  1993, \mnras, 262, 627

\bibitem[\protect\citeauthoryear{{Laureijs}, {Amiaux}, {Arduini},
  {Augu{\`e}res}, {Brinchmann}, {Cole}, {Cropper}, {Dabin}, {Duvet} \& et
  al.}{{Laureijs} et~al.}{2011}]{euclidredbook}
{Laureijs} R.,  {Amiaux} J.,  {Arduini} S.,  {Augu{\`e}res} J.~.,  {Brinchmann}
  J.,  {Cole} R.,  {Cropper} M.,  {Dabin} C.,  {Duvet} L.,    et al. 2011,
  ArXiv e-prints

\bibitem[\protect\citeauthoryear{{LSST Science Collaborations}, {Abell},
  {Allison}, {Anderson}, {Andrew}, {Angel}, {Armus}, {Arnett}, {Asztalos},
  {Axelrod} \& et al.}{{LSST Science Collaborations} et~al.}{2009}]{anderson01}
{LSST Science Collaborations} {Abell} P.~A.,  {Allison} J.,  {Anderson} S.~F.,
  {Andrew} J.~R.,  {Angel} J.~R.~P.,  {Armus} L.,  {Arnett} D.,  {Asztalos}
  S.~J.,  {Axelrod} T.~S.,    et al. 2009, ArXiv e-prints

\bibitem[\protect\citeauthoryear{{Macci{\`o}}, {Dutton} \& {van den
  Bosch}}{{Macci{\`o}} et~al.}{2008}]{maccio08}
{Macci{\`o}} A.~V.,  {Dutton} A.~A.,    {van den Bosch} F.~C.,  2008, \mnras,
  391, 1940

\bibitem[\protect\citeauthoryear{{Macci{\`o}}, {Dutton}, {van den Bosch},
  {Moore}, {Potter} \& {Stadel}}{{Macci{\`o}} et~al.}{2007}]{maccio07}
{Macci{\`o}} A.~V.,  {Dutton} A.~A.,  {van den Bosch} F.~C.,  {Moore} B.,
  {Potter} D.,    {Stadel} J.,  2007, \mnras, 378, 55

\bibitem[\protect\citeauthoryear{{Meneghetti}, {Argazzi}, {Pace}, {Moscardini},
  {Dolag}, {Bartelmann}, {Li} \& {Oguri}}{{Meneghetti}
  et~al.}{2007b}]{meneghetti07b}
{Meneghetti} M.,  {Argazzi} R.,  {Pace} F.,  {Moscardini} L.,  {Dolag} K.,
  {Bartelmann} M.,  {Li} G.,    {Oguri} M.,  2007b, \aap, 461, 25

\bibitem[\protect\citeauthoryear{{Meneghetti}, {Bartelmann} \&
  {Moscardini}}{{Meneghetti} et~al.}{2003}]{meneghetti03}
{Meneghetti} M.,  {Bartelmann} M.,    {Moscardini} L.,  2003, \mnras, 346, 67

\bibitem[\protect\citeauthoryear{{Meneghetti}, {Fedeli}, {Zitrin},
  {Bartelmann}, {Broadhurst}, {Gottl{\"o}ber}, {Moscardini} \&
  {Yepes}}{{Meneghetti} et~al.}{2011}]{meneghetti11}
{Meneghetti} M.,  {Fedeli} C.,  {Zitrin} A.,  {Bartelmann} M.,  {Broadhurst}
  T.,  {Gottl{\"o}ber} S.,  {Moscardini} L.,    {Yepes} G.,  2011, \aap, 530,
  A17+

\bibitem[\protect\citeauthoryear{{Meneghetti}, {Rasia}, {Merten}, {Bellagamba},
  {Ettori}, {Mazzotta}, {Dolag} \& {Marri}}{{Meneghetti}
  et~al.}{2010b}]{meneghetti10b}
{Meneghetti} M.,  {Rasia} E.,  {Merten} J.,  {Bellagamba} F.,  {Ettori} S.,
  {Mazzotta} P.,  {Dolag} K.,    {Marri} S.,  2010b, \aap, 514, A93+

\bibitem[\protect\citeauthoryear{{Metcalf} \& {Madau}}{{Metcalf} \&
  {Madau}}{2001}]{metcalf01}
{Metcalf} R.~B.,  {Madau} P.,  2001, \mnras, 563, 9

\bibitem[\protect\citeauthoryear{{Morandi}, {Pedersen} \& {Limousin}}{{Morandi}
  et~al.}{2010}]{morandi10}
{Morandi} A.,  {Pedersen} K.,    {Limousin} M.,  2010, \apj, 713, 491

\bibitem[\protect\citeauthoryear{{Navarro}, {Frenk} \& {White}}{{Navarro}
  et~al.}{1997}]{navarro97}
{Navarro} J.~F.,  {Frenk} C.~S.,    {White} S.~D.~M.,  1997, \apj, 490, 493

\bibitem[\protect\citeauthoryear{{Neto}, {Gao}, {Bett}, {Cole}, {Navarro},
  {Frenk}, {White}, {Springel} \& {Jenkins}}{{Neto} et~al.}{2007}]{neto07}
{Neto} A.~F.,  {Gao} L.,  {Bett} P.,  {Cole} S.,  {Navarro} J.~F.,  {Frenk}
  C.~S.,  {White} S.~D.~M.,  {Springel} V.,    {Jenkins} A.,  2007, \mnras,
  381, 1450

\bibitem[\protect\citeauthoryear{{Oguri}, {Bayliss}, {Dahle}, {Sharon},
  {Gladders}, {Natarajan}, {Hennawi} \& {Koester}}{{Oguri}
  et~al.}{2012}]{oguri12}
{Oguri} M.,  {Bayliss} M.~B.,  {Dahle} H.,  {Sharon} K.,  {Gladders} M.~D.,
  {Natarajan} P.,  {Hennawi} J.~F.,    {Koester} B.~P.,  2012, \mnras, 420,
  3213

\bibitem[\protect\citeauthoryear{{Oguri}, {Takada}, {Umetsu} \&
  {Broadhurst}}{{Oguri} et~al.}{2005}]{oguri05}
{Oguri} M.,  {Takada} M.,  {Umetsu} K.,    {Broadhurst} T.,  2005, \apj, 632,
  841

\bibitem[\protect\citeauthoryear{{Okabe}, {Okura} \& {Futamase}}{{Okabe}
  et~al.}{2010a}]{okabe10}
{Okabe} N.,  {Okura} Y.,    {Futamase} T.,  2010a, \apj, 713, 291

\bibitem[\protect\citeauthoryear{{Okabe}, {Takada}, {Umetsu}, {Futamase} \&
  {Smith}}{{Okabe} et~al.}{2010b}]{okabe10b}
{Okabe} N.,  {Takada} M.,  {Umetsu} K.,  {Futamase} T.,    {Smith} G.~P.,
  2010b, \pasj, 62, 811

\bibitem[\protect\citeauthoryear{{Perlmutter}, {Aldering}, {della Valle},
  {Deustua}, {Ellis}, {Fabbro}, {Fruchter}, {Goldhaber}, {Groom} \& et
  al.}{{Perlmutter} et~al.}{1998}]{perlmutter98}
{Perlmutter} S.,  {Aldering} G.,  {della Valle} M.,  {Deustua} S.,  {Ellis}
  R.~S.,  {Fabbro} S.,  {Fruchter} A.,  {Goldhaber} G.,  {Groom} D.~E.,    et
  al. 1998, Nature, 391, 51

\bibitem[\protect\citeauthoryear{{Perlmutter}, {Aldering}, {Goldhaber}, {Knop},
  {Nugent}, {Castro}, {Deustua}, {Fabbro}, {Goobar} \& et al.}{{Perlmutter}
  et~al.}{1999}]{perlmutter99}
{Perlmutter} S.,  {Aldering} G.,  {Goldhaber} G.,  {Knop} R.~A.,  {Nugent} P.,
  {Castro} P.~G.,  {Deustua} S.,  {Fabbro} S.,  {Goobar} A.,    et al. 1999,
  \apj, 517, 565

\bibitem[\protect\citeauthoryear{{Rasia}, {Meneghetti}, {Martino}, {Borgani},
  {Bonafede}, {Dolag}, {Ettori}, {Fabjan}, {Giocoli}, {Mazzotta}, {Merten},
  {Radovich} \& {Tornatore}}{{Rasia} et~al.}{2012}]{rasia12}
{Rasia} E.,  {Meneghetti} M.,  {Martino} R.,  {Borgani} S.,  {Bonafede} A.,
  {Dolag} K.,  {Ettori} S.,  {Fabjan} D.,  {Giocoli} C.,  {Mazzotta} P.,
  {Merten} J.,  {Radovich} M.,    {Tornatore} L.,  2012, ArXiv e-prints

\bibitem[\protect\citeauthoryear{{Rix}, {de Zeeuw}, {Cretton}, {van der Marel}
  \& {Carollo}}{{Rix} et~al.}{1997}]{rix97}
{Rix} H.-W.,  {de Zeeuw} P.~T.,  {Cretton} N.,  {van der Marel} R.~P.,
  {Carollo} C.~M.,  1997, \apj, 488, 702

\bibitem[\protect\citeauthoryear{{Sereno} \& {Zitrin}}{{Sereno} \&
  {Zitrin}}{2012}]{sereno12}
{Sereno} M.,  {Zitrin} A.,  2012, \mnras, 419, 3280

\bibitem[\protect\citeauthoryear{{Shaw}, {Weller}, {Ostriker} \& {Bode}}{{Shaw}
  et~al.}{2006}]{shaw06}
{Shaw} L.~D.,  {Weller} J.,  {Ostriker} J.~P.,    {Bode} P.,  2006, \apj, 646,
  815

\bibitem[\protect\citeauthoryear{{Sheth}, {Mo} \& {Tormen}}{{Sheth}
  et~al.}{2001}]{sheth01b}
{Sheth} R.~K.,  {Mo} H.~J.,    {Tormen} G.,  2001, \mnras, 323, 1

\bibitem[\protect\citeauthoryear{{Sheth} \& {Tormen}}{{Sheth} \&
  {Tormen}}{1999}]{sheth99b}
{Sheth} R.~K.,  {Tormen} G.,  1999, \mnras, 308, 119

\bibitem[\protect\citeauthoryear{{Sheth} \& {Tormen}}{{Sheth} \&
  {Tormen}}{2004}]{sheth04a}
{Sheth} R.~K.,  {Tormen} G.,  2004, \mnras, 349, 1464

\bibitem[\protect\citeauthoryear{{Springel}, {White}, {Jenkins}, {Frenk},
  {Yoshida}, {Gao}, {Navarro}, {Thacker}, {Croton}, {Helly}, {Peacock}, {Cole},
  {Thomas}, {Couchman}, {Evrard}, {Colberg} \& {Pearce}}{{Springel}
  et~al.}{2005}]{springel05b}
{Springel} V.,  {White} S.~D.~M.,  {Jenkins} A.,  {Frenk} C.~S.,  {Yoshida} N.,
   {Gao} L.,  {Navarro} J.,  {Thacker} R.,  {Croton} D.,  {Helly} J.,
  {Peacock} J.~A.,  {Cole} S.,  {Thomas} P.,  {Couchman} H.,  {Evrard} A.,
  {Colberg} J.,    {Pearce} F.,  2005, Nature, 435, 629

\bibitem[\protect\citeauthoryear{{Springel}, {White}, {Tormen} \&
  {Kauffmann}}{{Springel} et~al.}{2001b}]{springel01b}
{Springel} V.,  {White} S.~D.~M.,  {Tormen} G.,    {Kauffmann} G.,  2001b,
  \mnras, 328, 726

\bibitem[\protect\citeauthoryear{{Tormen}, {Diaferio} \& {Syer}}{{Tormen}
  et~al.}{1998}]{tormen98b}
{Tormen} G.,  {Diaferio} A.,    {Syer} D.,  1998, \mnras, 299, 728

\bibitem[\protect\citeauthoryear{{Umetsu}, {Birkinshaw}, {Liu}, {Wu},
  {Medezinski}, {Broadhurst}, {Lemze}, {Zitrin}, {Ho}, {Huang}, {Koch}, {Liao}
  et~al.,}{{Umetsu} et~al.}{2009}]{umetsu09}
{Umetsu} K.,  {Birkinshaw} M.,  {Liu} G.-C.,  {Wu} J.-H.~P.,  {Medezinski} E.,
  {Broadhurst} T.,  {Lemze} D.,  {Zitrin} A.,  {Ho} P.~T.~P.,  {Huang}
  C.-W.~L.,  {Koch} P.~M.,  {Liao}   et~al., 2009, \apj, 694, 1643

\bibitem[\protect\citeauthoryear{{van den Bosch}, {Tormen} \& {Giocoli}}{{van
  den Bosch} et~al.}{2005}]{vandenbosch05}
{van den Bosch} F.~C.,  {Tormen} G.,    {Giocoli} C.,  2005, \mnras, 359, 1029

\bibitem[\protect\citeauthoryear{{Wang}, {Li}, {Kauffmann} \& {De
  Lucia}}{{Wang} et~al.}{2006}]{wang06}
{Wang} L.,  {Li} C.,  {Kauffmann} G.,    {De Lucia} G.,  2006, \mnras, 371, 537

\bibitem[\protect\citeauthoryear{{Wechsler}, {Bullock}, {Primack}, {Kravtsov}
  \& {Dekel}}{{Wechsler} et~al.}{2002}]{wechsler02}
{Wechsler} R.~H.,  {Bullock} J.~S.,  {Primack} J.~R.,  {Kravtsov} A.~V.,
  {Dekel} A.,  2002, \apj, 568, 52

\bibitem[\protect\citeauthoryear{{White} \& {Rees}}{{White} \&
  {Rees}}{1978}]{white78}
{White} S.~D.~M.,  {Rees} M.~J.,  1978, \mnras, 183, 341

\bibitem[\protect\citeauthoryear{{Zhao}, {Jing}, {Mo} \& {Bn{\"o}rner}}{{Zhao}
  et~al.}{2009}]{zhao09}
{Zhao} D.~H.,  {Jing} Y.~P.,  {Mo} H.~J.,    {Bn{\"o}rner} G.,  2009, \apj,
  707, 354

\bibitem[\protect\citeauthoryear{{Zitrin}, {Broadhurst}, {Umetsu}, {Coe},
  {Ben{\'{\i}}tez}, {Ascaso}, {Bradley}, {Ford}, {Jee}, {Medezinski},
  {Rephaeli} \& {Zheng}}{{Zitrin} et~al.}{2009}]{zitrin09}
{Zitrin} A.,  {Broadhurst} T.,  {Umetsu} K.,  {Coe} D.,  {Ben{\'{\i}}tez} N.,
  {Ascaso} B.,  {Bradley} L.,  {Ford} H.,  {Jee} J.,  {Medezinski} E.,
  {Rephaeli} Y.,    {Zheng} W.,  2009, \mnras, 396, 1985

\end{thebibliography}
\label{lastpage}
\end{document}